\documentclass[12pt]{article}
\usepackage{mathrsfs}
\usepackage{cite}
\usepackage{hyperref}
\usepackage{verbatim}
\hypersetup{colorlinks=true,backref=true,linkcolor=black,anchorcolor=black,citecolor=black,filecolor   =black,menucolor=black,pagecolor=black,urlcolor=black}
\usepackage[english]{babel}
\newlength{\abstractwidth}
\usepackage{epsf}
\usepackage{color}
\usepackage{graphicx}
\usepackage{amssymb}
\usepackage{latexsym}
\usepackage{epstopdf}

\usepackage{amsmath}
\usepackage[normalem]{ulem}

\usepackage{color}
\definecolor{dgreen}{rgb}{0,0.70,0.30}
\definecolor{gold}{rgb}{0.85,.66,0}
\definecolor{purple}{rgb}{1.0,0.3,0.6}
\definecolor{red}{rgb}{1.0,0.0,0.0}

\usepackage{tikz}
\usetikzlibrary{calc,decorations.pathreplacing}

\def\cA{{\cal A}}

\def\cC{{\cal C}}
\def\cD{{\cal D}}

\def\cF{{\cal F}}
\def\cG{{\cal G}}
\def\cH{{\cal H}}

\def\cJ{{\cal J}}

\def\cL{{\cal L}}
\def\cM{{\cal M}}
\def\cN{{\cal N}}
\def\cO{{\cal O}}

\def\cS{{\cal S}}
\def\cT{{\cal T}}
\def\cU{{\cal U}}
\def\cV{{\cal V}}

\def\cY{{\cal Y}}

\def\mF{\mathfrak{F}}

\def\mJ{\mathfrak{J}}

\def\mN{\mathfrak{N}}

\def\mS{\mathfrak{S}}
\def\mT{\mathfrak{T}}

\def\me{\mathfrak{e}}
\def\mg{\mathfrak{g}}

\def\mm{\mathfrak{m}}

\def\mpp{\mathfrak{p}}

\def\p{\partial}
\def\half{{1\over 2}}

\def\ep{\varepsilon}
\def\veps{\varepsilon}
\def\om{\omega}
\def\pp{ \prime \prime}

\def\hp{{\hat +}}
\def\hm{{\hat -}}
\def\hr{{\hat r}}
\def\hu{{\hat u}}
\def\hv{{\hat v}}
\def\hmu{{\hat \mu}}

\def\no{\nonumber}
\def\sm{\smallskip}

\def\CC{{\mathbb C}}
 
\def\RR{{\mathbb R}}

\def\ZZ{{\mathbb Z}}

\renewcommand{\thefootnote}{\fnsymbol{footnote}}
\renewcommand{\thanks}[1]{\footnote{#1}}
\newcommand{\starttext}{
\setcounter{footnote}{0}
\renewcommand{\thefootnote}{\arabic{footnote}}}

\newcommand{\bea}{\begin{eqnarray}}
\newcommand{\eea}{\end{eqnarray}}

\newcommand{\<}{\langle}
\renewcommand{\>}{\rangle}

\begin{document}
\starttext
\setcounter{footnote}{0}

\begin{flushright}
2016 June 14 \\
\end{flushright}

\bigskip

\begin{center}

{\Large \bf Emergent Super-Virasoro on Magnetic Branes\footnote{This research has been supported in part by National Science Foundation grant PHY-13-13986.}}

\vskip 2cm

{\large \bf Eric D'Hoker and Bijan Pourhamzeh}

\vskip 0.1in

{ \sl  Department of Physics and Astronomy }

{\sl University of California, Los Angeles, CA 90095, USA} 

\vskip 0.1in

{\tt  dhoker@physics.ucla.edu; bijan@physics.ucla.edu}

\end{center}

\vskip -1 in

\begin{abstract}

The low energy limit of  the stress tensor, gauge current, and supercurrent two-point correlators 
are calculated in the background of the supersymmetric magnetic brane solution to gauged  
five-dimensional supergravity constructed by Almuhairi and Polchinski. The resulting correlators 
provide evidence for the emergence of an $\cN=2$ super-Virasoro algebra of right-movers, in addition to a bosonic  Virasoro algebra and a $U(1)\oplus U(1)$-current algebra of left-movers (or the parity transform of left- and right-movers depending on the sign of the magnetic field), in the holographically dual strongly interacting two-dimensional effective field theory of the lowest Landau level.
 
\end{abstract}

\newpage
\setcounter{tocdepth}{2}

\newpage     


\section{Introduction}
\setcounter{equation}{0}
\label{sec:Intro}

Holography provides a powerful method for the study of strongly interacting gauge theories with
fermionic matter. It allows for a geometric interpretation of renormalization group (RG) flow in the 
dual gravity theory in terms of motion along a holographic coordinate. The dual geometry of a UV 
fixed point in the  gauge theory is asymptotic to an $AdS$ space-time, while that of an IR fixed point is asymptotic to  another $AdS$. The dimensions of the UV and IR asymptotic $AdS$ geometries 
need not be the same,  and often differ from one another in concrete solutions. For reviews on holographic methods, see  for example \cite{Aharony:1999ti,D'Hoker:2002aw,Hartnoll:2009sz,Herzog:2009xv}.

\sm

The case of four-dimensional $\cN=4$ supersymmetric Yang-Mills in the presence of an external magnetic field provides a non-trivial illustration of an RG flow between two fixed points which is physically relevant. The external magnetic field is associated with the gauging of a  $U(1)$ subgroup of the $SU(4)$ R-symmetry group of $\cN=4$ super Yang-Mills,  and couples to the scalars and gauginos of the theory, but not to its gauge fields.  In the low energy limit, only fermions in the lowest Landau level contribute, and their dynamics is confined to the spatial dimension along  the magnetic field. The IR fixed point theory thus consists of an effective two-dimensional  conformal field theory (CFT) of strongly interacting fermions of the Luttinger-liquid type (see for example \cite{Giamarchi} on strongly interacting fermion systems in one spacial dimension). 

\sm

The holographic dual to the above field theory set-up is a magnetic brane, which was constructed in 
\cite{D'Hoker:2009mm} (see also \cite{D'Hoker:2012ih} for a review) as a solution to  minimal 
five-dimensional gauged supergravity. The fact that minimal five-dimensional supergravity is  a consistent truncation of Type~IIB supergravity was established in \cite{Gauntlett:2007ma},
building on earlier results in \cite{Gauntlett:2006ai}, and   guarantees that the 
solutions of \cite{D'Hoker:2009mm} can be lifted up to the UV completion, namely  Type~IIB string 
theory. The magnetic brane is a smooth solution which interpolates between an asymptotic 
$AdS_5$ in the UV and an asymptotic $AdS_3 \times T^2$ in the IR.  The torus  $T^2$ occupies the two spatial dimensions perpendicular to the magnetic field, and may be represented by $\CC/\Lambda$ for a lattice $\omega_1 \ZZ + \omega_2 \ZZ$ with arbitrary period $\omega _1, \omega _2 \in \CC$.  This geometric  picture indeed reflects the expected dual RG flow from four-dimensional $\cN=4$ super Yang-Mills 
to a two-dimensional CFT. The qualitatively different IR behavior which occurs in superconductors in the presence of an external magnetic field  has been studied by holographic 
methods as well, for example, in \cite{Albash:2008eh,Albash:2009ix}. 

\sm

The asymptotic symmetry of $AdS_3$ is enhanced from the $SO(2,2)$ isometry of $AdS_3$ 
to left- and right-moving copies of the Virasoro algebra \cite{Brown:1986nw}, characteristic 
of a dual two-dimensional CFT.  A holographic calculation of two-point correlators of the $U(1)$ 
current and stress tensor in the IR reveals the presence of a single chiral current algebra as well 
as left-and right-moving Virasoro algebras \cite{D'Hoker:2010hr}. The coordinate transformations 
on $AdS_{3}$ by which  these Virasoro symmetries act in the IR originate in the UV from physical 
deformations on $AdS_{5}$ which are not pure coordinate transformations.  This effect provides a
holographic realization for the {\sl emergence of symmetries in the IR} which were not present in the UV.

\sm

The magnetic brane solution discussed above preserves no supersymmetry, and {\sl minimal }
five-dimensional supergravity has no magnetic solutions that do. Correspondingly, the 
supersymmetry of the $\cN=4$ theory is completely broken in the IR limit, as the energy levels 
of scalars and gauginos are split by the magnetic field. As a result, the low energy behavior is 
entirely in terms of fermions.

\sm

A generalization of the magnetic brane was proposed in \cite{Almuhairi:2010rb} within the 
framework of  a {\sl non-minimal} gauged five-dimensional supergravity in which the gauged 
$SU(4)$ is truncated to its $U(1)^3$ Cartan subgroup \cite{Gunaydin:1983bi,Gunaydin:1984ak}
(see also \cite{Cacciatori:2003kv} for domain wall solutions in this theory). 
In addition to the fields of the minimal five-dimensional supergravity, this non-minimal supergravity 
further contains two Maxwell super-multiplets, thereby adding a pair of Maxwell gauge
fields, two real scalars, and two gauginos. Embedding the magnetic field into 
the truncated $U(1)^3$ gauge group leads to a supersymmetric  magnetic brane \cite{Almuhairi:2011ws}.
More precisely, the supersymmetric magnetic brane is actually a two-parameter family 
of solutions, one parameter being the magnitude of the magnetic field, the other parametrizing 
its  embedding  into $U(1)^3$. A smooth supersymmetric magnetic brane solution was shown to 
exist via numerical methods in \cite{Donos:2011pn}  for a special choice of  embedding with enhanced symmetry. To realize the corresponding low energy supersymmetry in the dual gauge theory, 
it suffices to turn on a suitable constant background auxiliary $D$-field in addition to the constant 
background magnetic field, as was shown in \cite{Almuhairi:2011ws}. 

\sm

The supersymmetric magnetic brane solution is again asymptotic to an $AdS_3 \times T^2$ 
space-time, and the IR fixed point of the dual theory is again a two-dimensional CFT. 
However, the universality classes in the IR of the duals to the supersymmetric and non-supersymmetric magnetic branes are different. The dual to the non-supersymmetric magnetic brane contains only fermions in the IR, while the dual to the supersymmetric brane contains both fermions and bosons in the IR, and exhibits supersymmetry. 

\sm

In the present paper, we shall argue that the supersymmetric magnetic brane solution has an asymptotic symmetry governed  by a unitary chiral $\cN=2$ super Virasoro algebra for one chirality, and a purely bosonic unitary chiral Virasoro algebra plus two unitary chiral $U(1)$ current algebras for the other chirality. To do so, we shall compute the two-point functions for the stress tensor, the $U(1)^3$ currents, and the supercurrent  in the low energy limit. In the supergravity theory, these correlators may be extracted from the perturbations
of the metric,  the Maxwell gauge fields, and the gravitinos and gauginos respectively. We shall solve the linearized field equations for the perturbations, and use the method of overlapping expansions to extract the correlators. 

\sm

We shall then show that the functional form of these correlators is consistent with the emergence in the IR of the symmetries, including the $\cN=2$ super Virasoro algebra,  announced earlier in this paragraph. In addition, the overall normalizations of the identity operator in the OPE of two stress tensors, and of two supercurrents, are accessible from the calculation of the two-point correlators of these operators, and are shown to match precisely with the form required by the $\cN=2$ superconfomal algebra. 
The corresponding calculation of the absolute normalization for two $U(1)$ currents is significantly complicated by the mixing effects of the three $U(1)$ gauge fields by the Chern-Simons term, and a derivation of the absolute normalization of the current will not be achieved here, but will be left for future work. 

\sm

The calculations of these correlators generally follow the procedures used in 
\cite{D'Hoker:2010hr} for the minimal supergravity. For the case of non-minimal supergravity 
of interest  here, however,  they become considerably more involved, especially for the correlators of
the gauge currents and supercurrent. We shall take this opportunity to present the derivations of the  proper normalizations of the holographically renormalized supercurrent in some detail.

\subsection{Organization}

The present paper is organized as follows. In Section~\ref{sec:2} we briefly review the essentials 
of the non-minimal five-dimensional supergravity theory and the formalism for the holographic 
calculation of stress tensor and current correlators. We discuss the structure of the supersymmetric 
magnetic brane solutions and demonstrate their existence numerically for a wide range of parameters.  
In Section~\ref{sec:3}, we compute the correlators for the stress tensor in the IR limit, 
following closely the methods used in \cite{D'Hoker:2010hr}. In Section~\ref{sec:4}
we compute the correlators for the $U(1)^3$ currents in the IR limit, and disentangle their chirality 
dependence on the embedding parameters. In Section~\ref{sec:5} we review the formalism for 
the holographic calculation of the fermionic fields in supergravity, and extract the supercurrent
two-point function in the IR limit. In Section~\ref{sec:6} we discuss the emergence of the super Virasoro symmetry in the IR limit, by putting together the information gathered from the preceding correlator calculations. A brief discussion of our results and outlook to future work is presented in section \ref{sec:7}. 
In Appendix~\ref{sec:A}, a comprehensive overview is presented of non-minimal five-dimensional supergravity, in which we pay careful attention to the various normalizations used in the existing literature. 
The construction and renormalization of the holographic supercurrent for this theory  is presented in detail in Appendix \ref{sec:F}. The asymptotic expansion of the Fermi fields is relegated to Appendix  \ref{sec:C}.

\section{Supersymmetric magnetic brane solution
\label{sec:2}}
\setcounter{equation}{0}

In this section, we shall give a synopsis of non-minimal five-dimensional  gauged supergravity \cite{Gunaydin:1983bi,Gunaydin:1984ak}, and discuss the supersymmetric magnetic brane solutions including their symmetries and asymptotic behavior.\footnote{A detailed review of non-minimal five-dimensional supergravity, including the notations and conventions used in this paper,  is relegated to Appendix \ref{sec:A}. In particular, summation over repeated indices will be assumed throughout, unless explicitly stated otherwise.} We shall also present  numerical evidence confirming the existence of the supersymmetric magnetic brane as a regular global solution interpolating between $AdS_5$ in the UV and $AdS_3 \times T^2$ in the IR for a wide range of parameters.

\subsection{Five dimensional supergravity synopsis}

The starting point is the $U(1)^{3}$ truncation of gauged five-dimensional supergravity with gauge group $SU(4)$. This supergravity is a truncation of the holographic dual to $\cN=4$ four-dimensional super-Yang Mills. The bosonic fields are the space-time metric $g_{MN}$ where $M,N=0,1,2,3,4$ denote Einstein indices, three Maxwell fields $A_M ^I$ labelled by $I=1,2,3$, and two neutral scalars $\phi^A$ with coordinate index $A=1,2$.  The fermionic fields are the gravitino $\psi _M$ and the gaugino $\lambda ^a$ with frame index  $a=1,2$, each of which is a doublet under the $SU(2)$ R-symmetry, and is subject to the symplectic-Majorana condition.

\sm

The complete supergravity action $S_{{\rm sugra}}$  will be given by,
\bea
\label{2a1}
S_{{\rm sugra}} = { 1 \over 8 \pi G_5} \int d^5x \sqrt{g} \Big ( \cL_0 + \cL_2 + \cL_4 \Big ) + S_{{\rm bndy}} + S_\text{ct}
\eea
Here, $G_5$ is Newton's constant in five space-time dimensions, $g=-\det (g_{MN})$,  while
$\cL_0, \cL_2$, $\cL_4$ refer to those parts of the classical Lagrangian density which are homogeneous  
in Fermi fields of degrees zero, two, and four respectively. For the purpose of holographic 
calculation and renormalization the space-time of interest will ultimately be asymptotically 
$AdS_5$ and will require a regularization cut-off near the boundary of $AdS_5$. These
holographic procedures will require the addition of a boundary term $S_{{\rm bndy}}$ and 
a counter-term $S_\text{ct}$  needed for holographic renormalization \cite{Balasubramanian:1998sn, Balasubramanian:1999re,Balasubramanian:1999jd,deHaro:2000vlm,Skenderis:2002wp}, which are computed in Appendix \ref{sec:F}.  

\subsubsection{Bosonic part}

The bosonic part of the Lagrangian density is given by,
\bea
\cL_0 & = &
-\frac{1}{2}R_g
-\frac{1}{4}G_{IJ}F_{MN}^{I}F^{JMN}
-\frac{1}{2}\cG_{AB}\partial_M \phi^{A}\partial^M \phi^{B}  - \mg^{2} P 
\no \\ &&
+ { 1 \over 48} \frac{\ep^{MNPQS}}{ \sqrt{g} } \, 
C_{IJK}F_{MN}^{I}F_{PQ}^{J}A_S^{K}
\label{2a2}
\eea
Here $\ep ^{MNPQS}$ is the totally anti-symmetric symbol in five dimensions,
$F^I_{MN}= \p_M A_N^I - \p_N A^I_M$ is the field strength of $A_M ^I$, 
and $\mg$ is the gauge coupling constant.   The rank three totally symmetric tensor 
$C_{IJK}$ is  constant by $U(1)^3$ gauge invariance. With the above normalization 
in the Lagrangian, its only non-zero  component is $C_{123}=1$ and permutations 
thereof with all other components vanishing \cite{Cvetic:1999xp}. The potential $P$ is given by, 
\bea
P= - 6 \left(X_{1}+X_{2}+X_{3}\right)
\eea
while the metrics $G_{IJ}$ and $\cG_{AB}$ take the form, 
\bea 
G_{IJ}=\frac{\delta_{IJ}}{2\left(X^{I}\right)^{2}}
\hskip 0.6in
\cG_{AB} = \half \delta _{AB}
\label{eq:V_G}
\eea
Both metrics are flat, a result which is special to the $U(1)^3$ case, as was 
shown in \cite{Cvetic:1999xp}.
The real scalar fields $X^{I} (\phi )$ satisfy the constraint $X^{1}X^{2}X^{3}=1$. 
A convenient parametrization of $X^I$ in terms of $\phi^A$ (on the branch where $X^I >0$
for all $I=1,2,3$)  is as follows,
\bea
\label{Xphi}
X^I= e^{- a^I _A \phi^A} 
& \hskip 0.6in &
a_1^I=  (1,1,-2)^I /\sqrt{6}
\no \\ && 
a_2^I =  (1,-1,0)^I / \sqrt{2}
\eea
The field equations for the metric $g_{MN}$, the Maxwell fields $A_M ^I$, and the scalars $\phi^A$
in the presence of vanishing Fermi fields are as follows,  
\bea
\label{fieldeqs}
0 & = & R_{MN}+G_{IJ} \left( g^{PQ} F_{MP}^ I F_{NQ}^{J}
-\frac{1}{6} g_{MN} F_{PQ}^{I} F^{ J PQ} \right)
+\frac{1}{2}\delta_{AB} \partial_M \phi^{A} \partial_N \phi^{B}
+\frac{2}{3}\mg^2 g_{MN}P 
\no \\
0 & = & \partial_M \left(\sqrt{g} \, G_{IJ}
F^{J MS}\right)+{ 1 \over 16} \ep^{MNPQS}
C_{IJK}F_{MN}^{J}F_{PQ}^{K} 
\no \\
0 & = & \delta _{AB} \Delta _g \phi^B + 12 \, \mg^2 a_A ^I X_I
-\frac{9}{4} \sum _{I=1}^3 F_{MN}^{I}F^{IMN} \partial_A \left(X_{I}\right)^{2}
\eea
where $\partial_A$ are the partial derivative with respect to $\phi^A$, 
$a_A^I$ are given in (\ref{Xphi}),  and $\Delta_g$ is the scalar Laplacian 
for the space-time metric $g_{\mu \nu}$  defined by 
$\Delta_g \phi = \sqrt{g^{-1}} \, \p_M ( \sqrt{g} g^{MN} \p_N \phi )$.

\subsubsection{Fermionic part}

The Lagrangian densities $\cL_2$ and $\cL_4$ were derived in \cite{Gunaydin:1984ak}.
The terms bilinear in the fermions $\psi_{M}$ and $\lambda ^a$ have been collected in $\cL_2$ and
are reviewed in (\ref{A18}) of Appendix \ref{sec:A}, while $\cL_4$ will not be needed for the 
calculations of the correlators, and will not be presented here.

\sm
 
The fermion field equations, to linear order in $\psi_{M}$ and $\lambda ^a$, may be 
found in (\ref{A23a}) and (\ref{A23b}), where the $SU(2)$ R-symmetry doublets $\psi_M$ 
and $\lambda ^a$ have been decomposed into pairs of single-component Dirac spinors 
$\psi_{M\pm}$ and $\lambda ^a_\pm $. The field equations for the $+$ components of the gravitino 
$\psi_{M}= \psi _{M+} $ and of the gaugino $\lambda^a= \lambda ^a _+$ are given by,
\bea
\label{2a6}
\Psi^{M}=\Lambda^{a}=0
\eea
where we have defined,
\bea
\label{2a7}
\Psi^{M} & = & 
\Gamma^{MNP}\mathcal{D}_{N}\psi_{P}
+\frac{3i}{8}X_{I}\left(\Gamma^{MNPR}\psi_{N}F_{PR}^{I}+2\psi_{N}F^{I\hspace{1pt} MN}\right)
-\frac{i}{2}\Gamma^{N}\Gamma^{M}\lambda^{a}f_{A}^{a}\partial_{N}\phi^A
\no \\ && 
-\frac{1}{4}\sqrt{\frac{3}{2}}X_{I}^{a}\Gamma^{NP}\Gamma^{M}\lambda^{a}F_{NP}^{I}
+\frac{3}{2} \mg \, \Gamma^{MN}\psi_{N}V_{I}X^{I}
-\frac{3i}{\sqrt{6}} \mg \, \Gamma^{M}\lambda^{a}V_{I}X^{Ia}
\no \\
\Lambda^{a} & = & 
\Gamma^{M}\mathcal{D}_{M}\lambda^{a}
+\frac{i}{2}\Gamma^{M}\Gamma^{N}\psi_{M}f_{A}^{a}\partial_{N}\phi^A
-\frac{1}{4}\sqrt{\frac{3}{2}}X_{I}^{a}\Gamma^{M}\Gamma^{NP}\psi_{M}F_{NP}^{I}
\no \\ && 
-\frac{i}{2}\left(\frac{1}{4}\delta^{ab}X_{I}+T^{abc}X_{I}^{c}\right)\Gamma^{MN}\lambda^{b}F_{MN}^{I}
-\frac{3i}{\sqrt{6}} \mg \, \Gamma^{M}\psi_{M}V_{I}X^{Ia}
-\frac{1}{\sqrt{6}} \mg \, \lambda^{b}P^{ab}
\hskip 0.5in 
\eea
The corresponding equations for the components $\psi _{M-} $ and $\lambda ^a _-$
of the $SU(2)$ doublets are given by equations (\ref{2a6}) and (\ref{2a7}) with 
the sign of $\mg$ reversed  $\mg \to -\mg $. The covariant derivative $\cD_M$ in (\ref{2a7})
is defined in (\ref{A7}) and (\ref{A8}) of Appendix \ref{sec:A}, while the frame $f_A^a$, 
the variables $X^{Ia}$, and the  tensor $P^{ab}$  are defined respectively in  
(\ref{A28}), (\ref{A22}), and (\ref{A21}).

\subsubsection{Supersymmetry transformations and the BPS equations}

The supersymmetry transformations, to lowest order in the Fermi fields, are as follows,
\bea
\delta\psi_{M} & = & \left ( \cD_{M}+\frac{i}{8}X_{I} F_{NP}^{I} \left ( \Gamma_{M}{} ^{NP}
-4\delta_{M}{} ^{N} \Gamma^P \right )-\frac{1}{2} \mg \, V_{I}X^{I}\Gamma_{M}\right ) \epsilon
\no \\
\delta\lambda_A & = & 
\left  ( -\frac{i}{2}\cG _{AB}\Gamma^{M}\partial_{M}\phi^{B}
+\frac{3}{8}\partial_{A}X_{I} F_{MN}^{I} \Gamma^{MN}
-\frac{3i}{2} \mg \, V_{I}\partial_{A}X^{I}\right ) \epsilon
\label{2a8}
\eea
Here $V_I$ is a constant vector which governs the $U(1)$ gauging specified in (\ref{A6}).
We are exhibiting the supersymmetry transformation on $\lambda_{A}= f_A ^a \lambda ^a$ 
in (\ref{2a8}) rather than 
on $\lambda^{a}$ in order to match the notations of  \cite{Almuhairi:2010rb,Cacciatori:2003kv}.
The full supersymmetry transformations, including all orders in the Fermi fields, were derived in \cite{Gunaydin:1984ak}.

\sm

The action $S_{{\rm sugra}}$ is invariant under the supersymmetry transformations (\ref{2a8})
on the fermions, along with the supersymmetry transformations on the Bose fields (which we
are not exhibiting here as we do not need them),  provided variations trilinear in the Fermi fields 
$\psi _M$ and $\lambda ^a$ are neglected. The Fermi field equations to linear order in the Fermi fields (\ref{2a7}) are, however,  invariant under (\ref{2a8}) to leading order in the Fermi fields without 
transforming the Bose fields.

\sm

The BPS equations are obtained by enforcing the conditions, 
\bea
\label{2a10}
\delta \psi _M = \delta \lambda ^a=0
\eea
on a configuration with vanishing Fermi fields. A bosonic field configuration is referred to as being BPS 
provided the BPS  equations (\ref{2a10}) admit a non-zero supersymmetry transformation $\epsilon$ 
subject to mild asymptotic conditions on $\epsilon$.

\subsection{Holographic asymptotics, stress tensor, current correlators}

The maximally symmetric solution to the  field equations for this non-minimal gauged 
supergravity is $AdS_5$ space-time obtained by setting $A_M ^I= \phi^A=0$. The only 
remaining non-trivial equation is then $R_{MN} = 4 \mg^2 \, g_{MN}$ whose 
maximally symmetric solution is an $AdS_5$  with radius $1/| \mg |$. $AdS_5$ admits 
the maximal number of 8 real supersymmetries.

\sm

We shall seek solutions which are asymptotically $AdS_{5}$ in the sense that they satisfy the  
Fefferman-Graham expansion. We shall choose the corresponding holographic coordinate 
$r=x^4$ and use the decomposition  $x^M = (x^\mu, r)$ with $\mu=0,1,2,3$ the four-dimensional 
Einstein index.  The asymptotic $AdS_5$ is chosen to be located at $r=+\infty$.  In these 
Fefferman-Graham coordinates, the metric admits the following 
expansion,\footnote{A more familiar choice of holographic Fefferman-Graham coordinate is 
given by $\rho = e^{-r}$ so that the boundary of $AdS_5$ is located at $\rho=0$, and the 
metric is   $ds^2 = d\rho^2/\rho^2 +g_{\mu\nu}\left(x, - \ln \rho \right)dx^{\mu}dx^{\nu}$.}
\bea
ds^{2} & = & dr^{2}+g_{\mu\nu}\left(x,r\right)dx^{\mu}dx^{\nu}
\no \\
g_{\mu\nu}\left(x,r\right) & = & e^{2r}g_{\mu\nu}^{\left(0\right)}(x)
+g_{\mu\nu}^{\left(2\right)}\left(x\right)+e^{-2r}g_{\mu\nu}^{\left(4\right)}(x)
+re^{-2r}g_{\mu\nu}^{\left(\ln\right)}\left(x\right)
+\cO(e^{-4r})
\label{eq:feff_g}
\eea
while the asymptotic expansions for the gauge fields and scalars are given by,
\bea
A_{\mu}^{I} (x,r) & = &
A_{\mu}^{I \, (0)} (x) +e^{-2r} A_{\mu}^{I \, (2)} (x)
+ \cO(e^{-4r})
\no \\
\phi ^A (x, r) & = & \phi ^{A \, (0)} (x) + e^{-2r} \phi ^{A \, (2)} (x)+ r \, e^{-2r} \phi ^{A \, ({\rm ln} )} (x) 
+ \cO(e^{-4r})
\label{eq:feff_A}
\eea
Here, $x$ stands for the dependence on $x^\mu$, while Fefferman-Graham gauge is governed by 
$g_{\mu r}=g_{r \mu}=0$, $g_{rr}=1$, and $A_r=0$. The holographic source fields are 
$g^{(0)} _{\mu \nu}$, $A_\mu ^{I\, (0)}$ and $\phi^{A\, (0)}$.  Use of the field equations in 
(\ref{fieldeqs}) shows that the coefficients
$g_{\mu\nu}^{(2)}$, $g_{\mu\nu}^{(\ln)}$, the trace of $g_{\mu\nu}^{(4)}$, and $\phi ^{A \, ({\rm ln})}$ 
are local functionals of $g_{\mu\nu}^{(0)}$,  $A_\mu ^{I \, (0)}$ and  $\phi^{A\, (0)}$. 

\sm

The response of the action $S_{{\rm sugra}}$ to infinitesimal variations of the source fields 
is given by the expectation values of the dual operators in the field theory  \cite{Balasubramanian:1998sn, Balasubramanian:1999re,Balasubramanian:1999jd,deHaro:2000vlm,Skenderis:2002wp}. In the present case, 
the response to the  variation of the source fields $g_{\mu \nu} ^{(0)}$, $A_\mu ^{I (0)}$, 
and $\phi ^{A (0)}$ is given by the expectation values 
$T^{\mu \nu}$, $J^\mu _I$ and $Y _A$ respectively of the stress tensor $\cT^{\mu \nu}$,  
the gauge current $\cJ^\mu _I$, and scalar operator $\cY_A$, 
\bea
\delta S_{{\rm sugra}}
=\int d^{4}x \sqrt{g^{(0)}}
\left(\frac{1}{2}T^{\mu\nu} \, \delta g_{\mu\nu}^{(0)}
+J^{\mu}_I \, \delta A_{\mu}^{I \, (0)}
+ Y_A \, \delta \phi ^{A \, (0)} \right)
\label{eq:var_S}
\eea
The expectation values are given in terms of the boundary field data by,
\bea
\label{TJY}
4\pi G_{5}T_{\mu\nu} & = & g_{\mu\nu}^{\left(4\right)}+\rm {local}
\no \\
4\pi G_{5}J_{\mu} ^I ~ & = & A_{\mu}^{I \, (2)}+\rm {local}
\no \\
2 \pi G_5 Y^A & = & \phi ^{A\, (2)} + {\rm local}
\eea
The indices $\mu, \nu$ are lowered with the help of $g_{\mu \nu} ^{(0)}$,
while the indices $I$ and $A$ are lowered respectively with the help of the 
metrics $G_{IJ} (\phi)$ and $\cG_{AB}(\phi)=\delta _{AB}/2$ evaluated at the fields $\phi^{A \, (0)}$. 
In equations (\ref{TJY}) the ``local'' terms refers to local functionals of $g_{\mu\nu}^{\left(0\right)}$,
$A_{\mu}^{I \, (0)}$, and $\phi ^{A \, (0)}$ which will not contribute to 
two-point functions of local operators evaluated at distinct points, and will not be retained further.
 
\sm

The Fefferman-Graham expansion for the fermion fields $\psi _M$ and $\lambda ^a$ will involve more 
formalism and will be presented  in Section \ref{sec:5}.

\subsection{The supersymmetric magnetic brane solution}
\label{sec:23}

The magnetic brane solutions considered here are holographic duals to $\cN=4$ four-dimensional 
supersymmetric Yang-Mills theory in the presence of a constant uniform external magnetic field. 
The magnetic field is taken to be in the $1$-direction, perpendicular to the $23$-plane. 
The symmetries of this set-up are translation invariance along the four physical space-time directions 
$x^\mu $ with $\mu =0,1,2,3$, Lorentz invariance in the $01$-plane, and rotation invariance in the 
$23$-plane. The most general Ansatz, for the bosonic fields, which is consistent with  these 
symmetries in this supergravity theory is given as follows,
\bea
ds^{2} & = & dr^{2}+  e^{2W (r)} \eta _{mn} dx^m dx^n 
+e^{2U(r)} \delta _{ij} dx^i dx ^j
\no \\
F^{I} & = & F_{23}^{I} \, dx^{2}\wedge dx^{3}
\no \\
\phi^A & = & \phi^A(r)
\label{Ansatz}
\eea
where $\eta ={\rm diag} (-1,+1)$ is the flat Minkowski metric in the $01$-plane 
while $\delta _{ij}$ is the flat Euclidean metric in the $23$-plane, with $m,n=0,1$ and $i,j=2,3$.
It will often be convenient to parametrize the $01$-plane by light-cone coordinates $x^\pm$
and the $23$-plane by complex coordinates $x^u$ and $x^v = (x^u)^*$ defined as follows,
\bea
\label{lc}
\eta _{mn} dx^m dx^n = 2 dx^+ dx^- 
& \hskip 1in & 
x^\pm =  (\pm x^{0} + x^{1})/\sqrt{2}
\no \\
\delta _{ij} dx^i dx^j = 2 dx^u dx^v ~ &&
x^u \, =  (x^2 + i x^3) /\sqrt{2}
\eea  
The functions $U, W, \phi^A$ depend only on $r$ in view of translation invariance in $x^\mu$,
while the field strength components $F_{23}^I$ are  constant in view of the Bianchi identities. 
The constants $F^I_{23}$ may be parametrized by the magnitude of a magnetic field $B>0$ 
and a vector of charges $q^I$ which specifies the embedding of the magnetic field in $U(1)^3$ by setting, 
\bea
F_{23} ^I= q^I B
\eea
This parametrization is not unique, as $B$ and $q^I$ may be rescaled while leaving their product fixed. We shall shortly impose a normalization on $q^I$ to eliminate this  arbitrariness. Translation invariance of the Ansatz in the 23 directions allows us to consider solutions in which the topology of the 23-space is either flat $\RR^2$ or a compactification of $\RR^2$ to a flat torus $T^2$ which may be represented in $\RR^2=\CC$ as the quotient $\CC/\Lambda$ by a lattice $\omega_1 \ZZ + \omega_2 \ZZ$ with arbitrary period $\omega _1, \omega _2 \in \CC$.

\sm

Minimal five-dimensional supergravity may be obtained from non-minimal supergravity by 
setting $A^I _M = A_M$ for $I=1,2,3$, which amounts to setting all charges $q^I$ 
equal to one another. The scalars may then be set to zero, $\phi^A=0$,
so that $X^I=1$, which allows us to set the gaugino to zero $\lambda ^a=0$.
The magnetic brane solution constructed in \cite{D'Hoker:2009mm} for this {\sl minimal} five-dimensional 
Einstein-Maxwell-Chern-Simons theory  breaks all supersymmetries. 

\sm

Supersymmetric magnetic brane solutions exist if and only if the relation $q^{1}+q^{2}+q^{3}=0$
holds and $V_I$ satisfies $V_I q^I=0$. We shall set,
\bea
\label{2a18}
V_I = { 1 \over 3}   \hskip 0.5in I=1,2,3
\eea
This condition forces the composite $U(1)$-gauge field $\cA_M$  to vanish on the solution so that  the covariant derivative $\cD_M$  on a spinor $\epsilon$ reduces to the covariant derivative with the spin 
connection $\om _M$ given by (\ref{A8}), and takes the following form on the Ansatz (\ref{Ansatz}), 
\bea
dx^M \cD_M \epsilon = d \epsilon  - \half dx^m W' \, \Gamma^r \Gamma_m \epsilon 
- \half dx^i U' \, \Gamma^r \Gamma_i \epsilon
\eea
where ${}^\prime$ denotes differentiation in $r$.

\subsubsection{The reduced BPS equations}

The supersymmetric magnetic brane solution proposed  in \cite{Almuhairi:2010rb,Almuhairi:2011ws}, 
and further investigated in \cite{Donos:2011pn}, is a solution to the BPS equations (\ref{2a8}) and (\ref{2a10}) reduced to the Ansatz of (\ref{Ansatz}). These reduced BPS equations are invariant under Lorentz transformations in the $01$-plane and rotations in the $23$-plane  respectively generated by,\footnote{No hats are required on the indices in $\Gamma ^+{}_+=- \Gamma ^-{}_-$ and $i \Gamma ^2{}_3=- i \Gamma ^3{}_2$ as the lowering of one index absorbs the corresponding scale factor of the metric.}
\bea
\Gamma ^{\hp \hm} & = & \Gamma ^{\hat{0} \hat{1} } 
\hskip 0.23in  = \Gamma ^+{}_+ \hskip 0.1in =   - \Gamma ^-{}_-
\no \\
\Gamma ^{\hu \hv} \, & = & -i \Gamma ^{\hat{2} \hat{3}} ~ = - i \Gamma ^2{}_3= i \Gamma ^3{}_2
\eea
The generators $\Gamma ^{\hp \hm}$ and $\Gamma ^{\hu \hv}$ square to unity, mutually commute, and commute with $\Gamma ^\hr = \Gamma ^r$. Their  product $\Gamma ^{\hp \hm} \Gamma ^{\hu \hv}  \Gamma ^\hr  = - i \Gamma ^{\hat 0 \hat 1 \hat 2 \hat 3 \hat 4}$ equals $\pm I$. The two possible signs distinguish the two irreducible representations of the Clifford algebra in odd dimensions which, however, lead to equivalent representations of the Lorentz group, mapped into one another by parity. Using the convention adopted in Section \ref{sec:A1}, we choose, 
\bea
\label{2a21}
\Gamma ^{\hp \hm} \, \Gamma ^{\hu \hv} \, \Gamma ^\hr  =I
\eea
The BPS equations may be separated by simultaneously 
diagonalizing $\Gamma ^\hr$ and $\Gamma ^{\hu \hv}$,
\bea
\Gamma^{\hr} \, \epsilon = \gamma \, \epsilon
\hskip 1in
\Gamma^{\hu \hv} \, \epsilon = -\eta \, \gamma \, \epsilon
\label{BPSa}
\eea
where  $\gamma$ and $\eta$ are independent from one another and may take the values $\pm 1$.

\sm

The reduced BPS equation for the index $M =r$ is a differential equation for $\epsilon$ which we shall 
not  need here.  Assuming the existence of a non-vanishing spinor $\epsilon$, 
the reduced BPS equations of (\ref{2a10}) for $M=\mu=0,1,2,3$ are algebraic and given by, 
\bea
\label{BPSred}
0 & = &   W' - \mg \, \gamma V_{I}X^{I}+\frac{1}{2}\eta \, B \, q^{I}X_{I} \, e^{-2U}
\no \\
0 & = &  U' - \mg \, \gamma V_{I}X^{I}-\eta \, B \, q^{I}X_{I} \, e^{-2U}
\no \\ 
0 & = & \delta_{AB}  (\phi^{B})'
+6 \mg \, \gamma V_{I}\partial_A X^{I}
+3\eta \, B \, q^{I}\partial_A X_{I} \, e^{-2U}
\eea
The magnitude of $\mg$ may be scaled to 1 by rescaling $B$ and $r$.
The eigenvalue $\gamma$ is correlated with the sign of  $\mg$. 
To see this, note that the supersymmetric magnetic brane solution should reduce to the 
$AdS_5$  solution upon letting $B \to 0$. For this solution to exist, given that we have chosen 
the branch $X^I >0$ in (\ref{Xphi}), along with (\ref{2a18}),  we must have,
\bea
\label{gamma}
 \gamma =  \mg 
\eea
Having set $\gamma = \mg$ for $|\mg|=1$, the BPS equations are independent of the sign of $\mg$.
Similarly, the eigenvalue $\eta$ is given as follows, 
\bea
\label{eta}
\eta =  {\rm sign} ( q^{1}q^{2}q^{3} )
\eea
a relation which is required in order to have a solution  asymptotic to $AdS_3\times T^2$.

\subsubsection{The $AdS_3 \times T^2$ solution}

The reduced BPS equations, with a supersymmetric charge arrangement  $q^1+q^2+q^3=0$
and none of the charges $q^I$ vanishing, admit an exact $AdS_3 \times T^2$
solution \cite{Almuhairi:2010rb} given by,
\bea
W=\frac{r}{L}
\hskip 0.7in
e^{2U}=   \bar q  \, B  
\hskip 0.7in 
X^{I}={\left ( q^{I} \right)^2 \over 4 \bar q^2} 
\hskip 0.7in 
F_{23}^I = q^I B
\label{ADS3}
\eea
Recall our choice $B >0$, and the charges $q^I$ characterizing the embedding of the magnetic 
field in the $U(1)^3$ gauge group. The $AdS_{3}$ radius $L$ and the combination $\bar q$ are given by,
\bea
\label{Lbarq}
{1 \over L}=\frac{3}{2}V_{I}X^{I}
\hskip 1in
\bar q= \half | q^1 q^2 q^3|^{{1 \over 3}}
\eea
The above $AdS_3 \times T^2$ solution is regular, and preserves one of the four symplectic 
Majorana supersymmetries.  When one of the charges $q^I$ vanishes, the number of 
supersymmetry generators  is doubled  but, as is clear from the above expressions, 
there is no regular solution with an asymptotic $AdS_3 \times T^2$ behavior in the IR. 
Henceforth, we shall assume that none of the charges vanishes and, by suitably rescaling $B$,  
we shall choose, 
\bea
\label{barq}
\bar q =1
\eea
As a function of the three real charges $q^I$, subject to the condition $V_I q^I=0$, one readily 
establishes the allowed range of the $AdS_3$ radius $L$, which is $0 < L < L_0$
with $L_0 = 2^{2/3} /3$. The maximum value $L_0$ is uniquely attained when any two of the 
charges $q^I$ coincide.

\subsubsection{Asymptotic $AdS_{3} \times T^{2}$ behavior
 of the supersymmetric magnetic brane}
 \label{sub:AdS_3_Behavior}

The supersymmetric magnetic brane solution, for given magnetic field $B$ and embedding charges $q^I$, 
has $F_{23}^I = B q^I$ and the leading asymptotics for its remaining fields coincide with the exact $AdS_3 \times T^2$ solution given in the preceding subsection. The detailed $r\rightarrow-\infty$ asymptotics near 
$AdS_{3} \times T^{2}$, including the leading deviation away from the exact solution of (\ref{ADS3}), is found to be given as follows,
\bea
\label{2c31}
W\left(r\right) & = & 
 \frac{r}{L} + \frac{1}{\sigma} \left ( 2V_{I}\partial_{A}X_{(0)}^{I}c^{A}
-\frac{2}{3L}c^{0}\right) e^{\sigma r}+\mathcal{O}\left(e^{2\sigma r}\right)
\no \\
U\left(r\right) & = & \frac{1}{2}\ln  B 
+ c^{0} \, e^{\sigma r}+\mathcal{O}\left(e^{2\sigma r}\right)
\no \\
\phi^{1}\left(r\right) & = & -\sqrt{6}\ln\left(q^{1}q^{2} \right)
+ c^{1}\, e^{\sigma r}+\mathcal{O}\left(e^{2\sigma r}\right)
\no \\
\phi^{2}\left(r\right) & = & -\sqrt{2}\ln\left( {q^{1}  \over  q^{2} } \right)
+ c^{2}\, e^{\sigma r}+\mathcal{O}\left(e^{2\sigma r}\right)
\eea
The coefficients  $c^{0},c^1, c^2$ are components of an eigenvector, associated with eigenvalue $\sigma$, 
of a symmetric matrix $\mS$.  Explicitly, these relations are given by,
\bea
\mS \left(\begin{matrix}  c^{0} \cr   c^{A}  \end{matrix} \right)=
\sigma  \left(\begin{matrix}  c^{0} \cr   c^{A} \cr \end{matrix} \right)
\eea
where the indices $A,B$ take the values $1,2$, and $\mS$ is given by,
\bea
\mS^{0 0 } \, \, & = & \frac{4}{3L}
\no \\
\mS^{0 A} \, & = & -V_{I}\partial_{A}X^{I}
\no \\
\mS^{AB} & = & -6 V_{I}\p_{A}\p_{B}X^{I}
-3 \eta Bq^{I}\partial_{A}\partial_{B}X_I e^{-2U} 
\eea
Here, it is understood that the fields $X^I$ and $U$ are evaluated on the $AdS_{3}$ solution of (\ref{ADS3}), 
which is exclusively in terms of the charges $q^{I}$. Since $\mS$ is a symmetric matrix, 
its eigenvalues $\sigma$ are guaranteed to be real and they solve the characteristic equation, 
\bea
\sigma^{3}-\frac{4}{L^{2}}\sigma+16=0
\label{eq:char_near}
\eea
For $0< L< 1/\sqrt{3}$, the three roots are real, two being positive and one negative. The root chosen here is always the largest positive root. At $L=1/\sqrt{3}$, we have $\sigma =2$, and for $L < 1/\sqrt{3}$ the value of $\sigma$ monotonically increases with decreasing positive $\sigma$, reaching the asymptotic expression $\sigma \approx 2/L$ as $L \to 0$.  The range $0 < L < L_0 = 2^{2/3}/3$ established earlier for $L$ is strictly contained in this interval since $L_0 < 1/\sqrt{3}$, so that the two positive roots never become degenerate  for $0 < L <L_0$, and the largest root always satisfies $\sigma > 2$. 

\sm

The  overall magnitude of the vector $(c^{0},c^1, c^2)$ is not fixed by the local asymptotic expansion, 
but may be related, by numerical integration of the full supersymmetric magnetic brane solution which interpolates between $AdS_3 \times T^2$ and $AdS_5$, to the asymptotic behavior near $AdS_5$, to be given below.

\subsubsection{Asymptotic $AdS_{5}$ behavior of the supersymmetric magnetic brane}

Given the magnetic field $B$ and the embedding charges $q^I$, as well as the $AdS_3 \times T^2$ asymptotics of the solution spelled out in the preceding subsection, the $r\rightarrow\infty$ asymptotics 
of the metric  fields $U,W$ are as follows,
\bea
W\left(r\right) & = & r+\ln W_0+\cO \left(e^{-4r}\right)
\no \\
U\left(r\right) & = & r+\ln U_0 +\cO \left(e^{-4r}\right)
\eea
The constants $W_0$ and $U_0$ are functions of the  magnetic field $B$, the charges $q^I$,
and the overall magnitude of the coefficient vector $c^0, c^1, c^2$ in the $AdS_3 \times T^2$ 
asymptotics,  and can be read off from the numerical solution, where the metric at $r\rightarrow\infty$
takes the form, 
\bea
\label{2c29}
ds^{2}=dr^{2}+ W_0^2 \, e^{2r} \eta _{mn} dx^m dx^n +U_0^2 \, e^{2r}\delta _{ij} dx^i dx^j
\eea
The physical meaning of the constants $W_0$ and $U_0$ is to provide the constant rescaling factors between 
the coordinates of space-time $x^m, x^i$ between the IR region for $r \to - \infty$ and the UV region for $r \to + \infty$. Naturally, one could rescale the coordinates $x^m $ by $W_0$ and $x^i$ by $U_0$ to recover standard normalizations in the $AdS_5$ region, at the expense of rescaling the coordinates also in the $AdS_3 \times T^2$ region. The present choice of normalization will be the more convenient one for our purpose.

\sm

The leading asymptotic behavior of the scalar fields $\phi^A$ is given by (\ref{ADS3}) and  the second 
line in (\ref{eq:feff_A}). Its sub-leading asymptotics will not be presented here, as it will not be needed in the sequel.  The coefficients $g^{(4)}_{\mu \nu}$ and $\phi ^{A \, (2)}$ are not determined by the local expansion, but may again be determined by numerically integrating  the field equations.

\subsubsection{Global regular solutions obtained numerically}
\label{sec:num_soln}

The existence of a regular solution to the reduced BPS equations of (\ref{BPSred}) 
for the charge assignment $q^1=q^2$ was shown numerically in \cite{Donos:2011pn}. 
We shall supplement this result by exhibiting regular solutions to (\ref{BPSred}) 
which interpolate between $AdS_3 \times T^2$ and $AdS_5$
over a range of charge assignments, again by numerical integration. Without loss of generality, 
we permute the $q^I$ so that $q^1$ and $q^2$ have the same sign and $q^2 <q^1$. We introduce a single parameter $\alpha$ to characterize the solution, as follows,
\bea
\alpha = { q^2 \over q^1} \hskip 1in 0 < \alpha <1  \hskip 0.5in {\rm sign}(q^3)= \eta 
\eea
where $\eta$ is the sign factor introduced in (\ref{eta}).

\begin{figure}
\begin{centering}
\begin{minipage}[t]{0.45\columnwidth}%
\begin{center}
\includegraphics[scale=0.35]{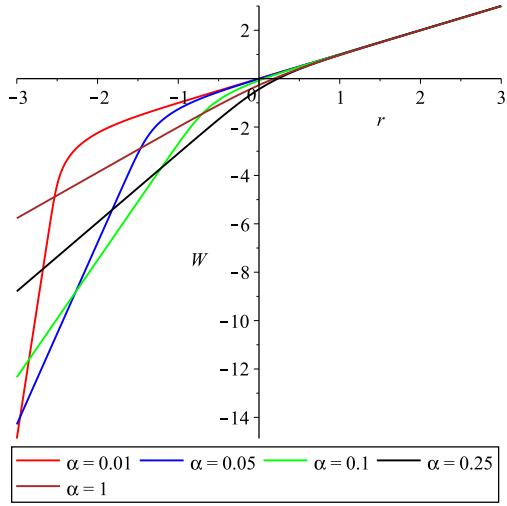}
\par\end{center}%
\end{minipage}\hfill{}%
\begin{minipage}[t]{0.45\columnwidth}%
\begin{center}
\includegraphics[scale=0.35]{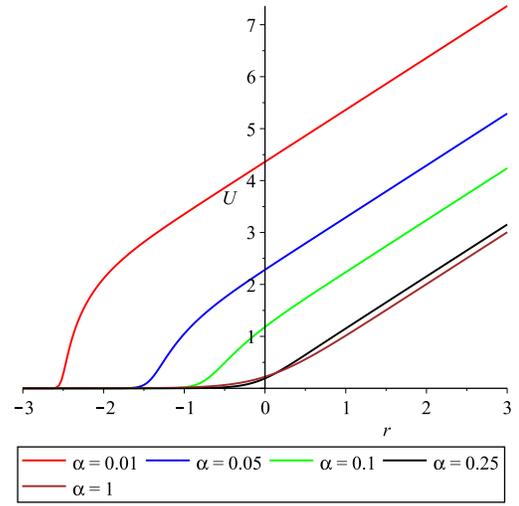}
\par\end{center}%
\end{minipage}\medskip{}
\end{centering}

\caption{The $r$-dependence of the metric functions $U$ and $W$,
obtained numerically for $\alpha =1, \, 0.25, \, 0.1, \, 0.05$, and $0.01$. }

\end{figure}

\begin{figure}
\begin{centering}
\begin{minipage}[t]{0.45\columnwidth}%
\begin{center}
\includegraphics[scale=0.35]{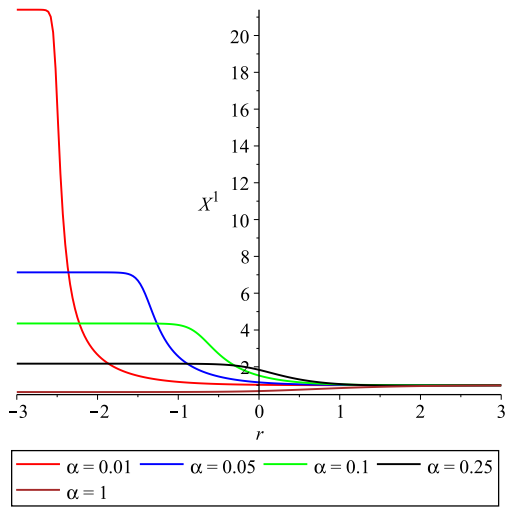}
\par\end{center}%
\end{minipage}\hfill{}%
\begin{minipage}[t]{0.45\columnwidth}%
\begin{center}
\includegraphics[scale=0.35]{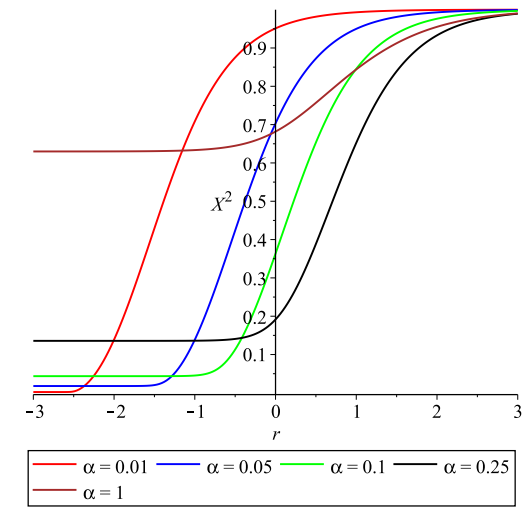}
\par\end{center}%
\end{minipage}
\par\end{centering}

\caption{The $r$-dependence of the scalar  functions $X^1$ and $ X^2$,
obtained numerically for $\alpha =1, \, 0.25, \, 0.1, \, 0.05$, and $0.01$. }

\end{figure}

\sm

The corresponding asymptotics of the metric as $r \to \infty$ is given by (\ref{2c29})
and $X^I \to 1$, while the asymptotics as $r \to - \infty$ for the metric function $U$ is 
constant, and is given for all functions in (\ref{2c31}). In particular, for the scalar fields $X^I$,
the asymptotics as $r \to - \infty$  is given by the $AdS_3 \times T^2$ solution in (\ref{ADS3})
and we have,
\bea
X^1 = \left ( { 1 \over \alpha  (1+\alpha)} \right ) ^{ 2 \over 3}
\hskip 0.6in 
X^2 = \left ( { \alpha ^2 \over 1+\alpha } \right )^{{2 \over 3}}
\hskip 0.6in 
X^3 = \left (  { (1 + \alpha )^2 \over \alpha } \right ) ^{{2 \over 3}}
\eea
We have verified that by using the largest positive root $\sigma$ of (\ref{eq:char_near}) 
in the initial conditions for the $AdS_3 \times T^2$ region, there always exists a solution that matches onto 
$AdS_{5}$ in the UV for the following values, 
\bea
\label{numvals}
\alpha =1, \, 0.5, \, 0.25, \, 0.1, \, 0.05, \, 0.025, \, 0.01, \, 0.005, \, 0.0025, \, 0.001
\eea
of which we have depicted a subset in figures 1 and 2. The dependence on  
$\alpha$ from one value to another appears to be smooth.

\section{Stress tensor correlators
\label{sec:3}}
\setcounter{equation}{0}

In this section, we shall compute the two-point correlators of the components in the $01$-plane of the stress
tensor in the presence of the supersymmetric magnetic brane solution,  in the IR limit. We follow the 
method of \cite{D'Hoker:2010hr} and solve the linearized Einstein equations for the corresponding 
components of the metric fluctuations $\delta g_{\mu \nu}$ with specified holographic boundary condition 
$\delta g_{\mu\nu}^{(0)}$. From this solution,  we obtain the induced expectation value 
$T^{\mu\nu} (x)$ of the stress tensor operator $\cT^{\mu \nu} (x)$ via the first equation of (\ref{TJY}) and read off the correlator 
from the linear response formula,
\bea
T^{\mu\nu} (x)=\frac{i}{2}\int d^{4}y \, \sqrt{g^{(0)}}
\left \langle \cT^{\mu\nu} (x) \cT ^{\rho \sigma } (y) \right \rangle 
\delta g_{\rho \sigma }^{(0)} (y)
\label{eq:T_expt}
\eea
We begin by isolating  the fluctuations needed to calculate the desired correlators.

\subsection{Structure of the perturbations}

In this section we shall determine the structure of the perturbations around the supersymmetric
magnetic brane solution needed to compute the two-point correlators of the components of
the stress tensor and the currents in the directions of the $01$-plane. 

\sm

Since the supersymmetric
magnetic brane solution is invariant under translations in $x^\mu$ for $\mu =0,1,2,3$ a general linear 
perturbation is a linear combination of plane waves, each with given momentum $p_\mu$.
Physically relevant to probing the dynamics of the effective low energy CFT in the $01$-plane 
is the dependence of the perturbations on the components $p_\pm$ only, so that we may set
$p_2=p_3=0$.   

\sm

Arbitrary perturbations of the metric around the supersymmetric magnetic brane 
will generally mix with gauge field and scalar perturbations. However, if we restrict the 
perturbations of the metric to the directions in the $01$-plane, namely if we turn on only 
the components $\delta g _{\pm \pm}$ and $\delta g_{+-}$ then it may be seen from the 
action that no mixing with the other components of metric fluctuations,  the  
gauge fields, and the scalar fields will occur as long as $p_2=p_3=0$. Key ingredients
in the argument are the invariances of the supersymmetric magnetic brane under translations 
along $x^\mu$ for $\mu =0,1,2,3$, Lorentz transformations in the $01$-plane, and rotations 
in the $23$-plane.

\sm

Consider, for example, the effect of turning on the fluctuation $\delta g ^{++}= g^{+-} g^{+-} \delta g_{--} $ 
on the gauge kinetic energy term proportional to $G_{IJ} (\phi) g^{MN} g^{PQ} F^I _{MP} F^J _{NQ}$. 
Since the gauge field strength of the supersymmetric brane solution is in the direction $F^I_{23}$
only, a fluctuation linear in $\delta g^{++}$ can turn on neither the fluctuation $\delta F^I _{+-}$ nor the fluctuation $\delta F_{23}$. It can also not turn on the fluctuations of the scalar field. The arguments for the 
other couplings in the action are similar.

\sm

Therefore, we consider the following plane wave perturbation $h_{mn}(r, p_\pm) e^{ip \cdot x}$ with momentum $p_\pm$ of the  supersymmetric magnetic brane, 
\bea
\label{3b1}
ds^{2} & = & ds_{B}^{2}
+ h_{mn} (p_\pm , r) \, e^{ip\cdot x} \, dx^m dx^n
\no \\
F^{I} & = & q^{I} B \, dx^{2}\wedge dx^{3}
\no \\
\phi ^A & = & (\phi _B)^A
\eea
where $ds_B^2$ and $(\phi_B)^A$ are respectively the metric and the scalar fields of the supersymmetric magnetic brane given by the Ansatz (\ref{Ansatz}) with $U,W, (\phi_B)^A$ provided by the numerical solution to (\ref{BPSred}). The indices $m,n$ take the values $0,1$ or equivalently $\pm$ and
we shall use the following notations throughout  for the inner product and norm in the $01$-plane, 
\bea
p\cdot x = p_{+}x^{+}+p_{-}x^{-}
\hskip 1in 
p^{2} = 2p_{+}p_{-} 
\label{eq:momenta}
\eea
Finally, we shall be interested only in momenta which are small compared with the inverse 
radius $|\mg|$ of $AdS_5$, which here has been set to 1, so that we shall work in the regime, 
\bea
0< p^2 \ll1
\label{eq:match_reg}
\eea
In this limit the equations for the metric perturbations $h_{\pm \pm}$ may be solved 
by matching the  asymptotic expansion valid in the near and far regions.  The near region is 
the range of $r$ where $AdS_3 \times T^2$ is a good approximation, namely $e^{2r} \ll 1$, 
while the far region is the range of $r$ for which we can neglect the momenta, namely $p^2  \ll e^{2r}$. 
In view of (\ref{eq:match_reg}), the overlap region $p^2 \ll e^{2r} \ll 1$  is parametrically large,
and matching the solutions in the near and far regions in the overlap region will produce 
a linearized solution valid for all $r$.

\sm

The linearized field equations for the perturbations (\ref{3b1}) of the metric are,
\bea
0 &=& 3h_{\pm\pm}^{\prime\prime}-6(W^{\prime}-U^{\prime})h_{\pm\pm}^{\prime}
+12(W^{\prime})^2h_{\pm\pm}- \mF \, h_{\pm\pm}
\no \\
0 &=& (p_\mp h_{\pm\pm}-p_\pm h_{+-})^\prime-2W^\prime(p_\mp h_{\pm\pm}-p_\pm h_{+-})
\no \\
0 &=& 3h_{+-}^{\prime\prime}+6U^\prime h_{+-}^\prime
+3e^{-2W}(p_-^2h_{++}+p_+^2h_{--}-2p_+p_-h_{+-})-\mF \, h_{+-}
\no \\
0 &=& h_{+-}^{\prime\prime}-2W^\prime h_{+-}^\prime-2W^{\prime\prime}h_{+-}
\label{3b4}
\eea
Here, the prime denotes differentiation with respect to $r$, the dependence on $r$ and $ p_\pm$ 
is understood, and we have introduced the following abbreviation, 
\bea
\mF = B^2e^{-4U}\left[\left(\frac{q^1}{X^1}\right)^2+\left(\frac{q^2}{X^2}\right)^2+\left(\frac{q^3}{X^3}\right)^2\right]+8\left(\frac{1}{X^1}+\frac{1}{X^2}+\frac{1}{X^3}\right)
\eea
We shall need of this function only its asymptotic values in the $AdS_5$ and $AdS_3 \times T^2$ regions,
which evaluate to $24$ and $12/L^2$, respectively.

\subsection{Near Region}

In the near region, where $e^{2r} \ll 1$, we set the background metric equal to the metric of the $AdS_{3}\times T^{2}$ solution of (\ref{ADS3}) given by,
\bea
\label{3c1}
ds_{B}^{2}=dr^{2}+e^{\frac{2r}{L}} \eta _{mn} dx^m dx^n + B \delta _{ij} dx^i dx^j
\eea
and the scalar fields $X^I$ equal to the values given in (\ref{ADS3}). All dependence on the charges $q^I$
and the magnetic field $B$  is through the $AdS_3$ radius $L$ only.
The linearized field equations derived from (\ref{3b4}) in the near region are given by,
\bea
0 & = & h_{\pm\pm}^{\pp}-\frac{2}{L}h_{\pm\pm}^{\prime} 
\no \\
0 & = & \left(p_{\pm}h_{+-}-p_{\mp}h_{\pm\pm}\right)^{\prime}-\frac{2}{L}\left(p_{\pm}h_{+-}-p_{\mp}h_{\pm\pm}\right)
\no \\
0 & = & h_{+-}^{\pp}-\frac{4}{L^{2}}h_{+-}+e^{-\frac{2r}{L}}\left(p_{-}^{2}h_{++}+p_{+}^{2}h_{--}-2p_{+}p_{-}h_{+-}\right)
\no \\
0 & = & h_{+-} '' -{ 2 \over L} h_{+-}' 
\label{eq:near_eqns}
\eea
where the prime denotes differentiation with respect to $r$. From the equations on the first and 
last lines of (\ref{eq:near_eqns}),  it is clear that the solutions for the components $h_{\pm \pm}$  
and $h_{+-}$ are all of the form,
\bea
h_{\mu\nu} (p_\pm, r) =s_{\mu\nu} (p_\pm) \, e^{\frac{2r}{L}}+t_{\mu\nu} (p_\pm )
\label{eq:h_near}
\eea
where the Fourier coefficients $s_{\mu \nu}$ and $t_{\mu \nu}$ depend on $p_\pm$, 
but are independent of $r$. The equations on the second and third lines in (\ref{eq:near_eqns}) 
impose the following relations between the Fourier coefficients $s_{\mu \nu}(p_\pm)$ and $t_{\mu \nu}(p_\pm)$,
\bea
t_{++} (p_\pm)& = & \frac{p_{+}}{p_{-}} \, t_{+-}(p_\pm )
\no \\
t_{--} (p_\pm )& = & \frac{p_{-}}{p_{+}} \, t_{+-}(p_\pm )
\no \\
t_{+-} (p_\pm )& = & \frac{L^{2}}{4} \Big (p_{-}^{2}s_{++} (p_\pm ) 
+p_{+}^{2}s_{--} (p_\pm ) -2p_{+}p_{-}s_{+-} (p_\pm ) \Big )
\label{eq:near_coeffs}
\eea
As is familiar from \cite{D'Hoker:2010hr}, we can identity the Fourier coefficients $s_{\mu\nu}$ and $t_{\mu\nu}$
as contributing to the Fourier transforms of the perturbation of the conformal boundary metric $\delta g_{\mu \nu}^{(0)}$ and the boundary stress tensor $\delta g_{\mu \nu}^{(4)}$, respectively. The top two lines of
(\ref{eq:near_coeffs}) express the linearized conservation equations of the stress tensor\footnote{Care is required in relating the $AdS_3$ stress tensor $\hat t _{\mu \nu}$ to $t_{\mu \nu}$ as their relation involves accounting for a trace term whose net effect is to reverse a sign as follows: $\hat t_{\pm\pm}=t_{\pm\pm}$ and $\hat t_{+-}=-t_{+-}$, as is explained for example in \cite{deHaro:2000vlm, Kraus:2006wn}.}
while the last line expresses the linearized trace anomaly of the stress tensor.

\subsection{Far region}

In the far region, where $p^2 \ll e^{2r}$, we can ignore the momentum dependent terms,
and we shall no longer exhibit the dependence on the momenta of the fluctuations $ h_{\mu \nu}$.
We will also take $h_{+-}=0$ in the far region, since this term will contribute to correlators 
involving $T_{+-}$ which contain only contact terms. 

\sm

The linearized field equations for $h_{\pm\pm}$ with the momentum
terms dropped are identical to the equations for $e^{2W}$ in the
Einstein equations (\ref{fieldeqs}) with Ansatz (\ref{Ansatz}).
Therefore, a first solution is given by,
\bea
h^{1}\left(r\right)=e^{2W\left(r\right)}\label{eq:h1}
\eea
where $W$ is the interpolating solution of the BPS equations. 
By analogy with \cite{D'Hoker:2010hr}, we find that another linearly
independent solution is given by,
\bea
h^{2} (r)
=e^{2W(r)} \int_{\infty}^{r} dr^{\prime} \, e^{-2W (r^{\prime})-2U(r^{\prime})}
\label{eq:h2}
\eea
Asymptotically, these functions have the following form. As $ r \to \infty$, we have, \footnote{The solution  $h^1(r)$ actually has a pre-factor of $W_0^{-2}$ which may be absorbed into the momenta, $p_\pm$,  because the momenta are defined as conjugate to coordinates $x^\pm$ on the  $AdS_5$ boundary with the conventional normalization. Therefore, we will not carry these factors around in the sequel.}
\bea
h^1(r) \sim e^{2r} \hskip 1in h^2(r) \sim - { 1 \over 4 U_0^2} \, e^{-2r}
\eea
while the asymptotics in the overlap region where $p^2 \ll e^{2r} \ll 1$, namely as $r \to - \infty$, 
is given as follows, 
\bea
h^1(r) \sim e^{{2r \over L}} \hskip 1in h^2(r) \sim -{L \over 2B}
\eea
Therefore, our solution in the far region is given by the linear combination,
\bea
h_{\pm\pm} (r)=
h^{1} (r) \, \delta g_{\pm\pm}^{(0)} -4U_0^2 h^{2} (r) \, \delta g_{\pm\pm}^{(4)}
\label{eq:h_far}
\eea
with coefficients chosen to obtain the following asymptotic form at
$r\rightarrow\infty$: 
\bea
h_{\pm\pm} (r) \sim e^{2r} \, \delta g_{\pm\pm}^{(0)}+e^{-2r} \, \delta g_{\pm\pm}^{(4)}
\label{eq:h_far_far}
\eea
The $r\rightarrow-\infty$ asymptotics of (\ref{eq:h_far}) then follows, 
by
\bea
h_{\pm\pm} (r) \sim e^{\frac{2r}{L}} \, \delta g_{\pm\pm}^{(0)}
+\frac{2U_0^2 L}{B} \, \delta g_{\pm\pm}^{(4)}
\label{eq:h_far_near}
\eea

\subsection{Matching and IR Correlators}

In the overlap region where $p^{2} \ll e^{2r}\ll1$, the solutions
(\ref{eq:h_near}) and (\ref{eq:h_far_near}) should match. Eliminating $s_{\mu \nu}$ and $t_{\mu \nu}$ 
between (\ref{eq:h_near}), (\ref{eq:h_far_near}), and (\ref{eq:near_coeffs}) gives
the following relations between $\delta g^{(0)}_{\pm \pm} $ and $\delta g^{(4)}_{\pm \pm}$,
\bea
\delta g_{++}^{\left(4\right)} & = & 
\frac{B L}{8U_0^2 }\left(\frac{p_{+}^{3}}{p_{-}} \, \delta g_{--}^{(0)}
+p_{+}p_{-} \, \delta g_{++}^{(0)}\right)
\no \\
\delta g_{--}^{\left(4\right)} & = & 
\frac{B L}{8 U_0^2}\left(\frac{p_{-}^{3}}{p_{+}} \, \delta g_{++}^{(0)}
+p_{+}p_{-} \, \delta g_{--}^{(0)}\right)
\eea
From (\ref{TJY}), the stress tensor is given by $4\pi G_{5}T_{\pm\pm}= \delta g_{\pm\pm}^{\left(4\right)}$
up to local terms. In order to normalize the stress tensor correlator to the conventional form suitable for two-dimensional CFTs, we define the two-dimensional stress tensor by  $\tilde{\cT}_{\pm\pm}=U_0^2 V_{2}\cT_{\pm\pm}$,
where $V_{2}$ denotes the volume of the compactified $23$-plane.
Writing $\tilde{T}_{\pm\pm}$ in terms of the Brown-Henneaux central
charge of the 1+1 dimensional CFT, given by, 
\bea
c=\frac{3L}{2G_{3}}=\frac{3LV_{2}}{2G_{5}} B
\label{eq:c}
\eea
we obtain,
\bea
\tilde{T}_{\pm\pm}=\frac{c}{48\pi}\frac{p_{\pm}^{3}}{p_{\mp}} \, 
\delta g_{\mp\mp}^{\left(0\right)}+\rm {local}\label{eq:T_hat}
\eea
If the $23$-plane is left uncompactified, $c$ should be viewed as the
central charge per unit area instead. Reading off the two-point functions
from (\ref{eq:T_expt}), we find,
\bea
\left\langle \tilde{\cT}_{\pm\pm}(p)\tilde{\cT}_{\pm\pm}(-p)\right\rangle 
=\frac{c}{24\pi}\frac{p_{\pm}^{3}}{p_{\mp}}\label{eq:TT}
\eea
up to contact terms. All other correlators involve only contact terms. 
Fourier transforming  this correlator to position space, we obtain, 
\bea
\left\langle\tilde{\cT}_{\pm\pm}(x)\tilde{\cT}_{\pm\pm}(0)\right\rangle 
=\frac{c}{8\pi^{2}}\frac{1}{\left(x^{\pm}\right)^{4}}\label{eq:TT_pos}
\eea
This is the standard formula for the stress tensor correlator in a
1+1 dimensional CFT with central charge $c$.

\section{Current-current correlators
\label{sec:4}}
\setcounter{equation}{0}

We now compute the two-point correlators for the $U(1)^3$ currents, following the method of the previous section. The results are qualitatively different from those of \cite{D'Hoker:2010hr} because we have three Maxwell fields  instead of one, and a corresponding dependence on the values of the charges $q^{I}$, 
and qualitatively on the signs of the charges. We  solve the linearized field equations (\ref{fieldeqs}) for the Maxwell fields with specified boundary condition $A_{\mu}^{I\left(0\right)} (x)$, read off the induced expectation value $J^{I \mu} (x)$ of the current operator $\cJ^{I\mu} (x)$ from the second equation in  (\ref{TJY}), and extract the correlators from the linear response formula,
\bea
J^{I\mu}\left(x\right)=i\int d^{4}y\sqrt{g^{(0)}}
\left\langle \mathcal{J}^{I\mu}(x)\mathcal{J}^{J\nu} (y)\right\rangle \delta A_{J\nu}^{(0)}(y)
\label{eq:J_expt}
\eea
We begin by isolating  the fluctuations needed to calculate the desired correlators.

\subsection{Structure of the perturbations}

We shall consider only the correlators of the components $\mathcal{J}_{\pm}^{I}$ of the currents 
along the $01$-directions, since we restrict here to probing the effective CFT that lives
in the $x^{\pm}$ space. As with the stress tensor correlators, translation invariance of the 
supersymmetric magnetic brane in $x^\mu$ for $\mu=0,1,2,3$ is used to Fourier decompose
the fluctuations into plane waves of given momentum $p_\mu$. Restricting to the correlators
of $\cJ_\pm ^I$, we retain dependence on $p_\pm$ only, and set $p_2=p_3=0$. The perturbed 
gauge field takes the form,
\bea
F^{I} & = & q^{I}Bdx^{2}\wedge dx^{3}+d A^{I}_p
\no \\
A^{I} _p& = & 
a_m^{I}\left(p_{\pm}, r\right) \, e^{ip\cdot x} \, dx^m
\eea
Turning on an arbitrary fluctuation of the gauge fields will generally induce perturbations of
the metric and of the scalar fields. But having set $p_2=p_3=0$, using translation invariance 
in $x^\mu$, Lorentz invariance in the $01$-plane, and rotation invariance in the $23$-plane,
we find that turning on perturbations of the gauge fields in the directions of only the $01$-plane 
will turn on perturbations of neither the metric nor the gauge field in the $23$-directions, 
nor the scalar fields. We can therefore consistently set all those perturbations to zero.

\sm

It will be convenient to define
$\veps_{\pm}^{I}\equiv p_{-}a_{+}^{I}\pm p_{+}a_{-}^{I}$. In this
notation, the linearized equations (\ref{fieldeqs}) for the Maxwell
fields reduce to, 
\bea
e^{2W}\left[e^{2U}G_{IJ}\left(\varepsilon_{-}^{J}\right)^{\prime}\right]^{\prime}
-\frac{B}{2}e^{2W} \cM_{IK} \left(\varepsilon_{+}^{K}\right)^{\prime}
-p^{2}e^{2U}G_{IJ}\varepsilon_{-}^{J} & = & 0\nonumber \\
e^{2U}G_{IJ}\left(\varepsilon_{+}^{J}\right)^{\prime}-\frac{B}{2} \cM_{IK} \varepsilon_{-}^{K} & = & 0
\label{eq:current_eq}
\eea
Throughout,  it will be convenient to define the following $3\times 3$ constant matrix,
\bea
\label{MM}
\cM_{IJ} = \sum _K C_{IJK} q^K 
\eea
We will solve  equations (\ref{eq:current_eq})  in the low energy limit given by (\ref{eq:match_reg}).

\subsection{Near region}

In the near region we have $e^{2r}\ll1$, the background metric and scalars are given by (\ref{ADS3}), 
and the metric $G_{IJ} = 8 \delta _{IJ} /(q^I)^4 $ is constant. Substituting these values into (\ref{eq:current_eq}) and simplifying by a factor of $B$, we obtain after some further rearrangements,
\bea
\left ( G_{IJ} \, \ep^J _- \right )'' -{ 1 \over 4} \cM_{IJ} G^{JK} \cM_{KL} \, \ep _- ^L 
- p^2 e^{{2r \over L}} \, G_{IJ} \, \ep _- ^J & = & 0
\no \\
\left ( G_{IJ} \ep ^J _+ \right ) ' - \half \cM_{IJ} \ep _- ^J & = & 0
\eea
To decouple this system of equations, we seek to diagonalize the matrices involved. 
While it may seem natural to multiply the first line to the left by $G^{-1}$, this would lead
to a matrix $G^{-1} \cM G^{-1} \cM$ in its second term, and this matrix is not generally symmetric. 
Instead,  we multiply on the left by $(q^I)^2$ (which is essentially the square root of $G_{IJ}$), 
and rearrange the equations as follows, 
\bea
\frac{\left(\veps_{-}^{I}\right)^{\pp}}{\left(q^{I}\right)^{2}}
-\sum_{JK} M^{IJ}M^{JK} \frac{\veps_{-}^{K}}{\left(q^{K}\right)^{2}} 
- p^{2}e^{-\frac{2r}{L}} \frac{\veps_{-}^I}{\left(q^ I \right)^{2}} & = & 0
\no \\
\frac{\left(\veps_{+}^{I}\right)^{\prime}}{\left(q^{I}\right)^{2}}-\sum_{J}M^{IJ}\frac{\veps_{-}^{J}}{\left(q^{J}\right)^{2}} & = & 0
\label{eq:near_lin}
\eea
where the matrix $M$ is defined by 
\bea
M^{IJ} = 4 \sum _K {C^{IJK} \over q^K}
\eea
In view of the normalization of the product of the charges $q^I$ adopted in (\ref{barq}), 
the matrix $M$ is related to the matrix $\cM$ of (\ref{MM}) by the diagonal matrix of charges $Q$, 
\bea
\label{MQM}
Q^2 \cM Q^2 = 16 M
\hskip 1in 
Q = {\rm diag} \left(q^{1},q^{2},q^{3}\right)
\eea
Since $M$ is manifestly symmetric its eigenvalues $m_I$ for $I=1,2,3$,  are real and $M$ can be 
diagonalized by a real orthogonal matrix $R$, so that we have $M=RDR^t$
where $D={\rm diag}\left(m_1,m_2,m_3\right)$. In terms of the new functions $\hat \ep _\pm ^I$,
defined in terms of $\ep^I_\pm$ and $R$ by,
\bea
{ \veps _\pm ^I \over (q^I)^2} = \sum _J R^{IJ} \hat \ep _\pm ^J
\eea
the set of equations (\ref{eq:near_lin}) decouples and we have,
\bea
\left(\hat{\veps}_{-}^{I}\right)^{\pp}
-\left [ \left ( m_I \right)^{2}+p^{2}e^{-\frac{2r}{L}} \right ] \hat{\veps}_{-}^{I} & = & 0
\no \\
\left(\hat{\veps}_{+}^{I}\right)^{\prime}- m_I\hat{\veps}_{-}^{I} & = & 0
\label{diag}
\eea
The first line in (\ref{diag}) is the modified Bessel equation in the variable
$\mpp L\, e^{-\frac{r}{L}}$ for index $L m_I$ and using the definition $\mpp = \sqrt{p^2}$. 
The solutions which are regular at the horizon are proportional to the modified Bessel function $K$ as follows, 
\bea
\hat{\veps}_{-}^{I}\left(r\right)
\sim K_{L m_I}\left(\mpp L\, e^{-\frac{r}{L}}\right)
\eea
In the low energy limit of (\ref{eq:match_reg}), we shall expand the above solutions in 
the limit $\mpp^{2}e^{-\frac{2r}{L}}\ll1$ where $r/L \gg 1$. Using the asymptotics of the 
modified Bessel function,  the asymptotic of $\hat{\veps}_{-}^{I}$ takes the following form,
\bea
\hat{\veps}_{-}^{I}\left(r\right) =  k_{+}^{I} \, e^{+ m_I r}-k_{-}^{I} \, e^{- m_I r}
\label{eq:eps_m_near}
\eea
The pre-factors $k^I _\pm$ are given by,
\bea
\label{k-eq}
k_{\pm}^{I} = \frac{C^I }{\Gamma\left(1\mp L m_I \right)}\left(\frac{\mpp L}{2}\right)^{\mp L  m_I }
\eea
where  $C^I$ are integration constants which do not depend on the subscript $\pm$.
Using this result, we obtain $\hat \ep ^I_+$ by integrating the second equation in (\ref{diag}) to get,
\bea
\hat{\veps}_{+}^{I}\left(r\right)  =  
 k_{+}^{I} \, e^{ + m_I r}+k_{-}^{I} \, e^{- m_I r} + \hat \ep _0^I
\label{eq:eps_p_near}
\eea
where $\hat \ep^{I}_0$ are integration constants which depend on $p_\pm$ and $q^I$,
but are independent of $r$. Converting back to $a_{\pm}^{I}$, we find,
\bea
a_{\pm}^{I}\left(r\right) = \frac{\left(q^{I}\right)^{2}}{p_{\mp}}
\sum_{J}R^{IJ} k_{\pm}^{J} \, e^{\pm m_J r}+p_{\pm} a_0 ^{I}
\label{eq:a_near}
\eea
where the constants $a_0 ^I$ are related to $\hat \ep_0^I$ as follows, 
$p^2 a_0 ^I = \sum_J (q^I)^2 R^{IJ} \hat \ep _0^J$.

\subsection{Far region}

In the far region, $p^2 \ll e^{2r}$, we neglect the momentum dependent terms in  (\ref{eq:current_eq}). 
The first equation may then be integrated exactly, and we obtain the first order system, 
\bea
e^{2U}G_{IJ}\left(\veps_{+}^{J}\right)^{\prime}-\frac{B}{2} \cM_{IJ} \veps_{-}^{J} & = & 0
\no \\
e^{2U}G_{IJ}\left(\veps_{-}^{J}\right)^{\prime}-\frac{B}{2} \cM_{IJ} \veps_{+}^{J} & = & \tilde{a}_{0I}
\eea
where $\tilde{a}_{0I}$ is a set of integration constants. Since the matrix $\cM_{IK}$
is constant and invertible, we can absorb $\tilde{a}_{0I}$ by a constant shift in 
$\veps_{+}^J$, which we shall denote by $p^{2}\alpha^{I}$. 
Converting back to $a_{\pm}^{I}$, the equations reduce to the following form, 
\bea
\left(a_\pm^I -p_\pm \alpha^I \right)^{\prime} \mp \cH^I{}_J \left(a_\pm ^J -p_\pm \alpha^J \right) = 0
\label{eq:lin_far}
\eea
where $\cH$ is a $3\times 3$ matrix-valued function of $r$ defined by,
\bea
\label{4c3}
\cH^I{}_J (r) =\frac{B}{2}G^{IK} (r) \cM_{KJ} \, e^{-2U(r)}
\eea
The solutions of  equations (\ref{eq:lin_far}) are given by path ordered exponentials, defined  by, 
\bea
\cU_\pm (r, r') = P \exp \left \{ \pm \int ^{r} _{r'} d\rho \, \cH(\rho) \right \} 
\eea
where the ordering is such that $\cH(r)$ is to the left of $\cH(r')$ in the expansion of the 
exponential in powers of $\cH$ or, equivalently, that 
\bea
\label{Uder}
\p_r \, \cU_\pm (r,r') & = & \pm \, \cH(r) \, \cU_\pm (r,r')
\no \\
\p_{r'} \, \cU _\pm (r,r') & = & \mp \, \cU_\pm (r, r') \, \cH(r')
\eea 
The path ordered exponentials satisfy $\cU_\pm (r,r)=I$ and the composition law,
\bea
\label{split}
\cU_\pm (r, r') \, \cU_\pm (r', r'') = \cU _\pm (r, r'')
\eea
The solution to (\ref{eq:lin_far}) may then be expressed in matrix notation, as follows,
\bea
\label{asol}
\left(a_\pm -p_\pm \alpha \right) (r) = \cU_\pm (r, \infty) \left(a_\pm^{(0)}  -p_\pm \alpha \right)
\eea
Note that the $r' \to \infty$ limit of $\, \cU_\pm(r,r')$  is well-defined since $\cH(r')$
tends to zero exponentially due to the  $e^{-2U}$ factor in (\ref{4c3}),
while the metric $G^{IK}$ tends to a finite limit.

\subsubsection{Asymptotics of the far region solution for $r \to + \infty$}

The asymptotics of $a_\pm (r)$ as $r \to + \infty$ may be evaluated in terms of the asymptotics of $X^{I}$ 
and $e^{2U}$ by substituting  the $AdS_{5}$ solutions into the integral and keeping only the first two leading 
orders in the expansion, and we find, 
\bea
a_{\pm}^{I}\left(r\right) = a_{\pm}^{I\left(0\right)}+a_{\pm}^{I\left(2\right)} \, e^{-2r}
\label{eq:a_far_2}
\eea
where
\bea
a_{\pm}^{I\left(2\right)}  =  \mp\frac{B}{2U_0^2} 
\sum_{J,K} \delta^{IJ}\cM_{JK}\left(a_{\pm}^{K\left(0\right)}-p_{\pm}\alpha^{K}\right)
\label{eq:a2}
\eea
In the $AdS_5$ approximation which is valid here, we have $G^{IJ} = 2 \delta ^{IJ}$
which has allowed for further simplification in this formula. The unknown in this equation is the
constant $\alpha ^I$, which we shall now determine by matching with the solution in the near region.

\subsubsection{Asymptotics of the far region solution for $ r \to -\infty$}

To obtain the $r \to -\infty$ asymptotics in the far region we use (\ref{split}) 
to factorize the path-ordered exponential in (\ref{asol}) at an arbitrary 
point $r_0$ in the overlap region,
\bea
a_{\pm} (r) - p_{\pm} \alpha=
\cU_{\pm} (r,r_0) \, \cU_{\pm}(r_0,\infty) \left(a_{\pm}^{(0)}-p_{\pm}\alpha\right) 
\eea
When both $r$ and $ r_0$ are in the overlap region,  the matrix $\cH$ in $\cU_\pm (r,r_0)$ 
may be  evaluated on the $AdS_{3}\times\mathbb{R}^{2}$ solution and is constant. The 
corresponding path ordered exponential may then be readily evaluated, 
\bea
\cU_{\pm}\left(r,r_0\right)
= \exp\left\{ \pm { 1 \over 16} Q^{4}\mathcal{M}\left(r-r_0\right)\right\} 
\eea
where we recall that $Q = {\rm diag} \left(q^{1},q^{2},q^{3}\right)$. 
Next, we define the combinations,
\bea
\Omega_{\pm}  = \exp\left \{ \mp { 1 \over 16} Q^{4} \cM r_0 \right \} \, \cU_{\pm}(r_0,\infty)
\eea
Within the approximations made, the matrices 
$\Omega _\pm$ are independent of $r_0$ in the overlap region.
If need be, they may be evaluated numerically from the numerical
supersymmetric magnetic brane solution to the BPS equations.
Making use also of the relation $Q^{4}\mathcal{M}= 16 Q^{2}MQ^{-2}$ 
we obtain the following expression for the coefficients $a_\pm$, 
\bea
a_{\pm} (r) - p_{\pm} \alpha=
Q^2  \, e^{ \pm r  M  } \,  Q^{-2} \, 
\Omega _\pm  \left ( a_{\pm}^{(0)}-p_{\pm}\alpha\right) 
\eea
Finally, in order to match the behavior of the far and near region solutions in the 
overlap region, we shall need a decomposition of the solution onto the exponential modes,
analogous to the one we had obtained in (\ref{eq:a_near}) for the near region solution. 
This may be done by diagonalizing $M= R D R^t$ by an orthogonal matrix $R$, 
and we have, 
\bea
a_{\pm}^{I} (r) - p_{\pm}\alpha^{I} = 
\sum_{JK} \left(Q^{2} R \right)^{IJ} \, e^{\pm r \, m_J }
\left( R^t Q^{-2} \, \Omega_{\pm} \right)^{JK} \left(a_{\pm}^{(0)K}-p_{\pm}\alpha^{K}\right)
\label{eq:a_near_2}
\eea
 in component notation. By inspection, it may be verified that the functional behavior of (\ref{eq:a_near_2})
in the overlap region is via the exponentials $e^{\pm m_I r}$ and matches the functional behavior of the near region solution in (\ref{eq:a_near}).

\subsection{Matching}

In the overlap region, we relate the solutions of the near and far
regions by matching (\ref{eq:a_near}) and (\ref{eq:a_near_2}) as functions of $r$. This
allows us to solve for the constants $C^{I}$, $a_0^{I}$ and $\alpha^{I}$, though we
shall neither need nor evaluate $C^I$. Matching the constant terms in (\ref{eq:a_near}) 
and (\ref{eq:a_near_2}), we immediately find $a_0^{I}=\alpha^{I}$.
Matching the coefficients of the ratios of the exponential terms for indices $+$ and $-$
we obtain three relations labeled by the index $I=1,2,3$ for the three parameters $\alpha ^J$, 
\bea
{ p_+ \over p_-} \, \frac{k_{+}^{I}}{k_{-}^{I}}=
\frac{\sum_{J}\left(R^t \, Q^{-2} \, \Omega_{+} \right)^{IJ}
\left ( a_{+}^{J (0)} -p_+  \alpha^{J}\right)}
{\sum_{J}\left(R^t \, Q^{-2} \, \Omega_{-} \right)^{IJ}
\left( a_{-}^{J(0)}- p_- \alpha^{J}\right)}
\label{eq:alpha_match}
\eea
Note that the integration constants $C^I$ which arose in (\ref{k-eq}) drop out of these relations.
We shall solve these equations by introducing a $3 \times 3$ matrix $Z$,  defined by, 
\bea
Z^I{}_J 
\equiv
\sum_{K} \Big ( (\Omega_{+})^{-1} Q^{2}R \Big )^{IK} f \left (m_K \right ) 
\Big ( R^t Q^{-2} \, \Omega_{-} \Big )^K{}_J 
\label{eq:zeta}
\eea
Here, $f$ is a function obtained from the ratio $k^I_+/k^I_-$, and is given as follows,
\bea
\label{Psi}
f (x) = \frac{\Gamma\left(1+Lx \right)}{\Gamma\left(1-Lx\right)}
\left(\frac{p^{2}L^{2}}{4}\right)^{-Lx}
\eea
Equivalently $Z$ may be defined  by the corresponding matrix relation, 
\bea
\label{Z}
Z = (\Omega_{+})^{-1}  Q^{2}R \, f(D) \, R^t Q^{-2} \, \Omega_{-}
\eea
In terms of the matrix $Z$, we solve for $\alpha^{I}$ in (\ref{eq:alpha_match}) as follows,
\bea
\alpha = {I \over I-Z} \left ( { a_+^{(0)}  \over p_+} - Z \, { a_- ^{(0)} \over p_-} \right )
\eea
Substituting this result  into (\ref{eq:a2}) we obtain,
\bea
a_{+}^{I\left(2\right)} & = & 
\frac{B}{2U_0^2 }\sum_{JK} \delta ^{IJ} 
\left ( \cM \, { Z \over I - Z } \right ) _{JK}  
\left(a_{+}^{K \left(0\right)}-\frac{p_{+}}{p_{-}}a_{-}^{ K \left(0\right)}\right)
\no \\
a_{-}^{I\left(2\right)} & = & 
\frac{B}{2 U_0^2 }\sum_{JK} \delta ^{IJ} 
\left ( \cM \, { I \over I- Z} \right ) _{JK}
\left(a_{-}^{K \left(0\right)}-\frac{p_{-}}{p_{+}}a_{+}^{K \left(0\right)}\right)
\label{eq:a2_m}
\eea
where again in the $AdS_5$ approximation we have used $G^{IJ} = 2 \delta ^{IJ}$.

\subsection{Extracting the current-current correlators}

From (\ref{eq:a2_m}), the current in terms of the asymptotic data of
the gauge field is given by 
\bea
4\pi G_{5}J_{\mu}^{I}=a_{\mu}^{I\left(2\right)}
\label{eq:J_2}
\eea
Similar to the two-dimensional stress tensor, we can define the two-dimensional current by $\tilde \cJ_{\pm}^{I} \equiv U_0^2 V_{2}\cJ _{\pm}^{I}$. The modified current in terms of the gauge field perturbation is,
\bea
\tilde{J}_{\pm}^{I}=\frac{U_0^2 \, c}{6\pi BL} \, a_{\pm}^{I\left(2\right)}
\label{eq:J_hat}
\eea
Using (\ref{eq:J_expt}), we can read off the correlators, which are
given by
\bea
\left \langle \tilde{\mathcal{J}}_{+}^{I}\left(p\right)\tilde{\mathcal{J}}_{+}^{J}\left(-p\right)\right\rangle  
& = & 
- \frac{c}{6 \pi L} \, \frac{p_{+}}{p_{-}} \left( \cM \, { Z \over I - Z}  \right)_{IJ} 
\no \\
\left\langle \tilde{\mathcal{J}}_{-}^{I}\left(p\right)\tilde{\mathcal{J}}_{-}^{J}\left(-p\right)\right\rangle  
& = & 
- \frac{c}{6 \pi L} \, \frac{p_{-}}{p_{+}} \left( \cM \, { I \over I - Z}  \right)_{IJ} 
\no \\
\left\langle \tilde{\mathcal{J}}_{-}^{I}\left(p\right)\tilde{\mathcal{J}}_{+}^{J}\left(-p\right)\right\rangle  
& = & 
+\frac{c}{6 \pi L} \left( \cM \, { I \over I - Z}  \right)_{IJ} 
\label{eq:Jm_Jp}
\eea
where the $ \langle \tilde{\mathcal{J}}_{-}\tilde{\mathcal{J}}_{+} \rangle $
correlator was read off from the second line in (\ref{eq:a2_m}).  We would have obtained the same
result, up to contact terms, if we had instead read off the correlator from the first line in (\ref{eq:a2_m}).

\subsection{The axial anomaly}

The two-dimensional axial anomaly relations are obtained by forming the following linear
combinations from  (\ref{eq:Jm_Jp}), 
\bea
p_+ a_- ^{I(2)} + p_- a_+ ^{I(2)} = { B \over 2 U_0^2} \sum _{JK} \delta^{IJ} \cM_{JK} 
\left( p_+ a_{-}^{K (0)}- p_-a_{+}^{K (0)}\right)
\eea
which are independent of $Z$. Using the above definition of the currents $\tilde \cJ^I _\pm $, the anomaly equation may be recast as an operator relation in space-time coordinates, given by,
\bea
\label{4z1}
\p_+ \tilde \cJ_- ^I + \p_- \tilde \cJ^I_+ 
= { c \over 12 \pi L} \sum _{JK} \delta ^{IJ} \cM_{JK} ( \p_+ A^K _- - \p_- A_+ ^K)
\eea
We shall see below how the anomaly equation is saturated by massless states in unitary representations of $U(1)$-current algebras only.

\subsection{Bose symmetry}

Bose symmetry of the current correlators $\langle \tilde \cJ_\pm^I (p) \, \tilde \cJ_\pm ^J (-p) \rangle$
requires that they be symmetric under  the interchange of the internal indices $I$ and $J$, 
given that both correlators are even under $p_\pm \to - p_\pm$. Although the expressions 
given in (\ref{eq:Jm_Jp}) do not exhibit this symmetry manifestly, the correlators are actually 
symmetric, as we shall now show.

\sm

The following simple but fundamental relation,
\bea
\Big ( \cM \, \cU_\pm (r,r') \Big )^t = \cM \, \cU_\mp (r',r) = \cM \, \cU_\mp (r,r')^{-1}
\eea
may be established using the differential equations satisfied by $\cU_\pm$ in the variables
$r$ and $r'$, the boundary conditions $\cU_\pm (r,r)=I$, and the relation $(\cM \cH)^t = \cM \cH$.
Letting $r' \to \infty$ and setting $r=r_0$, and using the defining relations for $\Omega _\pm$,
we deduce the following relation, 
\bea
\label{omega_inv}
\Big ( \cM \, \Omega _\pm \Big ) ^t = \cM \, (\Omega _\mp)^{-1}
\eea
The expressions for $Z$ and for its inverse $Z^{-1}$ may be recast in terms of $\Omega _-$ and
$\Omega _+$ respectively, instead of in terms of both $\Omega _\pm$, and we have,
\bea
Z ~ & = & \cM^{-1} \Sigma _- ^t \, D f(D) \, \Sigma _-
\hskip 1in 
\Sigma _\pm = 4 R^t \, Q^{-2} \, \Omega _\pm
\no \\
Z^{-1}  & = & \cM^{-1} \Sigma _+ ^t \, D f(D)^{-1}  \, \Sigma _+
\eea
It is now manifest that the combinations $\cM Z^n$ are symmetric matrices for all integer $n$,
as are the combinations $\cM (I-Z)^{-1}$ and $\cM Z (I-Z)^{-1}$, thereby establishing Bose symmetry of the two-point correlators. The symmetry may be exhibited conveniently by re-expressing the correlators in terms of the currents $\hat J^I _\pm$ as follows,
\bea
\label{Jhat}
\tilde J_\pm ^I = \sum _J (\cM \Sigma _\pm ^{-1}) ^{IJ} \hat J_\pm ^J
\eea
The non-local correlators then take the following form, 
\bea
\label{corr1}
\left \langle \hat {\mathcal{J}}_{+}^{I} (p) \hat {\mathcal{J}}_{+}^{J}\left(-p\right)\right\rangle  
& = & 
+ \frac{c}{6 \pi L} \, \frac{p_{+}}{p_{-}} \Big ( \cA_+ - D f(D)^{-1}  \Big )_{IJ} ^{-1}
\no \\
\left\langle \hat {\mathcal{J}}_{-}^{I}\left(p\right)\hat{\mathcal{J}}_{-}^{J}\left(-p\right)\right\rangle  
& = & 
- \frac{c}{6 \pi L} \, \frac{p_{-}}{p_{+}} \Big ( \cA_- - D f(D)   \Big )_{IJ} ^{-1}
\eea
where we have defined,
\bea
\cA_\pm = (\Sigma _\pm ^{-1} )^t \cM \Sigma _\pm ^{-1}
\eea
The matrix $D$ being diagonal, and the matrices $\cA_\pm$ being symmetric by construction, Bose symmetry of the correlators in (\ref{corr1})  is now manifest.

\subsection{The IR limit of the current-current correlators}

The calculation of the IR limit of the current-current correlators may be carried out directly
on the expressions for the correlators presented in (\ref{corr1}).  To evaluate their IR limit as
$p^2 \to 0$ we note that all dependence on $p^2$ is concentrated in the function $f(D)$, 
and it will be convenient to decompose $D$ and $f(D)$ in terms of the rank one projection 
operators $\Pi _I$ onto the eigenspace with eigenvalue $m_I$ for $I=1,2,3$, 
\bea
D & = & m_1 \Pi _1 + m_2 \Pi _2 + m_3 \Pi_3
\no \\
f(D) \hskip 0.1in  & = & f(m_1) \, \Pi _1 + f(m_2) \, \Pi _2 + f(m_3) \, \Pi _3 
\no \\
f(D)^{-1}  & = &  f(m_1)^{-1}  \, \Pi _1 + f(m_2)^{-1}  \, \Pi _2 + f(m_3)^{-1} \, \Pi _3 
\eea
Since the eigenvalues $m_I$ are all real and distinct they may  be ordered such that $m_1<m_2<m_3$.
In view of the relation $m_1+m_2+m_3=0$, it follows that $m_3> 0$ while $m_1<0$.
The sign of $m_2$ is correlated with the sign of the charges $q^I$ as follows, 
\bea
{\rm sign} (m^2)  = - {\rm sign} (q^1 q^2 q^3) = - \eta
\eea
Given the expressions for $f(m_I)$ and $f$ in (\ref{Psi}), the asymptotic 
behavior as $p^2 \to 0$ is given by  $ f(m_1) \to 0$, $f(m_3) \to \infty$ for either value of $\eta$, 
while $f(m_2) \to \infty$ when $\eta <0$ and $f(m_2) \to 0 $ when $\eta >0$. 
We find the following limits,\footnote{The utmost right objects in (\ref{proj}) 
have been cast in a notation where the left and right multiplication by a 
projector $\Pi_I$  is to be understood as an instruction to invert the projected matrix on the subspace 
corresponding to the range of $\Pi_I$, and to set the inverse to zero on the kernel of $\Pi_I$.}
\bea
\label{proj}
\eta >0 &\hskip 0.6in & \lim _{p^2 \to 0} \Big ( \cA_+ - Df(D)^{-1} \Big )^{-1}  = { 1 \over (\cA_+)_{33} } \Pi _3
= \Big ( \Pi _3 \, \cA_+ \, \Pi _3 \Big )^{-1}
\no \\
\eta <0 &\hskip 0.6in & \lim _{p^2 \to 0} \Big ( \cA_-  - \, Df(D) \, \Big )^{-1}  \hskip 0.1in 
= { 1 \over (\cA_-) _{11} } \Pi _1 = \Big ( \Pi _1 \, \cA_- \, \Pi _1 \Big )^{-1}
\eea
where we have assumed that $(\cA_+)_{33}, (\cA_-)_{11}  \not= 0$. 
The remaining limits may be expressed in terms of the same notations, as follows, 
\bea
\eta < 0 &\hskip 0.6in &
\lim _{p^2 \to 0} \Big ( \cA_+ - Df(D)^{-1} \Big )^{-1} 
= \Big ( (I-\Pi _1) \,  \cA_+ \, (I- \Pi _1)  \Big )^{-1}
\no \\
\eta > 0 &\hskip 0.6in &
\lim _{p^2 \to 0} \Big ( \cA_-  - Df(D) \Big )^{-1} \hskip 0.15in 
= \Big ( (I- \Pi _3) \,  \cA_- \, (I-\Pi _3)  \Big )^{-1}
\eea
The final expressions for the correlators simplify and we find, for $\eta >0$, 
\bea
\label{eta_g0}
\left \langle \hat{\mathcal{J}}_{+}^{I}\left(p\right) \hat{\mathcal{J}}_{+}^{J}\left(-p\right) \right\rangle  
& = & 
+ \frac{c}{6 \pi L} \, \frac{p_{+}}{p_{-}}  \Big ( \Pi _3 \, \cA_+ \, \Pi _3 \Big )^{-1} _{IJ} 
\no \\
\left\langle \hat {\mathcal{J}}_{-}^{I}\left(p\right)\hat{\mathcal{J}}_{-}^{J}\left(-p\right)\right\rangle  
& = & 
- \frac{c}{6 \pi L} \, \frac{p_{-}}{p_{+}} \Big ( (I- \Pi _3) \,  \cA_- \, (I-\Pi _3)  \Big )^{-1}_{IJ} 
\eea
while for $\eta <0$ we have, 
\bea
\label{eta_l0}
\left \langle \hat{\mathcal{J}}_{+}^{I}\left(p\right)\hat{\mathcal{J}}_{+}^{J}\left(-p\right)\right\rangle  
& = & 
+  \frac{c}{6 \pi L} \, \frac{p_{+}}{p_{-}}  \Big ( (I-\Pi _1) \,  \cA_+ \, (I- \Pi _1)  \Big )^{-1}_{IJ} 
\no \\
\left\langle \hat{\mathcal{J}}_{-}^{I}\left(p\right)\hat{\mathcal{J}}_{-}^{J}\left(-p\right)\right\rangle  
& = & 
- \frac{c}{6 \pi L} \, \frac{p_{-}}{p_{+}}  \Big ( \Pi _1 \, \cA_- \, \Pi _1 \Big )^{-1}_{IJ} 
\eea
We note that these expressions are consistent under an overall reversal of the sign of the 
charges $q^I$. Indeed, under $q^I \to - q^I$, we have of course $\eta \to -\eta$, and $\cM \to -\cM$
so that $m_I \to -m_I$ and $f(m_I) \to f(m_I)^{-1}$. From these, we deduce that $\cU_\pm \to \cU_\mp$, 
and thus  $\Omega _\pm \to \Omega _\mp$,  $\Sigma _\pm \to \Sigma _\mp$, and $\cA_\pm \to - \cA_\mp$.
Combining these results, it is manifest in both (\ref{corr1}), (\ref{eta_g0}), and (\ref{eta_l0})
that an overall reversal of the sign of $q^I$ corresponds to a reversal of the chirality 
of the currents, namely $\hat \cJ_\pm \to \hat \cJ_\mp$.

\subsection{Unitarity of the IR current algebras}

In this section, we verify that the current correlators computed above are unitary by checking the sign 
of the position space correlators. We shall specialize to the case $\eta >0$, since the opposite case is 
simply related by a reversal of the chirality of the currents, as shown in the preceding section. Fourier 
transforming the correlators of (\ref{eta_g0}), we find, 
\bea
\label{JJ_pos}
\left\langle
\hat{\mathcal{J}}_+^I\left(x\right)\hat{\mathcal{J}}_+^J\left(0\right)
\right\rangle
& = &
-\frac{c}{12\pi^2L} \, \frac{1}{(x^+)^2}\Big ( \Pi _3 \, \cA_+ \, \Pi _3 \Big )^{-1}_{IJ}\\
\left\langle \hat{\mathcal{J}}_{-}^{I}\left(x\right)\hat{\mathcal{J}}_{-}^{J}\left(0\right)\right\rangle  
& = & 
+ \frac{c}{12 \pi^2 L} \, \frac{1}{(x^-)^2} \Big ( (I- \Pi _3) \,  \cA_- \, (I-\Pi _3)  \Big )^{-1}_{IJ} 
\no
\eea
As was shown in Appendix C of \cite{D'Hoker:2010hr}, the proper sign  of the current two-point  correlator in a unitary theory should be negative. Thus, to have unitarity in the IR, the non-zero entry of $\Pi_3 \cA_+ \Pi_3$ 
should be positive, while the non-zero $2 \times 2$ part of the matrix $(I- \Pi _3) \,  \cA_- \, (I-\Pi _3)$ should be negative.

\sm

To determine these signs from the explicit form of the correlators, we have computed the corresponding matrices numerically for  special values,  namely,\footnote{Recall that we may restrict $\alpha$ to the interval $0 < \alpha \leq 1$ since  permutations on  the charges induce the transformation $\alpha \to 1/\alpha$ and $\alpha \to - 1 - \alpha$.}
\bea
\alpha =1, \, 0.9, \, 0.8, \, 0.7, \, 0.6, \, 0.5, \, 0.4, \, 0.25,  \, 0.1, \, 0.05
\eea 
which form a subset of the numerical values where the supersymmetric magnetic brane solution was evaluated numerically in (\ref{numvals}). This calculation is done by solving (\ref{Uder}) numerically, extracting $\Omega_\pm$ numerically from the solutions, and using these ingredients to  compute $\cA_\pm$ and their projections. For each of the above values of $\alpha$, we have verified that the value $\lambda _3$ of $(\cA_+ )_{33}$ is positive, and that both non-zero eigenvalues $\lambda _1$ and $\lambda _2$ of $ (I- \Pi _3)  \cA_- (I-\Pi _3)$ are negative. Numerical results for $\alpha < 0.05$ become significantly less reliable. Thus, all numerically accessible signs are consistent with unitarity for all current-current correlators. 

\sm

Away from the above range of values, we can make a partially analytical argument that the signs will remain 
consistent with unitarity. In particular, the sign of a correlator cannot change by the correlator vanishing. 
This is because the coefficient of the correlator is given by the inverse of a combination of matrices $\cM$ 
and $\Sigma _\pm$ all of which are regular and finite at all values of the charges. Thus, the only other 
possibility left is that the signs of the eigenvalues could change by having the correlator diverge
for special values of the charges~$q^I$. While it is not yet clear how this possibility can be ruled out 
analytically, certainly our numerical evidence points to the contrary. 

\sm

Finally, we note that these signs are consistent with the ones obtained from considering the anomaly equation (\ref{4z1}) in the IR limit. The mixing matrix $\cM$ has the following characteristic polynomial, 
\bea
\mm^3 - {8 \over L} \mm - 16 \, \eta =0
\eea
satisfied by the three eigenvalues $\mm_1, \mm_2, \mm_3$ of ${\cal M}$. In particular, the sum of the eigenvalues vanishes, $\mm_1 + \mm_2 + \mm_3=0$, and the product of the eigenvalues satisfies $\mm_1 \mm_2 \mm_3= 16 \eta$. When $\eta >0$, as was assumed throughout this subsection, two of the eigenvalues of ${\cal M}$ must be negative, and one positive, in agreement with the counting obtained above
from studying the full current correlators in (\ref{JJ_pos}), and in agreement with the fact that the anomaly equation is saturated by the unitary part of the correlator. 

\sm

Therefore, we conclude that, for $\eta >0$, the IR limit of the two-point correlator of the operator $\tilde J_+^I$ 
corresponds to a single component of the three currents associated with a unitary current algebra. In section \ref{sec:6}, we shall see that the corresponding current operator fits into the emergent $\cN=2$ superconformal algebra of the IR limit. The other two components of $\cJ_+^I$ do not correspond to a current algebra, but will receive contributions from double-trace operators, as was shown in \cite{D'Hoker:2010hr} for the non-supersymmetric magnetic brane. The two unitary components of $\tilde \cJ_-^I$ generate unitary current algebras, but they are not part of any superconformal algebra.

\section{Supercurrent correlators}
\label{sec:5}
\setcounter{equation}{0}

In this section, we shall compute the two-point correlators of the supercurrent in the background of a
general magnetic brane solution, in the IR limit. We follow the methods of the preceding sections. We begin by
decoupling and solving the linearized field equations for the fermion fields $\psi_{M}$ and $\lambda_a$ 
subject to specified holographic boundary conditions $\psi_{\mu}^{(0)}$ and $\lambda^{(0)}_a$. 
We then extract the supercurrent two-point correlator. Since holographic calculations involving fermion 
fields in non-trivial backgrounds are somewhat less standard than those with bosons, our presentation
will include more details than the calculations for boson fields did.  Some of these details have been 
relegated  to Appendices  \ref{sec:F}, and \ref{sec:C}. Useful references to holographic calculations involving fermions in general, and the supercurrent in particular, may be found  in \cite{Kraus:2007vu,Iqbal:2009fd,Gauntlett:2011wm,Argurio:2014uca}.

\sm

The Fermi field equations for the $\pm$ components $\psi_{M\pm}$ and $\lambda _\pm ^a$ of the $SU(2)$ 
R-symmetry doublets under which the Fermi fields transform are related by reversing the sign of $\mg$ 
on the one hand, and by complex conjugation on the other hand (see Appendix \ref{sec:A}). As a result,  
we may analyze the field equations for the component corresponding to case $\mg=+1$,  with the field 
equation for the component corresponding to $\mg=-1$ being given by complex conjugation. 
Henceforth, we shall set $\mg=1$ without loss of generality.

\subsection{Holographic asymptotics}
\label{sec:51}

The asymptotic form of the gravitino field $\psi _{\hat M} (r,x)$ near the $AdS_{5}$ boundary where  $ r \to \infty$ is given by the following expansions, expressed in frame indices $\hat M=(\hat \mu, \hat r)$, 
\bea
\label{5a1}
&&
\psi_{\hmu} (x, r) = 
e^{-(\Delta -3) r } \psi _{\hmu}^{(0)}(x) + 
\cdots +
e^{- \Delta r } \psi _{\hmu}^{(3)}(x) + 
r e^{-\Delta  r } \psi _{\hmu}^{(\ln )}(x) + 
\cO(e^{-(\Delta +1) r } )
\\ &&
\psi_\hr (x,r) = 
e^{-(\Delta-2)r} \psi _{\hr}^{(1)}(x) + \cdots 
+ e^{-(\Delta+1)r} \psi _{\hr}^{(4)}
+re^{-(\Delta+1)r} \psi _{\hr}^{(\rm {ln})}(x)
+ \cO(e^{-(\Delta +2)r})
\hskip 0.5in 
\no
\eea
The  conformal dimension of the four-dimensional supercurrent is denoted  by  $\Delta=7/2$. 
The coefficients $\psi_{\hmu}^{(0)}$ and $\psi_{\hmu}^{(3)}$ are respectively the source and 
expectation value of the supercurrent, while  $\psi_\hr^{(1)}$ and $\psi_\hr^{(4)}$ are 
auxiliary fields without dynamical contents. 

\sm

The complete expansion of the gravitino field $\psi _{\hat M}$, including the terms with coefficients 
$\psi^{(\ell)}_\hmu$ and $\psi _\hr ^{(\ell+1)}$  for $\ell=1,2$, is presented in Appendix \ref{sec:C}, 
as is the expansion of the gaugino field $\lambda _a$,  and the interrelation between the coefficients 
in the expansion which result from the fermion field equations in the background of the supersymmetric 
magnetic brane solution. 
Of these results, we shall highlight here the following projection relations,
\bea
\label{5a2}
(I-\Gamma^{r}) \psi _{\hmu}^{(0)}=0 \quad\quad & 
(I+\Gamma^{r}) \psi _{\hmu}^{(3)}=0 \quad\quad & 
(I+\Gamma^{r}) \psi _{\hmu}^{(\rm {ln})}=0
\no \\
(I-\Gamma^{r}) \psi _{\hr}^{(1)}=0 \quad\quad & 
(I+\Gamma^{r}) \psi _{\hr}^{(4)}=0 \quad\quad & 
(I+\Gamma^{r}) \psi_{\hr}^{(\rm {ln})}=0
\eea
Alternatively, the asymptotic expansion may be cast in terms of the gravitino field $\psi _M$ 
expressed in Einstein indices. The relation is, of course, obtained in terms of the orthonormal frame 
$\me _M {}^{\hat M}$, and we have, $\psi _M = \me_M {}^{\hat M} \psi _{\hat M}$. The orthonormal frame itself admits a Fefferman-Graham expansion which must be consistent with that of the metric in (\ref{eq:feff_g}).
It will be convenient to choose a gauge for the frame structure group $SO(1,4)$ given by,
\bea
\me _r {}^{\hat r} =1 \hskip 1in \me _r {}^{\hat \mu} = \me _\mu {}^{\hat r} =0
\eea
while the remaining components have the following expansion, 
\bea
\me _\mu {}^{\hat \mu} (x,r) = 
e^r \, \me ^{(0)} \! {}_\mu {}^{\hat \mu} (x) 
+ e^{-r} \, \me ^{(2)} \! {}_\mu {}^{\hat \mu} (x) + 
e^{-3r} \, \me ^{(4)} \! {}_\mu {}^{\hat \mu} (x) 
+ r \, e^{-3r} \, \me ^{({\rm ln} )} \! {}_\mu {}^{\hat \mu} (x) + \cO(e^{-5r}) \quad
\eea
As a result, the expansion of $\psi _M$, expressed  in Einstein indices, is given as follows, 
\bea
\label{5a2_1}
&&
\psi_{\mu} (x, r) = 
e^{-(\Delta -4) r } \psi _{\mu}^{(0)}(x) + 
\cdots +
e^{- (\Delta -1)  r } \psi _{\mu}^{(3)}(x) + 
r e^{- (\Delta-1)   r } \psi _{\mu}^{(\ln )}(x) + 
\cO(e^{-\Delta  r } )
\\ &&
\psi_r (x,r) = 
e^{-(\Delta-2)r} \psi _r^{(1)}(x) + \cdots 
+ e^{-(\Delta+1)r} \psi _r^{(4)}
+re^{-(\Delta+1)r} \psi _r^{(\rm {ln})}(x)
+ \cO(e^{-(\Delta +2)r})
\hskip 0.5in 
\no
\eea
where $\psi _r (x,r) = \psi _{\hat r} (x,r)$ and $\psi _\mu (x,r) = \me_\mu {}^{\hat \mu} (x,r) \psi _{\hat \mu}(x,r)$.  By expanding each of these equations in powers of $e^r$ we derive relations between the expansion coefficients. For $\psi _\hr$ these are simply obtained by dropping the hat on $r$, while for the other components, they generate relations of the type
$\psi ^{(0)} _\mu (x) = \me ^{(0)} {}_\mu {}^{\hat \mu} (x) \psi ^{(0)}  _{\hat \mu}(x)$, and so on. 

\sm

Finally, we comment on gauge-fixing local supersymmetry. The choice of gauge affects our ability to separate variables in the solution of the supergravity equations for the Fermi fields, and must therefore be made with care. In pure $AdS$ space-time, natural gauge choices for local supersymmetry  include $D^M \psi _M=0$ and $\Gamma ^M \psi _M=0 $, since they preserve the symmetries of $AdS$, and these choices were indeed made,  for example in  \cite{Gauntlett:2011wm}. For the supersymmetric magnetic brane background, the symmetries are reduced, and the above gauge choices do not allow for a suitable separation of variables in the Fermi field equations. Therefore, we shall, for the time being, refrain from choosing a gauge, and retain all components of the Fermi fields. The natural choice of gauge will then be identified during the solution of the supergravity equations, and will include the fermionic counterpart of Fefferman-Graham gauge $\psi _r=0$, which was used earlier in \cite{Kraus:2007vu}. The residual gauge freedom left by this gauge choice will be fixed by setting a suitable projection (associated with the particular supersymmetric magnetic brane solution)  of the spinor-tensor $\Gamma ^\mu \psi _\nu$  to zero.

\subsection{Holographic supercurrent correlators}

The response of the on-shell action $S_{{\rm sugra}}$ to an infinitesimal variation of the source field 
$\psi ^{(0)}_\mu$ is given by the expectation values $S^\mu$ of the supercurrent operator $\cS^\mu$, 
\bea
\delta S_{{\rm sugra}} = \frac{1}{2}\int d^{4}x\sqrt{g^{(0)}}\,
\bar{S}^{\mu i} \delta\psi _{\mu i}^{(0)}
\label{eq:var_Su}
\eea
where $i=\pm$ is the $SU(2)$ index on the Fermi fields. The value of $S^\mu_i$ in terms of the boundary gravitino data is given by,
\bea
8\pi G_{5}S^{\mu}_+ & = & -\Gamma^{\mu\nu} \psi _{\nu +}^{(3)} + \rm{local}\no\\
8\pi G_{5}S^{\mu}_- & = & + \Gamma^{\mu\nu} \psi _{\nu -}^{(3)} + \rm{local}
\label{eq:S}
\eea
The normalizations of these formulas will be carefully derived in Appendix \ref{sec:F}. They will be obtained using a boundary action to ensure that the variational principle for the gravitino is well-defined, and a counter-term to cancel out UV divergences and regularize the action near the boundary of $AdS_5$.   The ``local'' terms depend locally on $\psi _\mu^{(0)}$; they will not contribute to the correlator at non-coincident points, and will be omitted here.

\sm

Finally, from the induced expectation value $S^{\mu i}(x)$ of the supercurrent operator $\cS^{\mu i} (x)$ we extract the supercurrent correlator using linear response theory,
\bea
\label{5a3}
S^{\mu i} (x) = \frac{i}{2}\int d^4y \sqrt{g^{(0)}} \left \< \cS ^{\mu i} (x) \bar \cS^{\nu j} (y) \right \> \delta \psi _{\nu j} ^{(0)} (y)
\eea
 The method used to compute the low energy correlators is as follows. We solve the linearized field 
equations in the near region $e^{2r} \ll 1$ where the geometry is effectively $AdS_3 \times T^2$, 
and in the far region $p^2 \ll e^{2r}$ where we can effectively set $p^2=0$. The solutions in the near and far  
regions are then matched in the overlap region  $p^2 \ll e^{2r} \ll 1$. Since we assume $p^2 \ll 1$, 
the matched solution is valid in this parametrically large region, and we use it to obtain the 
expectation value $S^{\mu}$ of the supercurrent $\cS^{\mu}$.

\subsection{Structure of the perturbations}

Since the supersymmetric magnetic brane solution is purely bosonic, 
linear fluctuations in the Fermi fields do not mix with bosonic fields. Thus, the bosonic fields are as follows, 
\bea
\label{5b1}
ds^2 & = &  dr^{2}+2e^{2W(r)}dx^{+}dx^{-}+2e^{2U(r)}dx^{u}dx^v
\no \\
F^I & = & q^I B \, dx^2 \wedge dx^3
\no \\
\phi ^A & = & \phi ^A(r)
\eea
where $U, W$ and $\phi$ are the functions of the supersymmetric magnetic brane solution,
given in Section \ref{sec:2}.
Translation invariance in $x^\mu$ of the brane solution is used to Fourier decompose the fluctuations 
in plane waves of given momentum $p_\mu$.  We shall consider only the correlators of the 
components $\cS^\pm$  of the supercurrent along the $01$-plane, 
and retain only their dependence on the coordinates $x^\pm$ of the $01$-plane. Thus, we may set to zero
the momentum components $p_2=p_3=0$, and retain only dependence on $p_\pm$, as follows, 
\bea
\label{5b2}
\psi_{M} (x,r) & = & \tilde \psi_{M} (p,r)\, e^{ip\cdot x}
\no \\
\lambda^{a} (x,r)  & = & \tilde \lambda^{a} (p,r) \, e^{ip\cdot x}
\eea
where the tildes indicate Fourier components, and we continue to use the notations of (\ref{eq:momenta}) for the inner product. Again we shall be interested in the IR limit, where $p^2 \ll 1$. Finally, in the following equations, we will denote the coordinates by $x^+,x^-$ with indices $m,n,\ldots$ and by $x^u,x^v$ with indices $\alpha,\beta,\ldots$.

\sm

The linear fluctuations of the fields $\psi _M$ and $\lambda ^a$ of (\ref{5b2})  in the presence of the supersymmetric magnetic brane solution satisfy the supergravity equations $\Psi ^M=\Lambda ^a=0$ given in (\ref{2a7}) 
to linearized order in the Fermi fields, and the bosonic fields are given in (\ref{5b1}).

\sm

To solve (\ref{2a7}), we decompose these equations according to their representation under the  symmetries 
of the supersymmetric magnetic brane solution, specifically the $SO(1,1)$ Lorentz symmetry in the
$01$-plane and the $SO(2)$ rotational symmetry of the $23$-plane. Under $SO(1,1) \times SO(2)$, 
the field equations  $\Psi ^r=0$ and $\Lambda^a=0$ are irreducible and transform under helicity $\pm \half$ for 
$SO(1,1)$ as well as $SO(2)$.
The remaining field equations $\Psi ^m = \Psi ^\alpha=0$ are, however, further reducible into 
helicity $\pm \half$ components $\Gamma _m \Psi ^m = \Gamma _\alpha \Psi ^\alpha$ 
components, and helicity $\pm {3 \over 2}$ components. The latter may be formulated in a 
variety of ways, such as by explicitly  implementing the subtraction of the helicity $\pm \half$ components,
\bea
\label{5b3}
2 \Psi^{m}- \Gamma^{m} \, \Gamma_{n}\Psi^{n} & = & 0 
\no \\
2 \Psi^{\alpha} \, - \Gamma^{\alpha} \, \Gamma_{\beta} \Psi^{\beta} & = & 0
\eea
An equivalent formulation, which will be more convenient in the present context, is by the explicit
use of the light-cone and complex coordinate indices $+, -, u, v$, for which we have,
\bea
\label{5b4}
\left(\Gamma^+\right)^{2}=
\left(\Gamma^-\right)^{2}=
\left(\Gamma^{u}\right)^{2}=
\left(\Gamma^v\right)^{2}=0
\eea
In summary, the Fermi field equations decompose into the following irreducible components.
The helicity $\pm {3 \over 2}$ components respectively under $SO(1,1)$ and $SO(2)$ are given by, 
\bea 
\label{5b5}
\Gamma ^+ \Psi ^+ =0 & \hskip 1in & \Gamma ^u \Psi ^u=0
\no \\
\Gamma ^- \Psi ^- =0 & \hskip 1in & \Gamma ^v \Psi ^v=0
\eea
The helicity $\pm \half$ components under both $SO(1,1)$ and $SO(2)$ are given by,
\bea
\label{5b6}
\Gamma _m \Psi ^m = 0 & \hskip 1in & \Psi ^r=0
\no \\
\Gamma _\alpha \Psi ^\alpha =0 && \Lambda ^a =0 
\eea
In the subsequent sections, we shall decouple these equations for the 
different helicity components of the fields $\Psi _M$ and $\lambda ^a$.

\subsection{Covariant derivatives  for the brane Ansatz}

Before we reduce the fermionic field equations in the next sections, we summarize the covariant derivatives on spinors in the background of the magnetic brane Ansatz,
	\bea
	ds^2 = dr^2 + 2 e^{2W} dx^+ dx^- + 2 e^{2U} dx^u dx^v
	\eea
where $U,W$ are functions of $r$ only. The associated frame fields and spin connection are given by (frame indices are hatted),
\bea
	\me_{r}{}^{\hat{r}}=1 
	& \hskip 0.5in &
	\me_{+}{}^{\hat{+}}=\me_{-}{}^{\hat{-}}=e^{W} 
	\hskip .85in  
	\omega_{+\hat{-}\hat{r}}=\omega_{-\hat{+}\hat{r}}=W^{\prime}e^{W}
	\no \\ &&
	\me_{u}{}^{\hat{u}} ~ =\me_{v}{}^{\hat{v}} ~ =e^{U} 
	\hskip 0.9in
	\omega_{u\hat{v}\hat{r}} \, =\omega_{v\hat{u}\hat{r}} ~ =U^{\prime}e^{U}
	\eea
and the covariant derivatives are given by,
	\bea
D_{M}\psi_{N} & = & 
\partial_{M}\psi_{N}+\frac{1}{4}\omega_{MRS}\Gamma^{RS}\psi_{N}-\Gamma_{MN}^{P}\psi_{P}
\no \\
D_{M}\lambda^{a} & = & 
\partial_{M}\lambda^{a}+\frac{1}{4}\omega_{MRS}\Gamma^{RS}\lambda^{a}
	\eea
Note that in the last term on the right side of the first line 	$\Gamma _{MN} ^P$ is the affine connection
(not to be confused with Dirac matrices represented by the same symbol), which will in fact cancel in the subsequent covariant derivatives since there $MN$ will enter only anti-symmetrically.
The covariant derivative terms in the field equations (\ref{2a7}), reduced from the magnetic brane Ansatz (\ref{Ansatz}), are given  by,
	\bea
	\Gamma^{M}D_{M}\lambda^{a} & = & \Gamma^{r}(\p_r + U'+W') \lambda^{a}+ip_{m}\Gamma^{m}\lambda^{a}
	\eea
	and 
	\bea
	2\Gamma_{r}\Gamma^{rNP}D_{N}\psi_{P} & = & 2i\Gamma^{+-}(p_{+}\psi_{-}-p_{-}\psi_{+})+(W^{\prime}+2U^{\prime})\Gamma^{rm}\psi_{m}
	\no \\
	&  & +(2W^{\prime}+U^{\prime})\Gamma^{r\alpha}\psi_{\alpha}+ip_{m}\Gamma^{m\alpha}\psi_{\alpha}
	\no \\
	\Gamma_{m}\Gamma^{mNP}D_{N}\psi_{P} & = & \Gamma^{rm} (\p_r + U'+W') \psi_{m}+(2\partial_{r}+3U^{\prime}+W^{\prime})\Gamma^{r\alpha}\psi_{\alpha}
	\no \\
	&  & +ip_{m}\Gamma^{m\alpha}\psi_{\alpha}+ip_{m}\Gamma^{mr}\psi_{r}-(W^{\prime}+2U^{\prime})\psi_{r}
	\no \\
	\Gamma_{\alpha}\Gamma^{\alpha NP}D_{N}\psi_{P} & = & 2i\Gamma^{+-}(p_{+}\psi_{-}-p_{-}\psi_{+})+(2\partial_{r}+3W^{\prime}+U^{\prime})\Gamma^{rm}\psi_{m}
	\no \\
	&  & +ip_{m}\Gamma^{m\alpha}\psi_{\alpha}+\Gamma^{r\alpha}\hat{\partial}\psi_{\alpha}+2ip_{m}\Gamma^{mr}\psi_{r}-(U^{\prime}+2W^{\prime})\psi_{r}
	\no \eea
where $m=+,-$ and $\alpha=u,v$.

\subsection{Reducing the helicity $\pm {3 \over 2}$  equations}

Decomposing the general Fermi field equations for $\Psi ^M$ of (\ref{2a7})  
into the helicity $\pm {3 \over 2}$ equations $\Gamma ^+ \Psi ^+=\Gamma ^- \Psi ^-=0$ 
of (\ref{5b5})  gives,  
\bea
\label{5c1}
\Gamma^+ \Gamma^{+ NP}D_{N}\psi_{P}
+\frac{3i}{4}X_{I}\Gamma^+ \Gamma^{+ N uv}F_{uv}^{I}\psi_{N} & = & 0
\no \\
\Gamma^- \Gamma^{- NP}D_{N}\psi_{P}
+\frac{3i}{4}X_{I}\Gamma^- \Gamma^{- N uv}F_{uv}^{I}\psi_{N} & = & 0
\eea
while decomposing  the general field equations for $\Psi ^M$ of (\ref{2a7})  
into the helicity $\pm {3 \over 2}$  equations $\Gamma ^u \Psi ^u=\Gamma ^v \Psi ^v=0$ of (\ref{5b5})  gives, 
\bea
\label{5c2}
\Gamma^{u}\Gamma^{uNP}D_{N}\psi_{P}
- \frac{3i}{4} \, e^{-2U}  X_{I}\Gamma^{u} F^{I}_ {uv}  \psi_v & = & 0
\no \\
\Gamma^{v}\Gamma^{vNP}D_{N}\psi_{P}
+\frac{3i}{4} \, e^{-2U} X_{I}\Gamma^{v} F^{I}_{uv}  \psi_{u} & = & 0
\eea
where $F_{uv}^{I}=iq^{I}B$ is constant. The contributions to the covariant derivative terms 
in (\ref{5c1}) vanish unless either $N$ or $P$ equals the index $-$ on the first line, and equals the index $+$ 
on the second line. Similarly, the contribution to the covariant derivative terms in (\ref{5c2}) 
vanishes unless $N$ or $P$ equal $v$ on the first line, and $u$ on the second line. 
The resulting simplified equations 
for the helicity $\pm {3 \over 2} $ fields under $SO(1,1)$ are as follows,
\bea
\left ( \Gamma^{r}\hat{\partial}_r
+\frac{3i}{4}X_{I}F_{uv}^{I}\Gamma^{uv}
+\frac{3}{2}V_{I}X^{I}\right ) \Gamma^{+}\psi_{-}
+ip_{-}\Gamma^{+}(\Gamma^{r}\psi_{r}
+\Gamma^{\alpha}\psi_{\alpha}) & = & 0
\no \\
\left ( \Gamma^{r}\hat{\partial}_r
+\frac{3i}{4}X_{I}F_{uv}^{I}\Gamma^{uv}
+\frac{3}{2}V_{I}X^{I}\right ) \Gamma^{-}\psi_{+}
+ip_{+}\Gamma^{-}(\Gamma^{r}\psi_{r}
+\Gamma^{\alpha}\psi_{\alpha}) & = & 0
\label{5c3}
\eea
Here and in the sequel, we use the following abbreviation,
\bea
\hat{\partial} _r \equiv \partial_{r}+U^{\prime}+W^{\prime}
\eea
Similarly, the  simplified equations 
for the helicity $\pm {3 \over 2} $ fields under $SO(2)$ are as follows,
\bea
\left ( \Gamma^{r}\hat{\partial}_r
+ip_{m}\Gamma^{m}
+\frac{3i}{4}e^{-2U}X_{I}F_{uv}^{I}
+\frac{3}{2}V_{I}X^{I}\right ) \Gamma^{u}\psi_{v} & = & 0
\no \\
\left ( \Gamma^{r}\hat{\partial}_r
+ip_{m}\Gamma^{m}
-\frac{3i}{4}e^{-2U}X_{I}F_{uv}^{I}
+\frac{3}{2}V_{I}X^{I}\right ) \Gamma^{v}\psi_{u} & = & 0
\label{5c4}
\eea
Note that in the presence of the supersymmetric magnetic brane solution, the helicity $\pm { 3 \over 2}$ spinors  $\Gamma^{u}\psi_{v}$ and $\Gamma^{v}\psi_{u}$ completely decouple 
from the rest of the equations, in both the near and far regions. Since our interest is in the 
components $\cS^\pm$ of the supercurrent only, we shall set the sources $\Gamma^{u}\psi_{v}^{(0)}$ 
and $\Gamma^{v}\psi_{u}^{(0)}$ to zero, so that the entire fields then vanish, 
\bea
\label{5c5}
\Gamma^{u}\psi_{v}= 
\Gamma^{v}\psi_{u}=0
\eea
Equivalently, the field components $\psi _u$ and $\psi_v$ may be expressed entirely in terms of the 
helicity $\pm \half$ combination $\Gamma ^\alpha \psi _\alpha$ by the relations, 
\bea
\psi _u & = & - \half \Gamma ^u {}_u \, \Gamma _u \, \Gamma ^\alpha \psi _\alpha
\no \\
\psi _v & = & - \half \Gamma ^v {}_v \, \Gamma _v \, \Gamma ^\alpha \psi _\alpha
\eea
where $\Gamma ^u{}_u = - \Gamma ^v {}_v$ is the chirality involution matrix for the group $SO(2)$.

\subsection{Reducing the helicity 1/2 equations}

The reduced helicity $\pm \half$ gravitino equations of (\ref{5b4})  may be written out as,
\bea
0 & = & 
\Gamma_{r}\Gamma^{rNP} D_{N} \psi_{P}
+\frac{3i}{4}X_{I}\Gamma_{r} \Gamma^{rN uv}\psi_{N}F_{uv}^{I}
-\frac{i}{2}\Gamma^{r}\lambda^{a}f_{A}^{a}\partial_{r}\phi^{A}
\no \\
 &  & 
 -\frac{1}{2}\sqrt{\frac{3}{2}}X_{I}^{a} \Gamma^{uv} \lambda^{a}F_{uv}^{I}
 +\frac{3}{2}\Gamma_{r}\Gamma^{rN}\psi_{N}V_{I}X^{I}-i\sqrt{\frac{3}{2}}\lambda^{a}V_{I}X_{a}^{I}
 \label{5d1}
 \\
0 & = & 
\Gamma_{m} \Gamma^{mNP} D_{N} \psi_{P}
+\frac{3i}{4}X_{I} \Gamma_{m} \Gamma^{mN uv} \psi_{N} F_{uv}^{I}
+i\Gamma^{r} \lambda^{a} f_{A}^{a} \partial_{r}\phi^{A}
\no \\
 &  & 
 -\sqrt{\frac{3}{2}} X_{I}^{a} \Gamma^{uv}\lambda^{a}F_{uv}^{I}
 +\frac{3}{2} \Gamma_{m} \Gamma^{mN} \psi_{\nu}V_{I}X^{I}
 -2i\sqrt{\frac{3}{2}} \lambda^{a}V_{I}X_{a}^{I}
 \label{5d2}
 \\
0 & = & 
\Gamma_{\alpha} \Gamma^{\alpha NP} D_{N}\psi_{P}
+\frac{3i}{4}X_{I}(F^{I})^{\alpha\beta} \Gamma_{\alpha} \psi_{\beta}
+i\Gamma^{r}\lambda^{a}f_{A}^{a}\partial_{r}\phi^{A}
\no \\
 &  & 
 +\sqrt{\frac{3}{2}}X_{I}^{a}\Gamma^{uv}\lambda^{a}F_{uv}^{I}
 +\frac{3}{2}\Gamma_{\alpha}\Gamma^{\alpha N}\psi_{N}V_{I}X^{I}
 -2i\sqrt{\frac{3}{2}}\lambda^{a}V_{I}X_{a}^{I}
 \label{5d3}
\eea
The gaugino field equations are given by,
\bea
0 & = & 
\Gamma^{M} D_{M}\lambda^{a}
+\frac{i}{2} \Gamma^{M} \Gamma^{N} \psi_{M}f_{A}^{a}\partial_{N}\phi^{A}
-\frac{1}{2}\sqrt{\frac{3}{2}} X_{I}^{a} F_{uv}^{I} \Gamma^{M} \Gamma^{uv}\psi_{M}
\no \\ &&
-i\left(\frac{1}{4}\delta_{ab}X_{I}+T_{abc}X_{I}^{c}\right) F_{uv}^{I} \Gamma^{uv}\lambda^{b}
-i\sqrt{\frac{3}{2}}\Gamma^{M}\psi_{M}V_{I}X_{a}^{I}
\no \\
 &  & 
 -2V_{I}\left(\frac{1}{4}\delta_{ab}X^{I}+T_{abc}X^{Ic}\right)\lambda^{b}
 \label{5d4}
\eea
When recast in the form of equations for the helicity $\pm { 3 \over 2} $ fields $\Gamma ^+ \psi _-$ and 
$\Gamma ^- \psi _+$ with vanishing fields $\Gamma ^ u \psi _v$ and $ \Gamma ^v \psi _u$ as
stated in (\ref{5c5}), and the helicity $\pm \half$ fields  
$\Psi ^r, \Gamma ^m \psi _m, \Gamma ^\alpha \psi_\alpha$ and $\lambda _a$, 
one shows by inspection that the equations decouple into the eigenspaces of $\Gamma ^{\hu \hv}$.

\subsubsection{Choice of an adapted basis of spinors}

To implement the decoupling of the helicity $\pm \half$ equations argued in the preceding subsection, 
we decompose the spinors onto a basis in which the generators $\Gamma ^{\hr}$ and 
$\Gamma ^{\hu \hv}$ are diagonal. 
This basis of spinors will be denoted $\chi_\pm$ and $\tilde \chi _\pm$, and are defined by
the relations,
\bea
\label{5d4a}
\Gamma ^\hr \, \chi _\pm = \pm \eta \, \chi _\pm 
& \hskip 1in &
\Gamma ^\hr \, \tilde \chi _\pm = \pm \eta \, \tilde \chi _\pm 
\no \\
\Gamma ^{\hu \hv} \, \chi _\pm = - \eta \, \chi _\pm 
&&
\Gamma ^{\hu \hv} \, \tilde \chi _\pm = + \eta \, \tilde \chi _\pm 
\eea
In view of the conventions adopted in Section \ref{sec:A1} and expressed in (\ref{2a21}) for the supersymmetric magnetic brane solution, $\Gamma ^{\hp \hm} \, \Gamma ^{\hu \hv} \, \Gamma ^\hr=I$,
we may read off the corresponding eigenvalues of $\Gamma ^{\hp \hm}$ on these basis spinors, 
\bea
\label{5d4b}
\Gamma ^{\hp \hm} \chi _\pm = \mp \chi _\pm
\hskip 1in 
\Gamma ^{\hp \hm} \tilde \chi _\pm = \pm \tilde \chi _\pm
\eea
From these relations, it readily follows that we have,
\bea
\Gamma ^\hp \chi _-  =  \Gamma ^\hm \chi _+  =  0
\no \\
\Gamma ^\hp \tilde \chi _+  =  \Gamma ^\hm \tilde \chi _-  =  0
\eea
The representation of $\Gamma ^{\hat{\pm}}$ on the basis spinors is then fixed,  
up to an simultaneous sign reversal of both $\Gamma ^{\hat{\pm}}$. We shall make the following
consistent choice,
\bea
\label{5d5}
\Gamma ^\hp \chi _+ = - \sqrt{2} \, \chi _- 
& \hskip 1in & 
\Gamma ^\hp \tilde \chi _- =  \sqrt{2} \, \tilde \chi _+ 
\no \\
\Gamma ^\hm \chi _- = - \sqrt{2} \, \chi _+ 
& \hskip 1in & 
\Gamma ^\hm \tilde \chi _+ =  \sqrt{2} \, \tilde \chi _- 
\eea

\subsubsection{Field decomposition onto the spinor basis (supersymmetric sector)}

Given the decoupling of the fermion equations into eigenspaces of $\Gamma ^{\hu \hv}$, 
the two sectors may be treated independently of one another. The two sectors are not equivalent
to one another, and in fact behave quite differently from a physical point of view. 

\sm

Given that we have set $\mg=1$, it follows from the analysis of the BPS equations that supersymmetry exists in the sector where the eigenvalue of $\Gamma ^\hr$ equals $\gamma =1$ in view of (\ref{gamma})
and (\ref{BPSa}), and where the eigenvalue of $\Gamma ^{\hu \hv}$ equals $-\eta$ in view of (\ref{BPSa}).
In the other three sectors, we have no supersymmetry. Since the supercurrent is the generator
of supersymmetry, its spinor properties must coincide with those of the supersymmetry parameter.
Hence the supercurrent correlator lives in the sector where the fields $\psi _m$ and $\lambda _a$
belong to the eigenspace of $\Gamma ^{\hu \hv}$ with eigenvalue $-\eta$. As a result of their 
$\Gamma$-matrix structure, so do the fields 
$\Gamma ^+\psi _-, \, \Gamma ^- \psi _+, \, \Gamma ^m \psi _m, \, \Gamma ^a \psi _a$  and $\psi _r$.
This in turn means that every field in the supersymmetric sector admits a decomposition onto
the spinors $\chi_\pm$ only, without components along $\tilde \chi_\pm$.

\sm

We use the following notation for the decomposition of the helicity $\pm {3 \over 2}$ fields,
\bea
\label{5d6}
\Gamma ^+ \psi _- & = & \om _- \, \chi _- 
\no \\
\Gamma ^- \psi _+ & = & \om _+ \, \chi _+
\eea
Note that the presence of $\Gamma ^+$ on the left side of the first line sets to zero the expansion coefficient onto the basis spinor $\chi_+$ on the right side, in view of (\ref{5d4}) and (\ref{5d5}),
and similarly sets to zero the coefficient of $\chi_-$ on the second line. The helicity $\pm \half$  gravitino components and the gaugino decompose a follows,
\bea
\label{5d7}
\Gamma ^r \psi _r & = & R_+ \, \chi _+ + R_- \, \chi _-
\no \\
\Gamma ^m \psi _m & = & M_+ \, \chi _+ + M_- \, \chi_-
\no \\
\Gamma ^\alpha \psi _\alpha & = & A_+ \, \chi_+ + A_- \, \chi_-
\no \\
\lambda ^a ~ & = & \ell _+^a  \, \chi_+ + \ell _- ^a \, \chi_-
\eea

\subsection{Supersymmetry transformations for the brane}
\label{sec:B1}

The supersymmetry transformations (\ref{2a8}) acting on the magnetic brane Ansatz reduce as follows.  
Expanding the supersymmetry parameter, $\epsilon$, in eigenspinors of $\Gamma^r$, $\Gamma^{\hat{u}\hat{v}}$, and $\Gamma^{\hat{+}\hat{-}}$, and in Fourier modes for the boundary coordinates,
\[
\epsilon=\epsilon_+(r)e^{ip\cdot x}\chi_+ + \epsilon_-(r)e^{ip\cdot x}\chi_-
\]
The supersymmetry transformations of the components of $\psi_M$ and $\lambda^a$ are given by,
\begin{align}
\label{Be7}
\delta\omega_+ &=  -i\sqrt{2}e^{-W}p_+\epsilon_- & \delta\omega_- &= -i\sqrt{2}e^{-W}p_-\epsilon_+
\no\\
\delta M_+ &=  -i\sqrt{2}e^{-W}p_-\epsilon_- & \delta M_- &= -i\sqrt{2}e^{-W}p_+\epsilon_+ - 2W^{\prime}\epsilon_-
\no\\
\delta R_+ &= \partial_r\epsilon_+ - \frac{1}{2}W^{\prime}\epsilon_+ & \delta R_- &= -\partial_r\epsilon_- - \frac{1}{2}W^{\prime}\epsilon_-\no\\
\delta A_+ &= 0 & \delta A_- &= -2U^{\prime}\epsilon_-
\no\\
\delta\lambda_+^a &= 0 & \delta\lambda_-^a &= if_A^a\phi^{A\prime}\epsilon_-
\end{align}
The field equations, reduced in the magnetic brane Ansatz, are invariant under these transformations.

\subsection{Reduced Fermi equations in the near region}

The near region is defined by the condition $e^{2r}\ll1$. We shall set $\eta =+1$ without loss of generality,
as the case $\eta =-1$ may be recovered by reversing the chiralities. With these assumptions, we solve equations (\ref{5c3}--\ref{5d4}) with the bosonic fields set to (\ref{ADS3}), in the $\chi_\pm $ sector.  
In terms of the components (\ref{5d6}--\ref{5d7}), the helicity $\pm {3 \over 2}$ components satisfy,
\bea
0 & = & \omega_{+}^{\prime}+\frac{3}{2L}\omega_{+}-i\sqrt{2}p_{+}e^{-\frac{r}{L}}(A_{-}+R_{-})
\no \\
0 & = & \omega_{-}^{\prime}+\frac{1}{2L}\omega_{-}+i\sqrt{2}p_{-}e^{-\frac{r}{L}}(A_{+}+R_{+})
\label{5e2}
\eea
The gaugino equations reduce to, 
\bea
\label{5e1}
&&
0 = (\ell _{+}^{a})' 
+ \frac{5}{6L} \ell _{+}^{a}
-i\sqrt{2}p_{-}e^{-\frac{r}{L}} \ell _{-}^{a}-2T_{abc}Bq^{I}X_{I}^{c}e^{-2U}\ell _{+}^{b}
-i\sqrt{\frac{3}{2}}Bq^{I}X_{I}^{a}e^{-2U}A_{+}
\no \\
&&
0 = (\ell _{-}^{a}) ' 
+\frac{7}{6L} \ell _{-}^{a}
+i\sqrt{2}p_{+}e^{-\frac{r}{L}} \ell _{+}^{a}
+2T_{abc}Bq^{I}X_{I}^{c}e^{-2U} \ell _{-}^{b}
+i\sqrt{\frac{3}{2}}Bq^{I}X_{I}^{a}e^{-2U}A_{-} 
\qquad \quad
\eea
Setting the following abbreviation,
\bea
l_{\pm}\equiv 2i\sqrt{\frac{3}{2}} \, q^{I}X_{I}^{a} \, \ell ^{a}_{\pm}
\eea
the equations for the helicity $\pm \half$ components of the gravitino simplify to, 
\bea
\label{5e3}
0 & = & 
-\frac{i}{\sqrt{2}}e^{-\frac{r}{L}}(p_{-} M_{-}
-p_{+}\omega_{-})+\frac{1}{L}(M_{+}+2A_{+})-i\sqrt{2}p_{-}e^{-\frac{r}{L}}A_{-}
\no \\
0 & = & 
\frac{i}{\sqrt{2}}e^{-\frac{r}{L}}(p_{+} M_{+}-p_{-}\omega_{+})+i\sqrt{2}p_{+}e^{-\frac{r}{L}}A_{+}
\no \\
0 & = & 
(M_{+}+2A_{+})^{\prime}
+\frac{3}{2L}(M_{+}+2A_{+})
-i\sqrt{2}p_{-}e^{-\frac{r}{L}}(A_{-}+R_{-})
\no \\
0 & = & 
(M_{-}+2A_{-})^{\prime}+\frac{1}{2L}(M_{-}-2A_{-}-4R_{-})+i\sqrt{2}p_{+}e^{-\frac{r}{L}}(A_{+}+R_{+})
\no \\
0 & = & (2M_{+}+A_{+})^{\prime}+\frac{1}{2L}(6M_{+}-5A_{+})+i\sqrt{2}p_{-}e^{-\frac{r}{L}}(A_{-}-2R_{-})-l_{+}
\no \\
0 & = & (2M_{-}+A_{-})^{\prime}+\frac{1}{2L}(A_{-}-8R_{-}+2M_{-})-i\sqrt{2}p_{+}e^{-\frac{r}{L}}(A_{+}-2R_{+})+l_{-}
\eea
We shall now proceed to further decouple these equations.

\subsubsection{Further decoupling}

Eliminating $M_+$ and its derivative $M_+'$ between the third and fifth lines of (\ref{5e3}) 
also eliminates $R_-$, and we obtain an equation involving only $A_+$ and $l_+$. Proceeding 
analogously for $M_-$ on the fourth and sixth lines also eliminates $R_\pm$, and we find, 
\bea
0 & = & 3A_{+}^{\prime}+\frac{17}{2}A_{+}-3i\sqrt{2}p_{-}e^{-\frac{r}{L}}A_{-}+l_{+}
\nonumber \\
0 & = & 3A_{-}^{\prime}-\frac{5}{2}A_{-}+3i\sqrt{2}p_{+}e^{-\frac{r}{L}}A_{+}-l_{-}
\label{5e4}
\eea
These equations are equivalent to the last two equations in (\ref{5e3}), and give $\l_\pm$ in terms of $A_\pm$. 
The remaining gravitino equations may be further simplified by defining,
\bea
\hat R_\pm & = & R_\pm + A_\pm
\no \\
\hat M_\pm & = & M_\pm + 2 A_\pm
\eea
The third and fourth equations in (\ref{5e2}) are dependent on the first two and (\ref{5e2}).
The remaining independent gravitino equations in the near region are thus given by,
\bea
\label{5e5}
0 & = & \omega_{+}^{\prime}+\frac{3}{2L}\omega_{+}-i\sqrt{2}p_{+}e^{-\frac{r}{L}} \hat R_{-}
\no \\
0 & = & \omega_{-}^{\prime}+\frac{1}{2L}\omega_{-}+i\sqrt{2}p_{-}e^{-\frac{r}{L}} \hat R_{+}
\no \\
0 & = & p_+ \hat M_+ - p_- \omega _+
\no \\
0 & = & 
-\frac{i}{\sqrt{2}} e^{-\frac{r}{L}} (p_{-} \hat M_{-} -p_{+}\omega_{-})+\frac{1}{L} \hat M_{+}
\eea
The equations for $A_\pm$ and $\ell_\pm ^a$ in (\ref{5e1}) and (\ref{5e4}) are manifestly decoupled
from the equations for $\om _\pm, \hat M_\pm$, and $\hat R_\pm$ in (\ref{5e5}). Since for the computation of the 
supercurrent it is only the modes $\om_\pm$ that are of interest, we see that turning on $\om_\pm$ 
requires turning on $\hat M_\pm$ and $\hat R_\pm$, but not $A_\pm$ and $\ell _\pm ^a$. Hence, we may 
consistently set,
\bea
\label{Aell}
A_\pm = \ell_\pm ^a=0
\eea
The only remaining equations are then those of (\ref{5e5}). To analyze them, we first discuss the choice of gauge.

\subsubsection{The choice of Fefferman-Graham gauge for the gravitino}

We have postponed making a choice of gauge for the gravitino field until now. At this point, it becomes clear that there is a natural and useful gauge choice to be made, namely Fefferman-Graham gauge for the gravitino field,
\bea
\psi _r =0
\eea
This gauge choice is natural because $\psi _M$ is a vector-spinor, and the gauge choice for its vector part is analogous to the Fefferman-Graham gauge for a gauge field $A_M$ for which the Fefferman-Graham gauge choice sets $A_r=0$. The gauge choice is also useful because, in view of (\ref{5d7}) it will imply $R_\pm =0$
which along with the results of (\ref{Aell}) implies that also $\hat R_\pm=0$. Therefore, in the near-region, we are left  with the following system of equations,
\begin{align}
0 & = \omega_{+}^{\prime}+\frac{3}{2L}\omega_{+} 
\no \\
0 & = \omega_{-}^{\prime}+\frac{1}{2L}\omega_{-} 
\no \\
0 & = p_+ M_+ - p_- \omega _+ 
\no \\
0 & = -\frac{i}{\sqrt{2}} e^{-\frac{r}{L}} (p_{-}M_{-} -p_{+}\omega_{-})+\frac{1}{L}M_{+}
\label{5e6}
\end{align}
which involve only $\omega _\pm$ and $M_\pm$.

\subsection{Reduced Fermi equations in the far region}

The far region is defined by $p^2 \ll e^{2r}$. For the fluctuations of the metric and the gauge field, we could
solve in the far region simply by dropping all dependence on the momenta $p_\pm$ in the reduced differential equations for the far region. For the fluctuations of the Fermi fields, additional care is needed. While it will indeed be permissible to omit the dependence on $p_\pm$ in the region $1 \ll e^{2r}$, the same will be true in the full overlap region for all reduced differential equations but two, and these will need to be analyzed with additional care.

\subsubsection{The helicity $\pm {3 \over 2}$ equations in the far region}

For the helicity $\pm {3 \over 2}$ equations, the dependence on $p_\pm$ may indeed be neglected in the reduced differential equations throughout the far region $p^2 \ll e^{2r} \ll 1$.  From (\ref{5c3}), it can be seen that by dropping the momentum dependent terms, the modes $\omega_\pm$ decouple from the other components of $\psi_\mu$. Expanding in the basis (\ref{5d4a}), equations (\ref{5c3}) reduce to
\bea
\left ( \partial_ r +U' + W' +\frac{3}{4}Bq^{I}X_{I}e^{-2U}+\frac{3}{2}V_{I}X^{I}\right ) \omega_{+} & = & 0
\no \\
\left ( \partial_r + U' +W' -\frac{3}{4}Bq^{I}X_{I}e^{-2U}-\frac{3}{2}V_{I}X^{I}\right ) \omega_{-} & = & 0
\label{5f1a}
\eea
Since the bosonic fields satisfy the BPS equations (\ref{BPSred}), the dependence on the scalars $X^I$ may be eliminated in favor of $U$ and $W$, and we obtain, 
\bea
\left ( \partial_ r +2 U' + {3 \over 2} W' \right ) \omega_{+} & = & 0
\no \\
\left ( \partial_r + {1 \over 2} W' \right ) \omega_{-} & = & 0
\label{5f1}
\eea
These equations are easily integrated to obtain,
\bea
\omega_{+}(r) & = & -\sqrt{2} \, b_+ U_0^2 \, e^{-\frac{3}{2}W(r)-2U(r)}
\no \\
\omega_{-}(r) & = & -\sqrt{2} \, b_{-} \, e^{-\frac{1}{2}W(r)}
\label{eq:far_sol_m}
\eea
The coefficients are chosen so that $\psi_\pm$ has the form given by (\ref{5a1}) in the $r\rightarrow\infty$ limit, 
\bea
\psi_{\hat\mu}=\psi _{\hat\mu}^{(0)}e^{-\frac{r}{2}}+\cdots+\psi _{\hat\mu}^{(3)}e^{-\frac{7r}{2}}+\cdots
\label{eq:psi_far}
\eea
with,
\[
\psi _{\hat -}^{(0)}=b_{-}\chi_{+} 
\quad\quad\quad
\psi _{\hat +}^{(3)}=b_{+}\chi _{-}
\]
The remaining components of $\psi_{\hat\mu}$ in the expression (\ref{eq:psi_far}), $\psi _{\hat +}^{(0)}$ and $\psi _{\hat -}^{(3)}$, are related to the helicity $\pm\frac{1}{2}$ components $M_\pm$.

\subsubsection{Helicity $\pm \half$ equations in the far region}

In terms of the asymptotic form of $M_\pm$, given in Appendix \ref{sec:C}, the fields $\psi_\pm$ take the form,
\bea
\psi_{\hat +} & = & -\frac{1}{\sqrt{2}}M_{-}^{(0)}
\chi_{+}e^{-\frac{r}{2}} + \cdots+b_{+}\chi_{-}e^{-\frac{7r}{2}} + \cdots
\no
\\
\psi_{\hat -} & = & b_{-}\chi_{+}e^{-\frac{r}{2}} + \cdots-\frac{1}{\sqrt{2}} 
M_{+}^{(0)} \chi _{-}e^{-\frac{7r}{2}} + \cdots\label{eq:psi_far2}
\eea
It remains to enforce the reduced helicity $\pm \half$ equations in the far region and establish the required remaining relation between $\omega _\pm$ and $M_\pm$. These equations decouple between the $+ \half $ helicity components $M_+, A_+, \ell ^a _+$, and $R_+$ on the one hand, and the  $- \half $ helicity components $M_-, A_-, \ell ^a _-$, and $R_-$ on the other hand. For both helicities we continue to use Fefferman-Graham gauge so that $R_\pm=0$, and we express the components $M_\pm$ is terms of $\hat M_\pm$ and $A_\pm$, as we had already done for the near region.

\subsubsection{The helicity $- {1 \over 2}$ equations in the far region}

Since we are only interested in the helicity $\pm\frac{3}{2}$ sources, we set the sources of all the helicity $\pm\frac{1}{2}$ fields to zero in the $AdS_5$ region. We begin analyzing the fields in the $\chi_-$ sector. In this sector, we can consistently neglect all terms involving the momenta $p_\pm$ throughout the full far region.  The resulting reduced equations are as follows, 
\bea
\label{5q1}
0 & = & \hat M_{-}^{\prime} + \frac{1}{2} W^{\prime} \hat M_{-}
+2 (U^{\prime}-W^{\prime})A_{-}
+2if_{A}^{a}\phi^{A\prime}\ell_{-}^{a}
\no\\
0 & = & A_{-}^{\prime}
+ { 1 \over 6} \left ( 8 U' -5 W' \right ) A_-  
+{2 \over 3} if_{A}^{a}\phi^{A\prime}\ell_{-}^{a}
-{2 \over 3} i\sqrt{\frac{3}{2}}Bq^{I}X_{I}^{a}e^{-2U}\ell_{-}^{a}
\no \\
0 & = & \ell_{-}^{a\prime}+\frac{1}{6}(7W^{\prime}+8U^{\prime})\ell_{-}^{a}+\sqrt{\frac{2}{3}}T_{abc}\left(2Bq^{I}X_{I}^{c}e^{-2U}-\sqrt{\frac{2}{3}}f_{A}^{c}\phi^{A\prime}\right)\ell_{-}^{b}
\no \\ &&
+i\sqrt{\frac{3}{2}}Bq^{I}X_{I}^{a}e^{-2U}A_{-}
\eea
Recall that the constraint equation, which corresponds to the $M=r$ component in the first line of (\ref{5d1}), is automatically satisfied on the supersymmetric brane solution. The source terms for these fields behave as follows in the $AdS_5$ region,  $\hat M_- ^{(0)} \sim e^{-r/2}$, $A_- ^{(0)} \sim e^{-r/2}$ and $(\ell ^{(0)}) ^a _- \sim e^{-3r/2}$.
Each one of these source terms must vanish in the solution we seek, so that the actual behavior of the fields must be suppressed at least by one power of $e^{-r}$. The actual suppression power is by $e^{-2r}$ since the bosonic field coefficients in the above equations all have an expansion in powers of $e^{-2r}$. It is now easy to see that the iterative expansion of $A_-$ and $\ell^a _-$ in the last two equations of (\ref{5q1})  implies that these fields must then vanish identically. The remaining equation for $\hat M_-$ then has a solution which is only the source term, and thus must vanish as well. In summary, we must have,
\bea
R_-=M_- = A_- = \ell ^a _-=0
\eea
Thus, in the far region, only the helicity $\pm { 3 \over 2}$ fields $\omega _\pm$ as well as the 
helicity $\pm { 1 \over 2}$ with subscript plus may be non-zero.

\subsubsection{The helicity $+ {1 \over 2}$ equations in the far region}

In the far region, all the reduced helicity +1/2 equations, except for the constraint equation in (\ref{5d1}), admit a smooth limit as $p^2 \ll 1$, and the corresponding limit may be taken as we did for the stress tensor and current
correlators, as well as for the helicity $\pm {3 \over 2}$ and $- \half$ components. The constraint of (\ref{5d1}) does not admit such a smooth limit and its analysis requires more care and will be handled separately.  

\sm

Setting the momenta $p_\pm$ equal to zero in equations (\ref{5d2}-\ref{5d4}) gives, 
\bea
\label{5s1}
0 & = & \left ( \partial_{r}+2U^{\prime}+\frac{3}{2}W^{\prime} \right ) \hat M_+
\no \\
0 & = & A_+' + { 1 \over 6} ( 17 W' + 4 U') A_+ + { 2 \over 3} (U'-W') \hat M_+ 
+ i\sqrt{\frac{2}{3}}Be^{-2U}q^{I}X_{I}^{a}\ell_+^a
\no \\
0 & = & \ell_{+}^{a\prime}+\frac{1}{6}(5W^{\prime}+4U^{\prime})\ell_{+}^{a}-\sqrt{\frac{2}{3}}T_{abc}\left(2Bq^{I}X_{I}^{c}e^{-2U}+\sqrt{\frac{2}{3}}f_{A}^{c}\phi^{A\prime}\right)\ell_{+}^{b}
\no \\
 &  & -if_{A}^{a}\phi^{A\prime}\hat M_{+}
+ if_{A}^{a}\phi^{A\prime} A_{+}
 -i\sqrt{\frac{3}{2}}Bq^{I}X_{I}^{a}e^{-2U}A_{+}
\eea
The remaining equation is for the constraint, which results from the component $M=r$. Omitting the dependence on $p_\pm$ in this equation is consistent in the region $1 \ll e^{2r}$, but not in the full far region. Thus, we shall keep all $p_\pm$-dependence here,   and obtain,
\bea
0 & = & (W'+2U') \hat M_+ - 3 U' A_+ - i f_A ^a \p_r \phi^A  \ell _+ ^a 
+ { i \over \sqrt{2}} e^{-W} p_+ \omega _-
\no \\
0 & = & p_+ (\hat M_+ -A_+) - p_- \omega _+ 
\eea
In the overlap region, where $U' =\p _r \phi^A=0$, and $W'=1/L$, we recover precisely the near-region equations
(\ref{5e5}) with $\hat M_-=0$, as should be expected.

\sm

The consistency of setting all the sources terms for the helicity $\pm \half$ fields to zero is now easily assured. First of all, the differential equation in (\ref{5s1}) for $\hat M_+$ guarantees that it contains no source terms. Next, the differential equation in (\ref{5s1}) for $A_+$ contains a term in $\hat M_+$ which behaves as $e^{-11r/2}$ and a term in $\ell^a _+$ which behaves as $e^{-7r/2}$ for the sources term of $\ell_+^a$ and $e^{-9r/2}$ for its vev term. In all these cases, the source term for $A_+$ must vanish by the second equation.

\sm

In the next section, we shall match the solutions for $\omega_\pm$ obtained in the far region with those obtained in the near region to obtain a solution valid in an overlap region defined by $p^2\ll e^{2r} \ll 1$. From the discussion above, we can consistency set $M_-=0$ throughout. The solutions to the near region equations (\ref{5e6}) are therefore 
\begin{align}
\omega_+(r) &= C_+e^{-\frac{3r}{2L}} & M_+(r) &= \frac{p_-}{p_+}C_+e^{-\frac{3r}{2L}}\no\\
\omega_-(r) &= \frac{i\sqrt{2}}{L}\frac{p_-}{p_+^2}C_+e^{-\frac{r}{2L}} & M_-(r)&=0
\label{eq:near_soln2}
\end{align}
where $C_+$ is an integration constant which remains to be determined.

\subsection{Matching and IR correlators}

To obtain a full solution in the overlap region, we match the solutions in the near and far region in the limit $p^2e^{-2r}\ll 1$. Matching equations (\ref{eq:far_sol_m}) in the $r\rightarrow -\infty$ limit to the left-hand column of (\ref{eq:near_soln2}), we get,
\begin{align}
C_+ &= -\frac{\sqrt{2}}{B}U_0^2b_+ & \frac{i\sqrt{2}}{L}\frac{p_-}{p_+^2}C_+ &=-\sqrt{2}b_-
\end{align}
From this, we can solve for $b_{+}$ in terms of $b_{-}$. Writing the
result in terms of the source and expectation value parts of the spinors, $\psi ^{(0)}$ and $\psi^{(3)}$,
we get,
\bea
\label{psi3psi0}
\psi_{+}^{(3)} & = & -\frac{iLB}{\sqrt{2}U_0^2}\frac{p_{+}^{2}}{p_{-}} \, 
\chi_-\chi_+^t\psi^{+(0)}
\eea
\sm
From general arguments based on super-conformal symmetry of the boundary theory, the gamma trace of the spinor $\psi_\mu^{(3)}$ should be composed of only local terms. This can be verified explicitly from the asymptotic expansion of $\psi_\mu$ in Appendix \ref{sec:C}. Therefore, up to these local terms, we can set $\Gamma^{\mu}\psi _{\mu}^{(3)}=0$. Then (\ref{eq:S}) can be written as
\[
8\pi G_{5}S_{\mu} = \psi _{\mu}^{(3)}+\rm {local}
\]
Using this expression and redefining $\tilde{\cS}_{\mu}=U_0^2 V_{2}\cS_{\mu}$,
where $V_{2}$ is the volume of the $T^{2}$ factor in the
$AdS_{3}\times T^{2}$ geometry, equation (\ref{psi3psi0}) can be rewritten as
\[
\tilde{S}_{+} = -\frac{ic}{12\sqrt{2}\pi}\frac{p_{+}^{2}}{p_{-}} \, \chi_-\chi_+^t\psi ^{+(0)}
\]
where the central charge is given by (\ref{eq:c}). Comparing this expression to (\ref{5a3}), the Euclidean momentum space two-point correlator of the supercurrent can be extracted. Restoring the $SU(2)$ index and including the result of analyzing the $i=-$ sector, we obtain
\begin{align}
\left\langle \mathcal{S}_{+,+}(p)\bar{\mathcal{S}}_{+,-}(-p)\right\rangle &=
\frac{ic}{6\sqrt{2}\pi}\frac{p_{+}^{2}}{p_{-}} \, \chi_{-}\chi_{+}^{t}
\no\\
\left\langle \mathcal{S}_{+,-}(p)\bar{\mathcal{S}}_{+,+}(-p)\right\rangle &=
-\frac{ic}{6\sqrt{2}\pi}\frac{p_{+}^{2}}{p_{-}} \, \tilde\chi_{+}\tilde\chi_{-}^{t}
\label{eq:SS_mom_p}
\end{align}
where the first index is the space-time index of the vector-spinor $\cS_\mu$ and the second is the $SU(2)$ index. All other correlators vanish. Note that these two correlators are related via conjugation with the charge conjugation matrix $C$,
$$
\left(C\left\langle \mathcal{S}_{+,+}(p)\bar{\mathcal{S}}_{+,-}(-p)\right\rangle\right)^t 
= -C\left\langle \mathcal{S}_{+,-}(p)\bar{\mathcal{S}}_{+,+}(-p)\right\rangle
$$
\sm
A similar analysis with $\eta=-1$ gives the same equations in the near and far region, 
but with the roles of $\chi_+$ and $\chi_-$ switched. To avoid repetition, we simply 
write down the final result for the two-point correlator, which is,
\begin{align}
\left\langle \mathcal{S}_{-,+}(p)\bar{\mathcal{S}}_{-,-}(-p)\right\rangle &=
\frac{ic}{6\sqrt{2}\pi}\frac{p_{-}^{2}}{p_{+}} \, \chi_{+}\chi_{-}^{t}
\no\\
\left\langle \mathcal{S}_{-,-}(p)\bar{\mathcal{S}}_{-,+}(-p)\right\rangle &=
-\frac{ic}{6\sqrt{2}\pi}\frac{p_{-}^{2}}{p_{+}} \, \tilde\chi_{-}\tilde\chi_{+}^{t}
\label{eq:SS_mom_m}
\end{align}
Fourier transforming the non-zero momentum space correlators to position space gives us the expected two-point correlators for the supercurrent,
\begin{align}
\left\langle \mathcal{S}_{+,+}(x)\bar{\mathcal{S}}_{+,-}(0)\right\rangle &=
\frac{c}{6\sqrt{2}\pi^2}\frac{1}{(x^+)^3} \, \chi_{+} \chi_{-}^{t} & \eta > 0
\label{eq:SS_pos}
\\
\left\langle \mathcal{S}_{-,+}(x)\bar{\mathcal{S}}_{-,-}(0)\right\rangle &=
\frac{c}{6\sqrt{2}\pi^2}\frac{1}{(x^-)^3} \, \tilde\chi_{+}\tilde\chi_{-}^{t} & \eta < 0
\label{eq:SS_neg}
\end{align}
\sm
We see from this result that the overall sign of the charges, given by $\eta$, determines whether the left- or right-movers have a non-vanishing correlators, similar to what we saw for the gauge current correlators. We will see in the next section how this fits with the stress tensor and gauge current correlators computed in the previous sections.

\section{Emergent Super-Virasoro Symmetry}
\label{sec:6}
\setcounter{equation}{0}

The presence of an asymptotic $AdS_3$ space-time in the near region signals the appearance of a Brown-Henneaux Virasoro algebra \cite{Brown:1986nw}, which carries over to a Virasoro symmetry in the IR limit of the dual field theory. For the magnetic brane solution without supersymmetry, the presence of a Virasoro algebra was derived directly from the structure of the stress tensor two-point correlators in \cite{D'Hoker:2010hr}. There is also an additional unitary $U(1)$ current algebra is generated by the Maxwell field on an asymptotically $AdS_3$ space-time, thereby producing an additional Kac-Moody symmetry in the IR limit of the dual field theory.

\sm

For the supersymmetric magnetic brane solution, discussed in the present paper, there again appears an asymptotic $AdS_3$ region, producing again a Virasoro algebra, but now with three extra $U(1)$ current algebras, as well as superconformal generators. This extended set of generators is responsible for extending the Virasoro algebra into an $\cN=2$ super-Virasoro algebra, as we shall argue below. The presence of  an asymptotic  $\cN=1$ super-Virasoro symmetry algebra  near the boundary of $AdS_3$ has been studied in the context of three-dimensional Chern-Simons supergravity in a number of earlier papers, including \cite{Banados:1998pi,Bautier:1999ds,Hyakutake:2012uv,Hyakutake:2015qua}.

\subsection{Matching correlators with superconformal central terms}

In this section, we shall assemble all the results of the calculations of two-point functions for the stress tensor $\cT$, the $U(1)^3$ current $\cJ^I$, and the supercurrents $\cS$ and $\bar \cS$, to support the emergence of an extended $\cN=2$ super-Virasoro algebra. To begin, we recall the structure of the low energy limit of the correlators of these operators in Table~1. Purely local contributions will  be omitted throughout. In Table 1, we collect the non-vanishing two-point correlators calculated in this paper, as a function of the sign of the charges $q^I$. Since the magnetic brane solution is supersymmetric, we expect the two-point correlator to reflect this supersymmetry. That is, we should find an equal number of bosonic and fermionic operators in the supersymmetric sector.  For both signs of $\eta$, we indeed find  this to be the case. The tilde on some of the indices denote a basis which diagonalizes the rank~2 projection matrices that are present in some of the gauge current correlators.

\begin{table}[htb]
	\centering
	\begin{tabular}{|c|c|c|c|}
		\hline\hline
	& helicity 	 & Left-movers & Right-movers\\
		\hline \hline
		& 1            & $\langle \tilde{\cJ}_+^3\tilde{\cJ}_+^3\rangle$ & $\langle \tilde{\cJ}_-^{\tilde{1}}\tilde{\cJ}_-^{\tilde{1}}\rangle$,
		           $\langle \tilde{\cJ}_-^{\tilde{2}}\tilde{\cJ}_-^{\tilde{2}}\rangle$ \\
		$\eta > 0$ &  ${ 3 \over 2} $ & $\langle \cS_+\bar{\cS}_+\rangle$     & \\
		           & 2 & $\langle \tilde{\cT}_{++}\tilde{\cT}_{++}\rangle$ & $\langle \tilde{\cT}_{--}\tilde{\cT}_{--}\rangle$\\
		\hline
		           & 1 & $\langle \tilde{\cJ}_+^{\tilde{2}}\tilde{\cJ}_+^{\tilde{2}}\rangle$, $\langle \tilde{\cJ}_+^{\tilde{3}}\tilde{\cJ}_+^{\tilde{3}}\rangle$ & $\langle \tilde{\cJ}_-^1\tilde{\cJ}_-^1\rangle$\\
		$\eta < 0$ &  ${ 3 \over 2}$ & & $\langle \cS_-\bar{\cS}_-\rangle$\\
		           & 2 & $\langle \tilde{\cT}_{++}\tilde{\cT}_{++}\rangle$ & $\langle \tilde{\cT}_{--}\tilde{\cT}_{--}\rangle$\\
		\hline
	\end{tabular}
		\caption{Non-zero correlators in the presence of the supersymmetric magnetic brane}
	\label{table:mult}
\end{table}

Next, we recall the part of the structure of the $\cN=2$ super-Virasoro algebra to which the two-point functions give access. These algebras enter chirally, and we shall concentrate here on the $+$ chirality part,
as is appropriate for the case $\eta >0$. The super Virasoro algebra is generated by the chiral stress tensor,  $\mT_{++}(z^+)$, a chiral $U(1)$-current $\mJ_+(z^+)$, and the chiral supercurrent components $\mS_+ (z^+)$ and $\bar \mS_+(z^+)$. The singular parts of their OPE relations are given as follows (see for example \cite{Polchinski:2005book}),
\bea
\label{6a1}
\mT_{++} (z^+) \, \mT_{++} (w^+) 
& \sim & 
{ {c \over 2} \over (z^+-w^+)^4} + { 2 \mT_{++}(w^+) \over (z^+-w^+)^2}
+ { \p_+ \mT_{++} (w^+) \over z^+-w^+} 
\no \\
\mS _+  (z^+) \, \bar \mS_+  (w^+) 
& \sim & 
{ {2c \over 3}  \over (z^+-w^+)^3} + { 2\mJ_+ (w^+) \over (z^+-w^+)^2}
+ {  2\mT_{++}(w^+) + \p_+ \mJ_+ (z^+) \over z^+-w^+}
\no \\
\mJ_+  (z^+) \, \mJ_+  (w^+) 
& \sim & 
 { {c \over 3} \over (z^+-w^+)^2} 
\eea
We have not included the OPEs  between distinct operators, as these are not accessible via the two-point functions, but require genuine three-point correlators. 
The terms beyond those proportional to the identity in (\ref{6a1}) are not accessible by our two-point function calculations either, but have been included here for the sake of completeness.

\subsubsection{Normalization of the stress tensor}

We have expressed the OPE relations of (\ref{6a1}) in terms of the customary Minkowski coordinates $z^\pm$  used to write down the super-Virasoro algebra, for example \cite{Polchinski:2005book}, namely $z^\pm = \pm x^0+x^1$ and $w^\pm = \pm y^0 + y^1$, while the normalization of coordinates used in the preceding sections of this paper was rather $x^\pm = (\pm x^0+x^1)/\sqrt{2}$ and $y^\pm = (\pm y^0 + y^1)/\sqrt{2}$. This change of variables amounts to a constant rescaling, which is conformal, and leaves the OPE for the stress tensor unchanged. As a  result, upon comparing the two-point function of $\cT_{++}$ in (\ref{eq:TT_pos}) with the term on the right side of the first line in (\ref{6a1}), and absorbing a standard factor of $(2\pi)^{-2}$ in the definition of the two-point correlator, we are led to set, 
\bea
\cT_{++} (x_+) (dx^+)^2= { 1 \over 2 \pi} \, \mT_{++}(z^+) (dz^+)^2
\eea
and we find perfect agreement between the predictions of our stress tensor correlators and the structure of the $\cN=2$ superconformal algebra.

\subsubsection{Normalization of the supercurrents}

The comparison for the supercurrent is slightly more tricky, for the following reason. In the current algebra of (\ref{6a1}), the operators $\mT_{++}$, $\mS_+$, $\bar \mS_+$, and $\mJ_+$ are viewed as conformal fields of respective weights $(2,0)$, $({3 \over 2}, 0)$, and $(1,0)$. This is also true for the operators $\tilde \cT_{++}$ and $\tilde \cJ_+^I$ with only Einstein indices arising from supergravity and holography. But the operators $\cS_+$ and $\bar \cS_+$ emerge from supergravity as a component of an Einstein vector tensored with a Lorentz spinor. As a result, the transformation law for conformal rescaling involves only the vector index, and we must identify the operators accordingly,
\bea
\cS_+ (x^+) dx^+ & = & { 1 \over 2 \pi} \, \mS _+ (z^+) dz^+
\no \\
\bar \cS_+ (x^+) dx^+ & = & { 1 \over 2 \pi} \, \bar \mS _+ (z^+) dz^+
\eea
With this relation, we find again perfect agreement between the result of the holographic calculations 
in (\ref{eq:SS_pos}) and the second line in (\ref{6a1}).

\subsubsection{Normalization of the currents}

Finally, the normalization of the $U(1)$ current that enters into the $\cN=2$ superconformal algebra poses a new challenge, which we have not resolved in the present paper, and leave for future work. The difficulty arises from the mixing of the three $U(1)$ gauge fields due to the Chern-Simons interaction in the presence of the supersymmetric magnetic brane solution. This mixing is in effect in both the near and far regions, as well as in the axial anomaly equation (\ref{4z1}) for the $U(1)^3$ currents. Disentangling which of the three $U(1)$ currents plays the role of $\mJ_+$ in (\ref{6a1}) appears to require the normalization of the $\mJ_+$ term in the OPE of $\mS_+$ and $\bar \mS_+$, which requires a three-point correlator and is not at present available.
This ambiguity of normalization is likely related to the fact that there is no natural prescription for identifying the graviphoton, $A_\mu$, that belongs to the supergravity multiplet in terms of the gauge fields, $A^I_\mu$, as was explained in \cite{Gunaydin:1984ak}.

\subsection{Virasoro generators in $AdS_{3}$ as physical modes in $AdS_{5}$}

An interesting issue addressed in \cite{D'Hoker:2010hr} is the compatibility of the 
infinite-dimensional Virasoro asymptotic  symmetry of the $AdS_{3} \times T^2$ geometry
with the finite-dimensional asymptotic symmetry $SO(2,4)$ of the $AdS_{5}$ geometry. 
For the non-supersymmetric magnetic brane solution, the pure coordinate transformations 
on the $AdS_{3}$ geometry were shown to become physical modes in the $AdS_{5}$ region. 
These physical modes cannot be undone by a coordinate transformation on $AdS_5$.
In this section, we will show that this is the case for the supersymmetric magnetic brane solution as well.

\sm

Starting with the metric (\ref{3c1}) of the $AdS_{3}\times\mathbb{R}^{2}$ geometry,
we consider an infinitesimal Brown-Henneaux coordinate transformation in which we
reparametrize $x^{+}$ by a transformation of the following form,
\bea
r & \rightarrow & r+\xi^{r} \, e^{ip_{+}x^{+}}
\no \\
x^{+} & \rightarrow & x^{+}+\xi^{+} \, e^{ip_{+}x^{+}}
\no \\
x^{-} & \rightarrow & x^{-}+\xi^{-} (r) \, e^{ip_{+}x^{+}}
\eea
where $\xi^{+}$ and $\xi^r$ are constant and related as follows,
\bea
\xi^{r} = -\frac{iL}{2}p_{+}\xi^{+}
\hskip 1in 
\xi^{-}\left(r\right) = \frac{L^{2}}{4} \, e^{-\frac{2r}{L}} \, p_{+}^{2}\xi^{+}
\eea
To first order in $\xi^{+}$ the metric takes the form,
\bea
ds^{2}=dr^{2}+2e^{\frac{2r}{L}}dx^{+}dx^{-}
+\frac{iL^{2}}{2}p_{+}^{3}\xi^{+} \, e^{ip_{+}x^{+}}\left(dx^{+}\right)^{2} + B \delta _{ij} dx^i dx^j
\label{eq:near_met_BH}
\eea
Comparing this to the perturbed metric in (\ref{3b1}) and
(\ref{eq:h_far_near}), we can read off, 
\bea
t_{++}=\frac{iL^{2}}{2}p_{+}^{3}\xi^{+}=\frac{2 U_0^2 L}{B} \, \delta g_{++}^{\left(4\right)}
\label{eq:t_g}
\eea
with $s_{++}=s_{--}=t_{--}=t_{+-}=p_{-}=0$. The second equality
in (\ref{eq:t_g}) gives the full asymptotically $AdS_{5}$ solution
with the near horizon behavior (\ref{eq:near_met_BH}). Specifically,
this component is given by
\bea
h_{++}(r)=-4U_0^2 h^{2}(r) \, \delta g_{++}^{(4)}
\eea
with $h^{2}(r)$ defined in (\ref{eq:h2}). This perturbation mode cannot be undone 
by a coordinate transformation on $AdS_5$, and is therefore physical. 

\sm

We can also show that the corresponding $AdS_{5}$ stress tensor transforms under 
the Brown-Henneaux coordinate reparametrization with a Schwarzian derivative
properly normalized for central charge $c$. This is expected since
(\ref{eq:t_g}) shows that the $AdS_{3}$ stress tensor is proportional
to the $AdS_{5}$ stress tensor and therefore transfers its Schwarzian
derivative transformation law. In particular, the $AdS_{5}$ stress
tensor is given by,
\bea
\tilde{T}_{++}=\frac{c}{24\pi}\partial_{+}^{3}\xi^{+}
\eea
where the right side is the Schwarzian derivative with the correct normalization for central charge $c$.
Interchanging the $+$ and $-$ indices, a similar computation gives the expression  
$\tilde{T}_{--}=c \, \partial_{-}^{3}\xi^{-}/ (24 \pi)$.

\subsection{Supercurrent generators in $AdS_{3}$ as physical modes in $AdS_{5}$}

Under a local supersymmetry transformation the supercurrent defined on the $AdS_5$ boundary transforms with a term analogous to the Schwarzian derivative for the stress tensor. We expect this to be the case for the same reason as the stress tensor, that is, the $AdS_3$ supercurrent was shown to be proportional to the $AdS_5$ supercurrent and so should transform with a similar term. This can be seen by applying the linear response formula (\ref{5a3}) to a local supersymmetry transformation. For simplicity, we use the momentum space version of (\ref{5a3}) given by,
\begin{align}
\tilde \cS_{+,+} &= -\frac{1}{2}\langle\cS_{+,+}(p)\bar{\cS}_{+,-}(-p)\rangle\delta\psi_{-,+}(p)\no\\
\tilde \cS_{+,-} &= \frac{1}{2}\langle\cS_{+,-}(p)\bar{\cS}_{+,+}(-p)\rangle\delta\psi_{+,-}(p)
\end{align}
Using the result for the two-point correlator of the supercurrent when $\eta>0$ and  a supersymmetry transformation with local supersymmetry parameter $\epsilon(r,x)$, we obtain,
\bea
\tilde \cS _{+i} = -\frac{c}{12\sqrt{2}\pi}\chi_-\chi_+^t\p_+^2\epsilon_i
\eea

\subsection{Composition of supersymmetry transformations}

The computation and matching of the two-point correlators by itself does not suffice to guarantee the existence of an $\cN=2$ superconformal algebra. In particular, we may ask whether the $U(1)$ current algebra that appears in the same sector as the supercurrents genuinely is a part of the superconformal algebra, or whether it is simply an additional current algebra as we had already in the case of the non-supersymmetric brane. In this last subsection, we shall provide additional arguments that demonstrate that indeed this current algebra is part of the superconformal algebra. 

\sm

The arguments are derived from the composition of two supersymmetry transformations. The action of a general 10-dimensional supersymmetry transformation $\ep _i$ on the frame $e_M {}^{\hat M}$, gauge fields $A_M ^I$, and gravitino $\psi _M ^i$ are as follows,
\bea
\delta e_M {} ^{\hat M}  & = & \half \bar \ep ^i \Gamma ^{\hat M} \psi _{Mi}
\no \\
\delta A_M ^I & = & i X^I \bar \psi _M ^i \ep _i + \cO(\lambda)
\no \\
\delta \psi _M^i & = & \cD_M \ep ^i + { i \over 8} X_I F^I _{NP} \left ( \Gamma _M{}^{NP} - 4 \delta _M {}^N \Gamma ^P \right ) \ep ^i + \half \mg V_I X^I \Gamma _M  \delta ^{ij} \ep _j
\eea
up to higher order terms in the fermi fields. The 2-dimensional conformal supersymmetry transformations which are asymptotic symmetries of the $AdS_3$ near region form a subset of  these supersymmetry transformations. The action on the bosonic fields of the composition of two supersymmetries may be easily read off from the above transformation rules, ignoring contributions involving the dilatino. Clearly, the composition of two supersymmetries produces a variation in the metric, which accounts qualitatively for the 
$\mT_{++}$ term in the OPE of two supercurrents on the second line in (\ref{6a1}), and  a variation in the gauge field proportional to $X^I$ which accounts for the $\mJ_+$ term in the second line in (\ref{6a1}). This provides confirmation that the $U(1)$ current algebra generated by $\cJ_+^I$ indeed is part of the superconformal algebra. 

\sm

The above arguments are clearly rather qualitative, and we shall leave a quantitative investigation of these issues for future work.

\section{Discussion}
\setcounter{equation}{0}
\label{sec:7}

There are several avenues along which the study of this paper could be extended. One immediate direction for future work, already mentioned in the previous section, is  to obtain a quantitative derivation of the superconformal algebra as an asymptotic symmetry algebra in the near region. 

\sm

Another direction is along the following lines. Magnetic branes may be dressed with an electric charge density and placed at finite temperature \cite{D'Hoker:2009bc,Donos:2011qt}.  For the non-supersymmetric brane, using a blend of analytical and numerical studies,  we were led to the discovery of a quantum critical point across a non-zero value of the magnetic  field \cite{Jensen:2010vd,D'Hoker:2010rz}.  Physics in the critical region may be explored completely by analytical methods alone \cite{D'Hoker:2010ij}. 

\sm

Finite temperature or chemical potential will of course break whatever supersymmetry existed at zero temperature.
Perhaps the most interesting question that can be exported from the  non-supersymmetric magnetic brane to its supersymmetric counterpart studied in this paper is the fate of the quantum phase transition, which was identified in \cite{Jensen:2010vd,D'Hoker:2010rz} for the non-supersymmetric magnetic brane.
Having shown here that the asymptotic symmetries of the supersymmetric and non-supersymmetric 
branes are different, we should expect the universality classes to which the corresponding dual 
CFTs belong to be different as well.  Therefore,  critical exponents and scaling functions 
should be different, and for the supersymmetric magnetic brane depend on the extra 
free parameter specifying the embedding of the magnetic field into $U(1)^3$. 

\sm

Another avenue of interest is the identification of a twisted super-Virasoro structure when a background electric charge density is turned out, extending the analysis of \cite{D'Hoker:2011xw} for the non-supersymmetric brane.

\section*{Acknowledgments}

\medskip

We are happy to thank Per Kraus for very helpful discussions throughout. 
BP gratefully acknowledges receipt of the Philip and Aida Siff Educational Foundation Scholarship.

\appendix

\section{Review of gauged five-dimensional supergravity}
\setcounter{equation}{0}
\label{sec:A}

Our starting point is five-dimensional $\cN=2$ supergravity with  $\mN$ Maxwell supermultiplets  
in which a $U(1)$ subgroup of the $SU(2)$  automorphism group of the supersymmetry algebra 
has been gauged \cite{Gunaydin:1983bi,Gunaydin:1984ak}. Einstein indices are denoted 
$M,N =0,1,2,3,4$, while space-time frame indices are denoted by 
$\hat M, \hat N=\hat{0},\hat{1},\hat{2},\hat{3},\hat{4}$. The orthonormal frame metric is given 
by $\eta _{\hat M \hat N} ={\rm diag}(-++++)_{\hat M \hat N}$, while the totally antisymmetric 
symbol in five-dimensional space-time will be denoted by $\ep^{MNPQR}$ and normalized to 
$\ep^{01234}= \ep^{\hat{0} \hat{1} \hat{2} \hat{3} \hat{4}} =1$.

\sm

The fields of the theory are the space-time metric $g_{MN}$, or equivalently the 
orthonormal frame $\me_M {}^{\hat M}$; $\mN+1$ Maxwell fields $A_M ^I$ with $I=1,\cdots, \mN+1$ 
(one Maxwell field arising from the supergravity multiplet); $\mN$ scalars $\phi^A$ with 
$A=1,\cdots, \mN$; one gravitino field $\psi _{\mu i} $ which is a doublet under $SU(2)$ 
labelled by $i=1,2$; and $\mN$ gaugino fields $\lambda _i ^a$ with $a=1,\cdots, \mN$,
which are doublets under $SU(2)$.

\subsection{Spinors}
\label{sec:A1}

We denote by $\Gamma ^M$ a basis of the Clifford algebra (written in Einstein indices),
\bea
\{ \Gamma ^M , \Gamma ^N \}=2 g^{MN} \, I
\eea
by $I$ the identity matrix, by $\Gamma ^{M_1 M_2 \cdots M_r}$ the rank $r$ 
antisymmetric product of $\Gamma$-matrices,  and by $C$ the charge conjugation 
matrix defined by $(\Gamma ^M)^t = C \Gamma ^M C^{-1}$ and $C^t=-C$.   
The charge conjugation relation on all Clifford generators is given by, 
\bea
(C \, \Gamma ^{M_1 M_2 \cdots M_r})^t =  t_r \, C \, \Gamma ^{M_1 M_2 \cdots M_r}
\eea
with $-t_0=-t_1=t_2=t_3=-t_4=-t_5=1$. Dirac matrices with frame indices are related as usual by   
$\Gamma ^{\hat M} = \Gamma ^M \me_M {}^{\hat M}$. Since the dimension of space-time is odd, 
we have the  relation, 
\bea
\label{A2}
\Gamma ^{\hat{M}_1 \hat{M}_2 \hat{M}_3 \hat{M}_4 \hat{M}_5} 
= \pm i \, \ep ^{\hat{M}_1 \hat{M}_2 \hat{M}_3 \hat{M}_4 \hat{M}_5 } \, I
\eea
The sign choice distinguishes between the two inequivalent irreducible  representations of the 
Clifford algebra, related by $\Gamma ^{\hat{M}} \to - \Gamma ^{\hat{M}}$, and which give rise to equivalent representations of the Lorentz algebra. Throughout, we shall choose the $+$ sign in (\ref{A2}).

\sm

All spinors are doublets under $SU(2)$, as is indicated by the label $i$ on $\lambda^a _i$ and $\psi _{Mi}$. They are subject to the symplectic-Majorana condition on a Dirac spinor $\chi_i$ (which may be either 
the fields $\psi _{M i}$,  $\lambda ^a_i$, or the supersymmetry generator $\epsilon_i$) which takes the form, 
\bea
\label{A3}
\bar \chi ^i \equiv (\chi_i )^\dagger \Gamma _{\hat 0} = (\chi ^i)^t C
\eea
The $SU(2)$-indices are raised and lowered by,
\bea
\label{A4} 
\chi^i = \ep^{ij} \, \chi_j \hskip 0.8 in \chi_j = \chi^i \, \ep _{ij} \hskip 0.8in \ep ^{12}=\ep_{12}=1
\eea
It will be convenient to introduce the following complex combinations of the real indices $i$,
\bea
\label{A5}
\chi _\pm =  { 1 \over \sqrt{2}} ( \chi _1 \pm i \chi _2  ) 
\hskip 1in
\chi ^\pm =  { 1 \over \sqrt{2}} ( \chi ^1 \mp i \chi ^2  ) 
\eea
In terms of these indices, the relations of (\ref{A4}) take the form, 
\bea
\label{A5_2}
\chi ^+ = i \chi_- \hskip 0.8in \chi^- = - i \chi_+ \hskip 0.8in \ep ^{+-} = - \ep _{+-} = i
\eea
The symplectic-Majorana condition of (\ref{A3}) then becomes, 
\bea
 ( \chi _+  )^\dagger \Gamma _{\hat{0}} = i  ( \chi _-  )^t C
\eea
Therefore, the symplectic-Majorana condition requires the components $\chi_+$ and $\chi_-$ of any spinor to be essentially complex conjugates of one another. As a result, we may just retain the analysis for one, that of the other being given by complex conjugation.

\subsection{Gauging $U(1)\subset SU(2)$}

Gauging a $U(1)$-subgroup of the $SU(2)$ automorphism group of the supersymmetry 
algebra  is achieved by coupling a linear combination $\cA_\mu$ of the Maxwell fields,
\bea
\label{A6}
\cA _M = { 3 \over 2} \, V_I \, A^I _M
\eea
to each $SU(2)$ doublet. Here, $V_I$ is a vector whose components are fixed numerical 
constants independent of the scalar fields $\phi^A$. Minimally coupling each $SU(2)$ 
doublet to $\cA_\mu$ is achieved by using the following covariant derivative, 
\bea
\label{A7}
(\cD _M \lambda^a  )^i = D_M \lambda ^{ia} +  \mg \, \cA_M \delta ^{ij} \lambda ^a _j
\eea
where $\mg$ is the $U(1)$-gauge coupling, $\delta ^{ij}$ acts as a (traceless) generator of $SU(2)$,
and $D_M$ is the covariant derivative with respect to the spin connection $\om_M $,
given by,
\bea
\label{A8}
D_M \lambda _i ^a = \p_M + { 1 \over 4} (\om_M)_{\alpha \beta } \Gamma ^{\alpha \beta} 
\eea
and affine connection when acting the the gravitino field $\psi _{M i}$.

\subsection{The bosonic part of the Lagrangian}

The bosonic part $\cL_0$ of the full gauged supergravity Lagrangian is given as follows,
\bea
 \cL_0 & = & - \half R_g - { 1 \over 4} G_{IJ} (\phi) F^I _{MN} F^{J MN} 
- \half \cG _{AB} (\phi) \p_M \phi ^A \p^M \phi ^B - \mg^2 P(\phi) 
\no \\
&& + { 1 \over 48 \, \sqrt{g} } C_{IJK} \, \ep^{MNPQS} F^I_{MN} F^J _{PQ} A^K _S
\eea
Here, $R_g$ is the Ricci scalar\footnote{Our conventions for the 
Riemann tensor, Ricci tensor,  and Ricci scalar are those of  
\cite{Gunaydin:1983bi,Gunaydin:1984ak}, namely
$R_{MN} {}^P {}_Q
= \p _M \Gamma ^P _{NQ } - \p _N \Gamma ^P _{MQ}
+\Gamma ^P _{MS} \Gamma ^S _{NQ} 
- \Gamma ^P _{NS} \Gamma ^S _{MQ} $ 
along with 
$R_{MQ} =  R_{MP}{}^P{}_Q$ and $R= g^{MQ} R_{MQ}$.} of the metric $g$; the volume form is given by $g=- \det (g_{MN} )$;
and $ \ep ^{MNPQS}/\sqrt{g}$ is the totally anti-symmetric tensor in five-dimensional space-time. 
Gauge invariance under the gauge transformations of the $\mN+1$ Maxwell fields 
requires the totally symmetric tensor $C_{IJK}$ to be constant, namely independent of the scalar fields $\phi^A$.

\sm

The remaining ingredients in the Lagrangian are functions of the scalars $\phi^A$ which 
parametrize an $N$-dimensional Riemannian manifold $\cM$. Given  the constant totally 
symmetric tensor $C_{IJK}$, all these data can be constructed uniquely, up to scalar field 
redefinitions. One embeds $\cM$ into an $\mN+1$-dimensional Riemannian manifold $\cC$  
parametrized by scalars $X^I$ with $I=1,\cdots, \mN+1$, and introduces an auxiliary potential,
\bea
\cV (X) = { 1 \over 6} C_{IJK} X^IX^JX^K
\eea 
The manifold $\cM$ is specified as a hypersurface in $\cC$ by the relation $\cV(X)=1$. 
The scalars $\phi^A$ are  local coordinates on $\cM$, independence of their choice  
being guaranteed by the tensorial structure of the Lagrangian.  The Riemannian metric 
$G_{IJ}$ on $\cC$, and the induced Riemannian metric $\cG_{AB}$ on $\cM$ are 
respectively given by,
\bea
G_{IJ} = - \half { \p^2 \ln \cV \over \p X^I \p X^J} 
\hskip 1in 
\cG_{AB} = G_{IJ} \, \p_A X^I \, \p_B X^J \bigg |_{\cV=1}
\eea
where $\p_A=\p / \p\phi^A$. The notation $G_{IJ} (\phi)$, used in the Lagrangian, 
indicates that $G_{IJ}$ is evaluated at points in the submanifold $\cM$ of $\cC$. 
Throughout, it will be useful to define  a variable $X_I$ dual to $X^I$ by,
\bea
X_I = {1 \over 6} C_{IJK} X^J X^K = { 1 \over 3} { \p \, \cV \over \p X^I} 
\eea
Restricted to  $\cM$ by the condition $\cV=1$, the vector $X_I$ is  normal 
to $\cM$ at the point $X^I |_{\cV=1}$. With the help of this notation $G_{IJ}$
may be calculated explicitly, and we have,
\bea
G_{IJ} =  { 9 \over 2} X_I X_J - \half C_{IJK} X^K 
\eea
It was shown in \cite{Gunaydin:1983bi} that the requirement of positive definiteness of the 
metric $G_{IJ}$ on $\cM$ imposes restrictions on the allowed choices for the constant tensor 
$C_{IJK}$, so that of its $(\mN+1)(\mN+2)(\mN+3)/6$ entries only $\mN(\mN+1)(\mN+2)/6$ can be chosen 
independently. For the case $\mN=2$, of interest to us in this paper, 
these constraints will be very simple, and may be solved by choosing $C_{123}=1$ along 
with its 5 permutations, and all other entries equal to 0. For the discussion of the general 
case, we refer to \cite{Gunaydin:1983bi,Gunaydin:1984ak}.

\sm

From the above considerations, it follows that $ { 3 \over 2}  X_{I} =  G_{IJ} X^J$, 
as well as 
\bea
C^{IJK} & = & \left ( { 3 \over 2} \right )^3 G^{II'} G^{JJ'} G^{KK'} C_{I'J'K'}  =  C_{IJK}  
\no \\
G^{IJ} & = & 2 X^I X^J - 6 C^{IJK} X_K
\eea
Note that due to the last equality on the first line $C^{IJK}$ is a constant symmetric 
tensor just as $C_{IJK}$ is. Finally, the scalar potential $P(\phi)$ occurring in the 
Lagrangian is given by,
\bea
P = -27 \, C^{IJK} \, V_I \, V_J \, X_K 
\eea
which is again a function of $\cM$ in view of the implicit restriction $\cV=1$.

\subsection{Relation with the notations of \cite{Gunaydin:1984ak}}

For completeness, we spell out the relation of our notations with those of 
\cite{Gunaydin:1983bi,Gunaydin:1984ak}, where the metrics $G_{IJ}$ and $\cG_{AB}$ 
are respectively denoted by $\hat a_{IJ}$
and $g_{xy}$, the indices $x,y$ playing the role of the indices $A,B$ here. Furthermore, we have,
\bea
\label{nota}
V_I = {2 \over 3} \, V_I ^{{\rm GST}} \hskip 0.25in & \hskip 1in & X^I  =  \sqrt{{3 \over 2}} \, h^I 
\no \\
C_{IJK} = 4 \, \sqrt{{2 \over 3}} \, C_{IJK}^{{\rm GST}} & \hskip 1in &  X_I = \sqrt{{2 \over 3}}  \, h_I
\eea
where $V_I ^{{\rm GST}}, \, C_{IJK} ^{{\rm GST}}, \, h^I$, and $ h_I$ are the notations used in 
\cite{Gunaydin:1983bi,Gunaydin:1984ak}.

\subsection{The fermionic part of the Lagrangian}

The part of the Lagrangian involving the fermion fields $\psi _{M i}$ and $\lambda _i^a$ 
contains terms bilinear in the fermion fields, and terms of higher order. For the purpose 
of this paper, only the bilinear terms will be needed, and we shall henceforth specialize to 
those, and denote the part of the Lagrangian bilinear in fermions by $\cL_2$. 
From \cite{Gunaydin:1984ak}, it is given as follows,\footnote{In the last term on the last line below, 
we have corrected for a factor of $\mg$ which was missing in \cite{Gunaydin:1984ak}.}
\bea
\label{A18}
\cL_2
& = & 
-\frac{1}{2}\bar{\psi}_M ^{i}\Gamma^{MNP}\cD_N \psi_{P i}
-\frac{1}{2}\bar{\lambda}^{ia} \Gamma ^M \left( \delta^{ab} \cD_M 
+\Omega_{A}^{ab} \p_M \phi^A \right)\lambda_{i}^{b}
-\frac{i}{2}\bar{\lambda}^{ia}\Gamma^M \Gamma^N \psi_{M i}f_A{}^{a}\partial_N \phi^A
\no \\ && 
+ \frac{1}{4}h_{I}^{a}\bar{\lambda}^{ia}\Gamma^M \Gamma^{NP}\psi_{M i}F_{NP}^{I}
+\frac{i}{8 \sqrt{6}}\left( \delta^{ab}h_{I}+ 4 T^{abc} h_{I}^{c}\right ) \bar{\lambda}^{ia}
\Gamma^{MN}\lambda_{i}^{b}F_{MN}^{I}
\no \\ && 
-\frac{3i}{8\sqrt{6}}h_{I} \Big ( \bar{\psi}_M ^{i} F_{PQ}^{I}  \Gamma^{MNPQ}\psi_{N i}
+2\bar{\psi}^{M i}\psi_{i}^N F_{MN}^{I} \Big ) 
\no \\ && 
-\frac{i\sqrt{6}}{8} \mg \, \bar{\psi}_M^{i}\Gamma^{MN}\psi_N^{j}\delta_{ij}P_{0}
-\frac{1}{\sqrt{2}} \, \mg \, \bar{\lambda}^{ia}\Gamma^M \psi_M ^{j}\delta_{ij}P^{a}
+\frac{i}{2\sqrt{6}} \, \mg \, \bar{\lambda}^{ia}\lambda^{jb}\delta_{ij}P^{ab}
 \eea
where we use the index $A$ instead of $x$ used in \cite{Gunaydin:1984ak}. 
Here, $f_A{}^a$ and $\Omega _A ^{ab}$ are respectively the $SO(n+1)$ frame and connection of $\cM$,
and  $P_0, P^a, P^{ab}$ are given as follows, 
\bea
P_{0}=2h^{I} V^{{\rm GST}} _{I}
\quad\quad 
P^a=\sqrt{2}h^{Ia} V^{{\rm GST}}_{I}
\quad\quad 
P^{ab}=\frac{1}{2}\delta^{ab}P_{0}+2\sqrt{2}T^{abc}P^{c}
\eea
where $T^{abc}$ is a covariantly constant tensor on $\cM$. Its proper definition and detailed properties
were analyzed in \cite{Gunaydin:1984ak}, and will not be needed here beyond the case $\mN=2$, 
for which its explicit formulas are given in (\ref{A30}). 

\sm

In terms of the notations of (\ref{nota}), the fermionic 
part of the Lagrangian reads, 
\bea
\label{A20}
\cL_2
& = & 
-\frac{1}{2}\bar{\psi}_M^{i}\Gamma^{MNP}\cD_N \psi_{P i}
-\frac{1}{2}\bar{\lambda}^{ia} \Gamma ^M \left( \delta^{ab} \cD_M 
+\Omega_{A}^{ab} \p_M \phi^A \right) \lambda_{i}^{b}
-\frac{i}{2}\bar{\lambda}^{ia}\Gamma^M \Gamma^N \psi_{M i}f_A{}^{a}\partial_N \phi^A
\no \\ && 
+ \frac{\sqrt{3}}{4\sqrt{2}}X_{I}^{a}\bar{\lambda}^{ia}\Gamma^M
\Gamma^{PQ}\psi_{M i}F_{PQ}^{I}
+\frac{i}{16}\left( \delta^{ab}X_{I}+ 4 T^{abc}X_{I}^{c}\right)\bar{\lambda}^{ia}
\Gamma^{MN}\lambda_{i}^{b}F_{MN}^{I}
\no \\ && 
-\frac{3i}{16}X_{I}\left ( \bar{\psi}_M^{i}\Gamma^{MNPQ}\psi_{N i}
F_{PQ}^{I}+2\bar{\psi}^{M i}\psi_{i}^NF_{MN}^{I}\right ) 
\no \\ && 
-\frac{3i}{4} \, \mg \, \bar{\psi}_M^{i}\Gamma^{MN} \psi_N ^{j}\delta_{ij}V_{I}X^{I}
-\frac{3}{\sqrt{6}} \, \mg \, \bar{\lambda}^{ia}\Gamma^M \psi_M ^{j}\delta_{ij}V_{I}X^{Ia}
+\frac{i}{2} \, \mg \, \bar{\lambda}^{ia}\lambda^{jb}\delta_{ij}P^{ab}
\eea
where,
\bea
\label{A21}
P^{ab}=\frac{\sqrt{6}}{2}\delta^{ab}V_I X^I +2\sqrt{6} V_I T^{abc} X^{Ic} 
\eea
The fields $X_{aI}$ and $X_a^I$ are tangent to $\cM$ and defined as follows, 
\bea
\label{A22}
X_{Ia} = f_a {}^A X_{AI} & \hskip 1in & X_{IA} = + \sqrt{{3 \over 2}} \, \p _A X_I  
\no \\
X_a^I = f_a {}^A X_A^I ~ && X_A^I = - \sqrt{{3 \over 2}} \, \p _A X^I  
\eea
The relations $X_I X^I_A=X^I X_{IA}=0$ follow directly by differentiating $\cV$ along the manifold where
$\cV=1$, and we have the following further relations  as well as,
\bea
G_{IJ} X^I_a X^J_b = { 3 \over 2} \delta _{ab}
\hskip 1in
G_{IJ}  =  { 2 \over 3} \, \left ( X_I X_J + X_I ^a X_J^a \right )
\eea
with analogous relations for $G^{IJ}$. 
For a detailed discussion, we refer to Section 3 of \cite{Gunaydin:1984ak}.
\sm

Lastly, we note that in the Lagrangian (\ref{A20}), one can use either the $i=1,2$ or the $i=+,-$ basis for the $SU(2)$ indices. In the former, we use the standard Kronecker delta, $\delta_{ij}=\text{diag}(1,1)_{ij}$, but in the $i=+,-$ basis we must use the rotated matrix
\begin{align}
\delta_{ij} &=
\begin{pmatrix}
0 & 1\\
1 & 0
\end{pmatrix}
_{ij}
\end{align}

\subsection{Fermion field equations}

The fermion field equations are deduced from the Lagrangian, using the symplectic 
Majorana restrictions $\bar \lambda ^i _a = (\lambda ^i _a)^t C$ and $\bar \psi_M  ^i  = (\psi _M ^i )^t C$. 
Expressing the result in terms of the fields $\lambda _\pm ^a$ and $\psi _{M \pm}$ in the $SU(2)$ basis 
of (\ref{A5}), the equations become $\Psi^M_\pm= \Lambda ^a _\pm=0$ with, 
\bea
\label{A23a}
\Psi_{\pm}^M & = & 
\Gamma^{MNP}\mathcal{D}_N \psi_{P \pm}
+\frac{3i}{8}X_{I}\left(\Gamma^{MNPQ}\psi_{N \pm} F_{PQ}^{I}
+2\psi_{N \pm } F^{I MN}\right)-\frac{i}{2}\Gamma^N \Gamma^M \lambda_{\pm}^{a}f_{A}^{a}\partial_N \phi^{A}
\no \\ &&
-\frac{1}{4}\sqrt{\frac{3}{2}}X_{I}^{a}\Gamma^{PQ}\Gamma^M \lambda _{\pm}^{a}F_{PQ}^{I}
\pm \frac{3}{2} \, \mg \, \Gamma^{MN}\psi_{N \pm}V_{I}X^{I}
\mp \frac{3i}{\sqrt{6}} \, \mg \, \Gamma^M \lambda_{\pm}^{a}V_{I}X^{Ia}
\hskip 0.8in
\eea
and
\bea
\label{A23b}
\Lambda_{\pm}^{a} 
& = & 
\Gamma^M \left( \delta^{ab} \cD_M +\Omega_{A}^{ab} \p_M \phi^A \right) \lambda_\pm^{b}
+\frac{i}{2}\Gamma^M \Gamma^N\psi_{M \pm}f_{A}^{a}\partial_N \phi^{A}
-\frac{1}{4}\sqrt{\frac{3}{2}}X_{I}^{a} F_{PQ}^{I} \Gamma^M \Gamma^{PQ}\psi_{M \pm}
\no \\ && 
-\frac{i}{8} \left(\delta^{ab}X_{I}+4 T^{abc}X_{I}^{c}\right) F_{MN}^{I} \Gamma^{MN}\lambda_{\pm}^{b}
\mp\frac{3i}{\sqrt{6}} \, \mg \, \Gamma^M \psi_{M \pm}V_{I}X^{Ia}
\mp\frac{1}{\sqrt{6}}\, \mg \, \lambda_{\pm}^{b}P^{ab}
\hskip 0.6in
\eea
Expressing the fields $\lambda _\pm ^a$ and $\psi _{M \pm}$ in the $SU(2)$-basis of (\ref{A5})
is responsible for decoupling the field equations with $SU(2)$-index $-$ from those with index $+$.
Furthermore, a reversal of the sign of $\mg$ reverses the indices on the field equations.  
Therefore, without loss of generality, we may restrict attention to the field equations for index $+$.

\subsection{Supersymmetry transformations on fermion fields}

The supersymmetry transformations also decouple in the $SU(2)$-basis of (\ref{A5}) and we get,
\bea
\delta \psi_{M \pm} & = & 
\cD_M \epsilon_{\pm}
+\frac{i}{8}X_{I} F_{NP}^{I} \left ( \Gamma_M {} ^{NP}-4\delta_M{}^N \Gamma^P \right) \epsilon_{\pm}
\mp \frac{1}{2}\, \mg  V_{I}X^{I}\Gamma_M \epsilon_{\pm}
\no \\
\delta \lambda_{\pm}^{a} & = & 
-\frac{i}{2}f_{A}^{a}\Gamma^M \partial_M \phi^{A}\epsilon_{\pm}
+\frac{1}{4}\sqrt{\frac{3}{2}}X_{I}^{a}\Gamma^{MN}F_{MN}^{I}\epsilon_{\pm}
\pm i\sqrt{\frac{3}{2}}\, \mg V_{I}X^{Ia}\epsilon_{\pm}
\eea
As we did  for the field equations, we restrict attention to the supersymmetry transformations with index $+$ and henceforth omit this index from the fields.

\subsection{The special case $\mN=2$}

The truncation to $\mN=2$ was studied in \cite{Cvetic:1999xp}. By performing constant linear transformations 
on the gauge fields $A^I_M$, the symmetric tensor $C_{IJK}$ may be reduced to $C_{123}=1$ 
along with its 5 permutations, all other components being zero. The auxiliary potential $\cV$ then 
reduces to $\cV(X)= X^1 X^2 X^3$ and the scalar manifold is defined by the embedding 
relation $X^1X^2X^3=1$. This relation may be solved explicitly by an exponential parametrization in terms 
of two unconstrained scalar fields $\phi^A=(\phi^1, \phi^2)$. A convenient choice is given by,
\bea
\label{A24}
X^I = e^{- a^I _A \phi ^A }
\hskip 0.6in 
a_1^I = { 1 \over \sqrt{6}}  ( 1, 1, -2 )^I 
\hskip 0.6in 
a_2^I = { 1 \over \sqrt{2}} (1, -1, 0)^I
\eea
Therefore, the manifold $\cM$ of the scalar fields is flat, the induced metric $\cG_{AB}$  is a multiple 
of the  Euclidean metric $\delta _{AB}$, the frame $f_A^a$ is constant, and the connection 
$\Omega _A ^{ab}$ vanishes, 
\bea
\label{A28}
\cG_{AB} = \half \delta _{AB} 
\hskip 0.7in 
f_A ^a = { 1 \over \sqrt{2}} \, \delta _A ^a
\hskip 0.7in 
\Omega _A ^{ab}=0
\eea
Furthermore, in terms of the above parametrization of $\cM$ we have $X_I = (3X^I)^{-1}$ and, 
\bea
G_{IJ} (\phi) & = & {9 \over 2} \,  \delta _{IJ}  (X_I)^2 = { \delta _{IJ} \over 2 (X^I)^2} 
\no \\
P (\phi) & = & - 6 \sum _{I=1}^3 X_I = - 2 \sum _{I=1}^3 { 1 \over X^I}
\eea
Finally, the covariantly constant tensor $T^{abc}$ becomes constant in the 
coordinates $\phi^A$ of (\ref{A24}), since the connection vanishes, and takes the following values,
\bea
\label{A30}
T^{111}=-T^{122}=-{ 1 \over \sqrt{2}}
\eea
along with permutations thereof, with all other components vanishing.

\section{Boundary and counter-terms for fermions}
\label{sec:F}
\setcounter{equation}{0}

In this section, we provide the details for computing the boundary terms required for a well-defined variational principle of the supergravity action with respect to the gravitino fields, and the counter-term required for a finite on-shell action. 
\sm

We begin by regulating the on-shell action by restricting the range of integration in the $r$ coordinate to $r\leq R$, and evaluating the boundary terms at $r=R$, where $R\gg1$ is a  large cutoff parameter. Now given with the most general asymptotic solution to the field equations for the gravitino, written below for convenience,
\begin{align}
\label{5a2_2}
\psi_{\mu} (x, r) &= 
e^{-(\Delta -4) r } \psi _{\mu}^{(0)}(x) + 
\cdots +
e^{- (\Delta -1)  r } \psi _{\mu}^{(3)}(x) + 
r e^{- (\Delta-1)   r } \psi _{\mu}^{(\ln )}(x) + 
\cO(e^{-\Delta  r } )
\end{align}
we expand the regulated on-shell action in a series which schematically takes the form
\begin{align}
S_\text{reg}&= \int\displaylimits_{r=R} d^4x\sqrt{g^{(0)}}
\Big ( e^{2R}a_0 + Ra_\text{ln} + \cO(R^0) \Big )
\end{align}
where $a_0$ and $a_\text{ln}$ are local functions of the sources $\psi_{\mu i}^{(0)}$, and the rest of the terms are finite  in the limit $R \to \infty$.
\sm
 
In the remainder of this Appendix, we shall derive $S_\text{reg}$. The ingredients in its construction will be the bulk action $S_2$ given in the previous Appendix,  plus a boundary term $S_\text{bndy}$ required for a well-defined variational principle, and counter-terms needed for holographic renormalization. The combination of these contributions will then be evaluated on-shell and expanded using (\ref{5a2_2}).  We will not include explicitly here, however,  the terms which cancel the logarithmic divergences  since they do not contribute to terms needed in the calculation of the supercurrent. Finally, we will compute the one-point function of the supercurrent from the finite on-shell action.

\subsection{Boundary terms}

To derive the boundary term, we focus on the kinetic term for the gravitino field in the supergravity action
regularized by a cutoff $r\leq R$, 
\bea
S_2 = \frac{1}{8\pi G_5}\int \displaylimits_{r \leq R} d^5x\sqrt{g}\left(-\frac{1}{2}\bar{\psi}_M^i\Gamma^{MNP}\cD_N\psi_{P i} + \cdots\right)
\eea
Here, the indices $i=\pm$ represent the  $SU(2)$ indices (see Appendix \ref{sec:A}),
and the ellipses stand for all the remaining terms in the supergravity action resulting from (\ref{A18}). 
The boundary contribution to the variation of this action results from the $N=r$ term,
\bea
S_2 = \frac{1}{8 \pi G_5}\int \displaylimits_{r \leq R} d^5x\sqrt{g}
\left( - \half \bar{\psi}_\mu^i\Gamma^{\mu r\rho}\p_r\psi_{\rho i} + \cdots\right)
\eea
Evaluating the action on-shell, its contribution is now entirely given by, 
\bea
\label{F5}
\delta S_2 &=&  
-\frac{1}{16\pi G_5}\int \displaylimits_{r=R} d^4x\sqrt{g}\,\bar{\psi}_\mu^i\Gamma^{\mu r\nu}\delta\psi_{\nu i}
\no\\
&=& 
-\frac{i}{16\pi G_5}\int \displaylimits_{r=R} d^4x\sqrt{g}
\Big ( \bar{\psi}_{\mu+}\Gamma^{\mu\nu r}\delta\psi_{\nu-} - \bar{\psi}_{\mu-}\Gamma^{\mu\nu r}\delta\psi_{\nu+}\Big ) 
\eea
where we have lowered all $SU(2)$ indices using (\ref{A5_2}).

\sm 

To define the supercurrent, $S^\mu_\pm$, we must vary the action with respect to the source of the supercurrent, $\psi_{\mu\pm}^{(0)}$. However, for a well-defined variational principle, we can only vary the action with respect to half of the components of  $\psi_{\mu\pm}^{(0)}$. From the symplectic-Majorana condition, (\ref{A5_2}), one can easily show that,
$$
(I \pm \Gamma^r)\chi_+ = 0 \quad\quad \Rightarrow \quad\quad (I \mp \Gamma^r)\chi_- = 0
$$
Therefore, we will vary the action with respect to the components of $\psi_{\mu\pm}^{(0)}$ satisfying,
\bea
(I-\Gamma^r)\delta\psi_{\mu+}^{(0)} &= & 0
\no\\
(I+\Gamma^r)\delta\psi_{\mu-}^{(0)} &= & 0
\eea
and remove the remaining variations with a boundary term. To cancel the unwanted variations in (\ref{F5}), we add a boundary term to the action given by,
\begin{align}
S_\text{bndy} &= 
\frac{i}{32\pi G_5}\int \displaylimits_{r=R} d^4x\sqrt{g}
\left(\bar{\psi}_{\mu+}\Gamma^{\mu\nu}\psi_{\nu-} + \bar{\psi}_{\mu-}\Gamma^{\mu\nu}\psi_{\nu+}\right)
\no\\
&= \frac{i}{32\pi G_5}\int \displaylimits_{r=R} d^4x\sqrt{g}\,\bar{\psi}_{\mu}^i\Gamma^{\mu\nu}\psi_{\nu}^j\delta_{ij}
\end{align}
where the last line is written to show that this term has the same structure as the mass term in the Lagrangian (\ref{A20}). Adding this to $S_2$, the variation is
\begin{align}
\delta(S_2 + S_\text{bndy}) &= \frac{i}{16\pi G_5}\int \displaylimits_{r=R} d^4x\sqrt{g}
\Big ( \bar{\psi}_{\mu+}\Gamma^{\mu\nu}(I-\Gamma^r)\delta\psi_{\nu-} + \bar{\psi}_{\mu-}\Gamma^{\mu\nu}(I+\Gamma^r)\delta\psi_{\nu+} \Big )
\end{align} 
It is now clear from the structure of this combined on-shell action that half of the variations are cancelled out
by the addition of the boundary term.

\subsection{Counter-terms}

Expanding the modified on-shell action, $S_\text{reg}=S_2+S_{\text{bndy}}$, using (\ref{5a2_2}) and regulating the integral at $r=R\gg 1$, we find that the bulk action $S_2$ vanishes and the leading order part is given by the boundary term,
\begin{align}
S_\text{reg} &= \frac{i}{16\pi G_5}\int\displaylimits_{r=R} d^4x\sqrt{g^{(0)}}\left(e^{2R}\bar{\psi}_{\mu}^{+(1)}\Gamma^{\mu\nu}\psi_{\nu}^{-(0)} +  e^{2R}\bar{\psi}_{\mu}^{-(1)}\Gamma^{\mu\nu}\psi_{\nu}^{+(0)} + \cO(R)\right)
\end{align} 
We will ignore the logarithmic, $\cO(R)$, divergences here because we only want to compute the finite terms of the variation of the full action, and the counter-terms which cancel the logarithmic terms do not contribute to the finite terms. To cancel the divergent terms above, we try a counter-term given by \cite{Argurio:2014uca}
\begin{align}
S_\text{ct}&= \frac{1}{32\pi G_5}\int\displaylimits_{r=R} d^4x\sqrt{g}\left(\bar{\psi}_\mu^i\Gamma^{\mu\nu\rho}\partial_\nu\psi_{\rho i} + R\cF[\psi,\lambda,\phi,\ldots]\right)\no\\
 &= \frac{i}{16\pi G_5}\int\displaylimits_{r=R} d^4x\sqrt{g^{(0)}} \left(-e^{2R}\bar{\psi}_{\mu}^{+(1)}\Gamma^{\mu\nu}\psi_{\nu}^{-(0)} -  e^{2R}\bar{\psi}_{\mu}^{-(1)}\Gamma^{\mu\nu}\psi_{\nu}^{+(0)} - \bar{\psi}_{\mu}^{+(1)}\Gamma^{\mu\nu}\psi_{\nu}^{-(2)}\right.\no\\
 &\hspace{.75in} \left.-  \bar{\psi}_{\mu}^{-(1)}\Gamma^{\mu\nu}\psi_{\nu}^{+(2)}  +  \bar{\psi}_{\mu}^{+(3)}\Gamma^{\mu\nu}\psi_{\nu}^{-(0)}  +  \bar{\psi}_{\mu}^{-(3)}\Gamma^{\mu\nu}\psi_{\nu}^{+(0)} + R\cF + \cO(e^{-2R})\right)
\end{align} 
where $\cF[\psi,\lambda,\phi,\ldots]$ are the terms needed to cancel out the logarithmic divergences in the full on-shell action. To go from the first equality to the second, we use the field equations, (\ref{A23a}), to rewrite the first term in terms of derivatives with respect to $r$ rather than with respect to boundary coordinates. We see in the first line of the second equality above that the $\cO(e^{2R})$ divergences cancel, and the full renormalized on-shell action is given by
\begin{align}
S_{\text{sugra}} &= \lim_{R\rightarrow\infty} S_\text{reg}
\no\\
 &= \frac{i}{16\pi G_5}\int d^4x\sqrt{g^{(0)}}
 \left(\bar{\psi}_{\mu}^{+(3)}\Gamma^{\mu\nu}\psi_{\nu}^{-(0)}  
 +  \bar{\psi}_{\mu}^{-(3)}\Gamma^{\mu\nu}\psi_{\nu}^{+(0)} + \text{local}\right)
\end{align}
All divergences are cancel out, higher order terms vanish in the $R\rightarrow\infty$ limit, and what remains are local terms proportional to the source, $\psi^{(0)}_{\mu i}$, as shown in Appendix \ref{sec:C}, and non-local terms proportional to $\psi^{(3)}_{\mu i}$.

\subsection{Extracting the supercurrent}

To obtain the supercurrent from the definition (\ref{eq:var_Su}), we vary the sources in the renormalized action $S_{\text{sugra}}$ to get
\begin{align}
\delta S_{\text{sugra}} &= \frac{i}{16\pi G_5}\int d^4x\sqrt{g^{(0)}}\left(\delta\bar{\psi}_{\mu}^{+(0)}\Gamma^{\mu\nu}\psi_{\nu}^{-(3)}  +  \delta\bar{\psi}_{\mu}^{-(0)}\Gamma^{\mu\nu}\psi_{\nu}^{+(3)} + \text{local}\right)
\end{align}
Comparing this to (\ref{eq:var_Su}), we get
\begin{align}
8\pi G_5 (S^{\nu+})^tC &= - (\psi_\mu^{+(3)})^tC\Gamma^{\mu\nu} + \text{local}\no\\
8\pi G_5 (S^{\nu-})^tC &= -(\psi_\mu^{-(3)})^tC\Gamma^{\mu\nu} + \text{local}
\end{align}
Solving for the supercurrents by using the transposition properties of the gamma matrices, we obtain
the final result of this Appendix, 
\begin{align}
8\pi G_5 S^{\mu}_+ &= -\Gamma^{\mu\nu}\psi_{\nu+}^{(3)} + \text{local}
\no\\
8\pi G_5 S^{\mu}_- &= + \Gamma^{\mu\nu}\psi_{\nu-}^{(3)} + \text{local}
\end{align}

\section{Asymptotic expansion of the Fermi fields}
\label{sec:C}
\setcounter{equation}{0}

In this section we will show some of the details of the asymptotic
expansion of the Fermi fields. The expansion is done similarly
to section 5 of \cite{Gauntlett:2011wm}. The fields at $r\rightarrow\infty$
take an asymptotic form given by,
\bea
\psi_{\hat\mu} & = & e^{-\frac{r}{2}}\psi_{\hat\mu}^{(0)} + e^{-\frac{3r}{2}}\psi_{\hat\mu}^{(1)} + e^{-\frac{5r}{2}}\psi_{\hat\mu}^{(2)} + e^{-\frac{7r}{2}}\psi_{\hat\mu}^{(3)} + re^{-\frac{7r}{2}}\psi_{\hat\mu}^{(\rm {ln})} + \cdots\nonumber\\
\psi_{\hat r} & = & e^{-\frac{3r}{2}}\psi_{\hat r}^{(1)}+e^{-\frac{5r}{2}}\psi_{\hat r}^{(2)}+e^{-\frac{7r}{2}}\psi_{\hat r}^{(3)}+e^{-\frac{9r}{2}}\psi_{\hat r}^{(4)}+re^{-\frac{9r}{2}}\psi_{\hat r}^{(\rm {ln})}+\cdots\nonumber\\
\lambda^{a} & = & e^{-\frac{3r}{2}}\lambda^{a(1)}+e^{-\frac{5r}{2}}\lambda^{a(2)}+re^{-\frac{5r}{2}}\lambda^{a(\rm {ln})}+e^{-\frac{7r}{2}}\lambda^{a(3)}+\cdots
\eea
where the frame indices $\hat\mu$ denote the boundary coordinates
$\{x^{+},x^{-},x^{u},x^{v}\}$. We can constrain the coefficients order
by order in equations (\ref{A23a}) and (\ref{A23b}),
where the bosonic fields are set to the interpolating magnetic brane solution to the BPS
equations described in section \ref{sec:23}. We denote their asymptotic behavior in the following way,
\bea
f_{A}^{a}\phi^{A\prime} & = & f_{0}^{a}e^{-2r}+g_{0}^{a}re^{-2r}+\mathcal{O}(e^{-4r})
\nonumber\\
X_{I}^{a} & = & X_{I}^{a(0)}+X_{I}^{a(1)}e^{-2r}+X_{I}^{a(\rm {ln})}re^{-2r}+\mathcal{O}(e^{-4r})
\nonumber\\
W & = & r+\ln{W_0}+\mathcal{O}(e^{-4r})
\nonumber\\
U & = & r+\ln{U_0}+\mathcal{O}(e^{-4r})
\eea
These expansion coefficients can be computed explicitly from the asymptotic form of $\phi^{A}$. Note
that we have not chosen a gauge for $\psi_{\mu}$.  To leading order, $e^{-\frac{r}{2}}$, we find that,
\bea
(I-\Gamma^{r})\psi_{\hat\mu}^{(0)}=0
\eea
At the next order, $e^{-\frac{3r}{2}}$, we have the projection conditions,
\begin{align}
(I+\Gamma^{r})\psi_{\hat\mu}^{(1)} &=0 & (I-\Gamma^{r})\psi_{r}^{(1)} &= 0 & (I-\Gamma^{r})\lambda^{a(1)} &= 0
\end{align}
and find that the coefficients are determined by the $\psi_{\mu}^{(0)}$ data,
\begin{align}
\Gamma_{\hat\pm}\psi_{\hat\pm}^{(1)} 
& =  -\frac{i}{2}p_{\pm}\Gamma_{\hat\pm}\Gamma^{\alpha}\psi_{\alpha}^{(0)}
\nonumber\\
\Gamma_{\hat u}\psi_{\hat u}^{(1)} 
& =  \frac{i}{2}p_{m}\Gamma^{\hat m}\Gamma_{\hat u}\psi_{\hat u}^{(0)}
\nonumber\\
\Gamma_{\hat v}\psi_{\hat v}^{(1)} 
& =  \frac{i}{2}p_{m}\Gamma^{\hat m}\Gamma_{\hat v}\psi_{\hat v}^{(0)}
\nonumber\\
\Gamma^{\alpha}\psi_{\alpha}^{(1)} 
& =  \frac{i}{3}\Gamma^{\hat + -}(p_{+}\psi_{-}^{(0)}-p_{-}\psi_{+}^{(0)})-\frac{i}{6}p_{m}\Gamma^{\hat m\alpha}\psi_{\alpha}^{(0)}
\nonumber\\
\Gamma^{m}\psi_{m}^{(1)} & =  -\frac{2i}{3}\Gamma^{\hat +-}(p_{+}\psi_{-}^{(0)}-\psi_{-}\psi_{+}^{(0)})-\frac{i}{6}p_{m}\Gamma^{\hat m\alpha}\psi_{\alpha}^{(0)}
\end{align}
where, as stated in the main text, $m=+,-$ and $\alpha=u,v$. 
At the next two orders, $e^{-\frac{5r}{2}}$ and $re^{-\frac{5r}{2}}$, we have,
\begin{align}
(I-\Gamma^{r})\psi_{\hat\mu}^{(2)} &= 0 & (I+\Gamma^{r})\psi_{r}^{(2)} &= 0
\no\\
(I+\Gamma^{r})\lambda^{a(2)} &= 0 & (I+\Gamma^{r})\lambda^{a(\rm {ln})} &= 0
\end{align}
The coefficients, up to a gauge transformation, are determined by
the source data, $\psi_{\hat\mu}^{(0)}$ and $\lambda^{a(1)}$.
Here we write them in terms of the data at the previous order,
\begin{align}
\Gamma^{\alpha}\psi_{\alpha}^{(2)}+\Gamma^{r}\psi_{r}^{(2)} 
& =  \frac{i}{3}\Gamma^{\hat +-}(p_{+}\psi_{-}^{(1)}-p_{-}\psi_{+}^{(1)}) - \frac{i}{3}p_{m}\Gamma^{\hat m\alpha}\psi_{\alpha}^{(1)}
\\
\Gamma^{m}\psi_{m}^{(2)}+\Gamma^{r}\psi_{r}^{(2)} 
& =  -\frac{2i}{3}\Gamma^{\hat +-}(p_{+}\psi_{-}^{(1)}-p_{-}\psi_{+}^{(1)}) + \frac{i}{6}p_{m}\Gamma^{\hat m}\psi_{r}^{(1)} - \frac{i}{2}p_{m}\Gamma^{\hat m\alpha}\psi_{\alpha}^{(1)}
\nonumber\\
\Gamma_{\hat\pm}\psi_{\hat\pm}^{(2)} 
& =  -\frac{i}{2}p_{\pm}\Gamma_{\hat\pm}(\psi_{r}^{(1)}+\Gamma^{\alpha}\psi_{\alpha}^{(1)})
\nonumber\\
\Gamma_{\hat u}\psi_{\hat u}^{(2)} 
& =  -\frac{i}{2}p_{m}\Gamma^{\hat m}\Gamma_{\hat u}\psi_{\hat u}^{(1)}
\nonumber\\
\Gamma_{\hat v}\psi_{\hat v}^{(2)} 
& =  -\frac{i}{2}p_{m}\Gamma^{\hat m}\Gamma_{\hat v}\psi_{\hat v}^{(1)}
\nonumber\\
\lambda^{a(\rm {ln})} 
& =  \frac{i}{}p_{m}\Gamma^{\hat m}\lambda^{a(1)} + \frac{i}{2}(\Gamma^{m}\psi_{m}^{(0)}+\Gamma^{\alpha}\psi_{\alpha}^{(0)})f_{0}^{a}
\nonumber\\
 &   -\frac{1}{2U_0^2}\sqrt{\frac{3}{2}}Bq^{I}X_{I}^{a(0)}\Gamma^{\hat u \hat v}(\Gamma^{m}\psi_{m}^{(0)}-\Gamma^{\alpha}\psi_{\alpha}^{(0)}) - i\sqrt{\frac{3}{2}}V_{I}X_{a}^{I(1)}(\Gamma^{m}\psi_{m}^{(0)}+\Gamma^{\alpha}\psi_{\alpha}^{(0)})
\end{align}
The coefficient $\psi_{r}^{(2)}$ can be gauged away to obtain a unique constraint on $\Gamma^{\alpha}\psi_{\alpha}^{(2)}$ and $\Gamma^{m}\psi_{m}^{(2)}$.


\newpage

\begin{thebibliography}{99}
\itemsep=0in

\bibitem{Aharony:1999ti}
  O.~Aharony, S.~S.~Gubser, J.~M.~Maldacena, H.~Ooguri and Y.~Oz,
  ``Large N field theories, string theory and gravity,''
  Phys.\ Rept.\  {\bf 323}, 183 (2000)
  [arXiv:hep-th/9905111].

\bibitem{D'Hoker:2002aw}
  E.~D'Hoker, D.~Z.~Freedman,
  ``Supersymmetric gauge theories and the AdS / CFT correspondence,''
  [hep-th/0201253].
  
\bibitem{Hartnoll:2009sz}
  S.~A.~Hartnoll,
  ``Lectures on holographic methods for condensed matter physics,''
  Class.\ Quant.\ Grav.\  {\bf 26}, 224002 (2009)
  [arXiv:0903.3246 [hep-th]].
  
\bibitem{Herzog:2009xv}
  C.~P.~Herzog,
  ``Lectures on Holographic Superfluidity and Superconductivity,''
  J.\ Phys.\ A A {\bf 42}, 343001 (2009)
  [arXiv:0904.1975 [hep-th]].
  
\bibitem{Giamarchi}
T.~Giamarchi, {\sl Quantum Physics in One Dimension}, Oxford University Press (2004).
  
\bibitem{D'Hoker:2009mm}
  E.~D'Hoker and P.~Kraus,
  ``Magnetic Brane Solutions in AdS,''
  JHEP {\bf 0910}, 088 (2009)
  [arXiv:0908.3875 [hep-th]].
  
\bibitem{D'Hoker:2012ih} 
  E.~D'Hoker and P.~Kraus,
 ``Quantum Criticality via Magnetic Branes,''
  Lect.\ Notes Phys.\  {\bf 871}, 469 (2013)
  [arXiv:1208.1925 [hep-th]].
  
\bibitem{Gauntlett:2007ma}
  J.~P.~Gauntlett and O.~Varela,
  ``Consistent Kaluza-Klein Reductions for General Supersymmetric AdS Solutions,''
  Phys.\ Rev.\  D {\bf 76} (2007) 126007
  [arXiv:0707.2315 [hep-th]].
  
\bibitem{Gauntlett:2006ai} 
J.~P.~Gauntlett, E.~O Colgain and O.~Varela,
  ``Properties of some conformal field theories with M-theory duals,''
  JHEP {\bf 0702}, 049 (2007)
  doi:10.1088/1126-6708/2007/02/049
  [hep-th/0611219].
  
\bibitem{Albash:2008eh} 
  T.~Albash and C.~V.~Johnson,
  ``A Holographic Superconductor in an External Magnetic Field,''
  JHEP {\bf 0809}, 121 (2008)
  [arXiv:0804.3466 [hep-th]].
  
\bibitem{Albash:2009ix} 
  T.~Albash and C.~V.~Johnson,
  ``Phases of Holographic Superconductors in an External Magnetic Field,''
  arXiv:0906.0519 [hep-th].
  
\bibitem{Brown:1986nw}
  J.~D.~Brown and M.~Henneaux,
  ``Central Charges in the Canonical Realization of Asymptotic Symmetries: An
  Example from Three-Dimensional Gravity,''
  Commun.\ Math.\ Phys.\  {\bf 104}, 207 (1986).
  
\bibitem{D'Hoker:2010hr}
  E.~D'Hoker, P.~Kraus and A.~Shah,
  ``RG Flow of Magnetic Brane Correlators,''
  JHEP {\bf 1104}, 039 (2011)
  [arXiv:1012.5072 [hep-th]].
  
\bibitem{Almuhairi:2010rb}
  A.~Almuhairi,
  ``AdS$_3$ and AdS$_2$ Magnetic Brane Solutions,''
  arXiv:1011.1266 [hep-th].

\bibitem{Gunaydin:1983bi} 
  M.~Gunaydin, G.~Sierra and P.~K.~Townsend,
  ``The Geometry of N=2 Maxwell-Einstein Supergravity and Jordan Algebras,''
  Nucl.\ Phys.\ B {\bf 242}, 244 (1984).

\bibitem{Gunaydin:1984ak} 
  M.~Gunaydin, G.~Sierra and P.~K.~Townsend,
  ``Gauging the d = 5 Maxwell-Einstein Supergravity Theories: More on Jordan Algebras,''
  Nucl.\ Phys.\ B {\bf 253}, 573 (1985).
  
\bibitem{Cacciatori:2003kv} 
  S.~L.~Cacciatori, D.~Klemm and W.~A.~Sabra,
  ``Supersymmetric domain walls and strings in D = 5 gauged supergravity coupled to vector multiplets,''
  JHEP {\bf 0303}, 023 (2003)
  [hep-th/0302218].
  
\bibitem{Almuhairi:2011ws}
  A.~Almuhairi and J.~Polchinski,
  ``Magnetic AdS$\times  R^2$: Supersymmetry and stability,''
  arXiv:1108.1213 [hep-th].
    
\bibitem{Donos:2011pn}
  A.~Donos, J.~P.~Gauntlett and C.~Pantelidou,
  ``Magnetic and electric AdS solutions in string- and M-theory,''
  arXiv:1112.4195 [hep-th].


\bibitem{Balasubramanian:1998sn} 
  V.~Balasubramanian, P.~Kraus and A.~E.~Lawrence,
  ``Bulk versus boundary dynamics in anti-de Sitter space-time,''
  Phys.\ Rev.\ D {\bf 59}, 046003 (1999)
  doi:10.1103/PhysRevD.59.046003
  [hep-th/9805171].
  
\bibitem{Balasubramanian:1999re}
  V.~Balasubramanian and P.~Kraus,
  ``A stress tensor for anti-de Sitter gravity,''
  Commun.\ Math.\ Phys.\  {\bf 208}, 413 (1999)
  [arXiv:hep-th/9902121].
  
  
  
\bibitem{Balasubramanian:1999jd}
  V.~Balasubramanian and P.~Kraus,
  ``Space-time and the holographic renormalization group,''
  Phys.\ Rev.\ Lett.\  {\bf 83}, 3605 (1999)
  [hep-th/9903190].

\bibitem{deHaro:2000vlm} 
  S.~de Haro, S.~N.~Solodukhin and K.~Skenderis,
  ``Holographic reconstruction of space-time and renormalization in the AdS / CFT correspondence,''
  Commun.\ Math.\ Phys.\  {\bf 217}, 595 (2001)
  doi:10.1007/s002200100381
  [hep-th/0002230].
  
\bibitem{Skenderis:2002wp} 
  K.~Skenderis,
  ``Lecture notes on holographic renormalization,''
  Class.\ Quant.\ Grav.\  {\bf 19}, 5849 (2002)
  doi:10.1088/0264-9381/19/22/306
  [hep-th/0209067].
  
\bibitem{Cvetic:1999xp} 
  M.~Cvetic, M.~J.~Duff, P.~Hoxha, J.~T.~Liu, H.~Lu, J.~X.~Lu, R.~Martinez-Acosta 
  and C.~N.~Pope {\it et al.},
``Embedding AdS black holes in ten-dimensions and eleven-dimensions,''
  Nucl.\ Phys.\ B {\bf 558}, 96 (1999)
  [hep-th/9903214].
  
\bibitem{Kraus:2006wn} 
  P.~Kraus,
``Lectures on black holes and the AdS(3) / CFT(2) correspondence,''
  Lect.\ Notes Phys.\  {\bf 755}, 193 (2008)
  [hep-th/0609074].


\bibitem{Kraus:2007vu} 
  P.~Kraus, F.~Larsen and A.~Shah,
  ``Fundamental Strings, Holography, and Nonlinear Superconformal Algebras,''
  JHEP {\bf 0711}, 028 (2007)
  [arXiv:0708.1001 [hep-th]].


\bibitem{Iqbal:2009fd} 
  N.~Iqbal and H.~Liu,
  ``Real-time response in AdS/CFT with application to spinors,''
  Fortsch.\ Phys.\  {\bf 57}, 367 (2009)
  [arXiv:0903.2596 [hep-th]].

\bibitem{Gauntlett:2011wm} 
  J.~P.~Gauntlett, J.~Sonner and D.~Waldram,
  ``Spectral function of the supersymmetry current,''
  JHEP {\bf 1111}, 153 (2011)
  [arXiv:1108.1205 [hep-th]].

  \bibitem{Argurio:2014uca} 
  R.~Argurio, M.~Bertolini, D.~Musso, F.~Porri and D.~Redigolo,
  ``Holographic Goldstino,''
  Phys.\ Rev.\ D {\bf 91}, no. 12, 126016 (2015)
  [arXiv:1412.6499 [hep-th]].

\bibitem{Banados:1998pi} 
  M.~Banados, K.~Bautier, O.~Coussaert, M.~Henneaux and M.~Ortiz,
 ``Anti-de Sitter / CFT correspondence in three-dimensional supergravity,''
  Phys.\ Rev.\ D {\bf 58}, 085020 (1998)
  doi:10.1103/PhysRevD.58.085020
  [hep-th/9805165].

\bibitem{Bautier:1999ds} 
  K.~Bautier,
  ``AdS(3) asymptotic (super)symmetries,''
  hep-th/9909097.
  
\bibitem{Hyakutake:2012uv} 
  Y.~Hyakutake,
  ``Super Virasoro Algebra From Supergravity,''
  Phys.\ Rev.\ D {\bf 87}, 045028 (2013)
  doi:10.1103/PhysRevD.87.045028
  [arXiv:1211.3547 [hep-th]].

\bibitem{Hyakutake:2015qua} 
  Y.~Hyakutake,
``Super Virasoro Algebras From Chiral Supergravity,''
  Universe {\bf 1}, no. 2, 292 (2015)
  doi:10.3390/universe1020292
  [arXiv:1507.00486 [hep-th]].

  \bibitem{Polchinski:2005book}
J.   Polchinski, {\sl  String Theory},  Vol. 2. 
 Cambridge University Press, 1998, page 47

\bibitem{D'Hoker:2009bc} 
  E.~D'Hoker and P.~Kraus,
  ``Charged Magnetic Brane Solutions in AdS (5) and the fate of the third law of thermodynamics,''
  JHEP {\bf 1003}, 095 (2010)
  doi:10.1007/JHEP03(2010)095
  [arXiv:0911.4518 [hep-th]].
  
\bibitem{Donos:2011qt}
  A.~Donos, J.~P.~Gauntlett and C.~Pantelidou,
  ``Spatially modulated instabilities of magnetic black branes,''
  JHEP {\bf 1201}, 061 (2012)
  [arXiv:1109.0471 [hep-th]].

\bibitem{Jensen:2010vd}
  K.~Jensen, A.~Karch and E.~G.~Thompson,
  ``A Holographic Quantum Critical Point at Finite Magnetic Field and Finite Density,''
  arXiv:1002.2447 [hep-th].

\bibitem{D'Hoker:2010rz}
  E.~D'Hoker and P.~Kraus,
  ``Holographic Metamagnetism, Quantum Criticality, and Crossover Behavior,''
  JHEP {\bf 1005}, 083 (2010)
  [arXiv:1003.1302 [hep-th]].

\bibitem{D'Hoker:2010ij}
  E.~D'Hoker and P.~Kraus,
  ``Magnetic Field Induced Quantum Criticality via new Asymptotically AdS$_5$ Solutions,''
  Class.\ Quant.\ Grav.\  {\bf 27}, 215022 (2010)
  [arXiv:1006.2573 [hep-th]].

\bibitem{D'Hoker:2011xw}
  E.~D'Hoker and P.~Kraus,
  ``Charged Magnetic Brane Correlators and Twisted Virasoro Algebras,''
  Phys.\ Rev.\ D {\bf 84}, 065010 (2011)
  [arXiv:1105.3998 [hep-th]].


\end{thebibliography}
\end{document}